\documentclass[final,1p,times]{elsarticle}

\usepackage[cmex10]{amsmath}
\usepackage{amsthm}

\usepackage{graphicx}
\usepackage{epstopdf}

\usepackage{algorithm} 
\usepackage{algpseudocode}

\usepackage{array}

\usepackage{stfloats}

\usepackage[caption=false,font=footnotesize]{subfig}

\usepackage{hyperref}

\journal{Journal of Parallel and Distributed Computing}

\begin{document}

\begin{frontmatter}



\title{A Framework for General Sparse Matrix-Matrix Multiplication on GPUs and Heterogeneous Processors\tnoteref{labeltitlenote1}
\tnoteref{labeltitlenote2}}
\tnotetext[labeltitlenote1]{This is an extended version of paper ``An Efficient GPU General Sparse Matrix-Matrix Multiplication for Irregular Data''~\cite{Liu:An} published at IPDPS '14.}
\tnotetext[labeltitlenote2]{The source code of this work is available at \url{https://github.com/bhSPARSE/Benchmark_SpGEMM_using_CSR}.}




\author[labeladd]{Weifeng Liu\corref{cor1}}
\cortext[cor1]{Corresponding author}
\ead{weifeng.liu@nbi.ku.dk}
\author[labeladd]{Brian Vinter}
\ead{vinter@nbi.ku.dk}
\address[labeladd]{Niels Bohr Institute, University of Copenhagen, Blegdamsvej 17, 2100 Copenhagen, Denmark}

\begin{abstract}
General sparse matrix-matrix multiplication (SpGEMM) is a fundamental building block for numerous applications such as algebraic multigrid method (AMG), breadth first search and shortest path problem. Compared to other sparse BLAS routines, an efficient parallel SpGEMM implementation has to handle extra irregularity from three aspects: (1) the number of nonzero entries in the resulting sparse matrix is unknown in advance, (2) very expensive parallel insert operations at random positions in the resulting sparse matrix dominate the execution time, and (3) load balancing must account for sparse data in both input matrices. 

In this work we propose a framework for SpGEMM on GPUs and emerging CPU-GPU heterogeneous processors. This framework particularly focuses on the above three problems. Memory pre-allocation for the resulting matrix is organized by a hybrid method that saves a large amount of global memory space and efficiently utilizes the very limited on-chip scratchpad memory. Parallel insert operations of the nonzero entries are implemented through the GPU merge path algorithm that is experimentally found to be the fastest GPU merge approach. Load balancing builds on the number of necessary arithmetic operations on the nonzero entries and is guaranteed in all stages.

Compared with the state-of-the-art CPU and GPU SpGEMM methods, our approach delivers excellent absolute performance and relative speedups on various benchmarks multiplying matrices with diverse sparsity structures. Furthermore, on heterogeneous processors, our SpGEMM approach achieves higher throughput by using re-allocatable shared virtual memory. 

\end{abstract}


\end{frontmatter}

\section{Introduction}

General matrix-matrix multiplication (GEMM) is one of the most crucial operations in computational science and modeling. The operation multiplies a matrix $A$ of size $m\times k$ with a matrix $B$ of size $k\times n$ and gives a resulting matrix $C$ of size $m\times n$. In many linear solvers and graph problems such as algebraic multigrid method (AMG)~\cite{Bell:Exposing}, breadth first search~\cite{Gilbert:High}, finding shortest path~\cite{Chan:More}, colored intersection~\cite{Kaplan:Colored} and sub-graphs~\cite{Vassilevska:Finding}, 
it is required to exploit sparsity of the two input matrices and the resulting matrix because their dense forms normally need huge storage space and computation cost for the zero entries. Therefore general sparse matrix-matrix multiplication (SpGEMM) becomes a common building block in these applications.

Compared to CPUs, modern graphics processing units (GPUs) promise much higher peak floating-point performance and memory bandwidth. Thus a lot of research has concentrated on GPU accelerated sparse matrix-dense vector multiplication~\cite{Bell:Implementing, Liu:CSR5, Liu:Speculative} and sparse matrix-dense matrix multiplication~\cite{Ortega:FastSpMM, Vazquez:Fast} and achieved relatively attractive performance. However, despite the prior achievements on these GPU sparse BLAS routines, massive parallelism in GPUs is still significantly underused for the SpGEMM algorithm, because it has to handle three more challenging problems: (1) the number of nonzero entries in the resulting matrix is unknown in advance, (2) very expensive parallel insert operations at random positions in the resulting matrix dominate the execution time, and (3) load balancing must account for sparse data in both input matrices with diverse sparsity structures.

Previous GPU SpGEMM methods~\cite{Bell:Exposing, Demouth:Sparse, NVIDIA:CUSPARSE, Dalton:CUSP, Dalton:Optimizing, Gremse:GPU} have proposed a few solutions for the above problems and demonstrated relatively good time and space complexity. However, the experimental results showed that they either only work best for fairly regular sparse matrices~\cite{Demouth:Sparse, NVIDIA:CUSPARSE, Gremse:GPU}, or bring extra high memory overhead for matrices with some specific sparsity structures~\cite{Bell:Exposing, Dalton:CUSP, Dalton:Optimizing}. Moreover, in the usual sense, none of these methods can constantly outperform well optimized SpGEMM approach~\cite{Intel:MKL} for multicore CPUs. 

Our work described in this paper particularly focuses on improving GPU SpGEMM performance for matrices with arbitrary irregular sparsity structures by proposing more efficient methods to solve the above three problems on GPUs and emerging CPU-GPU heterogeneous processors. 

In this paper, we make the following contributions:

\begin{itemize}

\item \textbf{\textit{A framework for fast SpGEMM}}. We design a 4-stage framework for implementing SpGEMM on manycore platforms including homogeneous GPUs and heterogeneous processors composed of CPU cores, GPU cores and shared virtual memory. This framework effectively organizes memory allocation, load balancing and GPU kernel launches. 

\item \textbf{\textit{A hybrid method for the resulting matrix pre-allocation}}. We present a hybrid method that initially allocates memory of upper bound size for short rows and progressively allocates memory for long rows. The experimental results show that our method saves a large amount of global memory space and efficiently utilizes the very limited on-chip scratchpad memory.

\item \textbf{\textit{Parallel insert operations through fast merging}}. We propose an efficient parallel insert method for long rows of the resulting matrix by using the fastest merge algorithm available on GPUs. We make an experimental evaluation and choose GPU merge path algorithm from five candidate GPU merge approaches.

\item \textbf{\textit{Heuristic-based load balancing}}. We develop a load balancing oriented heuristic method that assigns rows of the resulting matrix to multiple bins with different subsequent computational methods. Our approach guarantees load balancing in all calculation steps.


\end{itemize}

Our framework and corresponding algorithms delivers excellent performance in two experimental scenarios: (1) calculating triple matrix Galerkin products (i.e., $P^TAP$) in AMG for 2D and 3D Poisson problems, and (2) computing matrix squaring (i.e., $A^2$) on a benchmark suite composed of 23 sparse matrices with diverse sparsity structures. 

In the context of Galerkin products, our method constantly outperforms the state-of-the-art GPU SpGEMM methods in two vendor supplied libraries cuSPARSE and CUSP. Average speedups of 1.9x (up to 2.6x) and 1.7x (up to 2.7x) are achieved when compared to cuSPARSE and CUSP, respectively.

In the context of matrix squaring, more comparison methods are included. \textit{First}, on two nVidia GPUs (i.e., a GeForce GTX Titan Black and a GeForce GTX 980), compared with cuSPARSE and CUSP, our approach delivers on average 3.1x (up to 9.5x) and 4.6x (up to 9.9x) speedups, respectively. 
\textit{Second}, compared to a recently developed CUDA-specific SpGEMM method RMerge~\cite{Gremse:GPU}, our method offers on average 2.5x (up to 4.9x) speedup on the nVidia GeForce GTX 980 GPU.
\textit{Third}, compared to the SpGEMM method in the latest Intel Math Kernel Library (MKL) on a six-core Xeon E5-2630 CPU and quad-channel system memory, our method gives on average 2.4x (up to 5.2x) and 2.1x (up to 4.2x) speedups on the nVidia GeForce GTX 980 GPU and an AMD Radeon R9 290X GPU, respectively. 


Furthermore, our approach can utilize re-allocatable memory controlled by CPU-GPU heterogeneous processors. On an AMD A10-7850K heterogeneous processor, compared to merely using its GPU cores, our framework delivers on average 1.2x (up to 1.8x) speedup while utilizing re-allocatable shared virtual memory in the system. 

\section{Preliminaries}

\subsection{SpGEMM Overview}

For the sake of generality, the SpGEMM algorithm description starts from discussion of the GEMM and gradually takes sparsity of the matrices $A$, $B$ and $C$ into consideration. For the matrix $A$, we write $a_{ij}$ to denote the entry in the $i$th row and the $j$th column of $A$ and $a_{i*}$ to denote the vector consisting of the $i$th row of $A$. Similarly, the notation $a_{*j}$ denotes the $j$th column of $A$. In the GEMM, the $i$th row of the resulting matrix $C$ can be defined by
\[
c_{i*}=( a_{i*}\cdot b_{*1}, a_{i*}\cdot b_{*2}, \ldots , a_{i*}\cdot b_{*p}), 
\]
where the operation $\cdot$ is the dot product of the two vectors.

We first give consideration to the sparsity of the matrix $A$.
Without loss of generality, we assume that the $i$th row of $A$ only consists of two nonzero entries in the $k$th and the $l$th column, respectively.
Thus $a_{i*}$ becomes $(a_{ik}, a_{il})$. Since all other entries are zeros, we do not record them explicitly and ignore their influence on the dot products in the calculation of the $i$th row of $C$. Then we obtain
\[
c_{i*}=( a_{ik}b_{k1}+a_{il}b_{l1}, a_{ik}b_{k2}+a_{il}b_{l2}, \ldots, a_{ik}b_{kp}+a_{il}b_{lp}).
\]

We can see in this case, only entries in the $k$th and the $l$th row of $B$ have contribution to the $i$th row of $C$. Then row vector form instead of column vector form is used for the matrix $B$. So we obtain 
\[
c_{i*}=a_{ik}b_{k*}+a_{il}b_{l*}.
\]

Since the matrix $B$ is sparse as well, again without loss of generality, we assume that the $k$th row of $B$ has only two nonzero entries in the $r$th and the $t$th column, and the $l$th row of $B$ also has only two nonzero entries in the $s$th and the $t$th column.
So the two rows are given by $b_{k*} = (b_{kr}, b_{kt})$ and $b_{l*} = (b_{ls}, b_{lt})$. Then
\begin{align}
c_{i*}&=a_{ik}( b_{kr}, b_{kt} )+a_{il}( b_{ls}, b_{lt} ). \nonumber 
\end{align}

Because the matrix $C$ is also sparse and the $i$th row of $C$ only has three nonzero entries in the $r$th, the $s$th and the $t$th column, the row can be given by
\[
c_{i*}=(c_{ir}, c_{is}, c_{it}),
\]
where $c_{ir} = a_{ik}b_{kr}$, $c_{is} = a_{il}b_{ls}$ and $c_{it} = a_{ik}b_{kt}+ a_{il}b_{lt}$.

In general there are more nonzero entries per rows of the matrices $A$, $B$ and $C$. But from the above derivation we can see that the SpGEMM can be represented by operations on row vectors of the matrices. Therefore, in this work we store all sparse matrices in compressed sparse row (CSR) format. The CSR format of a matrix consists of three separate arrays: (1) row pointer array of size $n+1$, where $n$ is the number of rows of the matrix, (2) column index array of size $nnz$, where $nnz$ is the number of nonzero entries of the matrix, and (3) value array of size $nnz$. Hence the overall space complexity of the CSR format is $O(n+nnz)$. Actually compressed sparse column (CSC) format is also widely used for sparse matrices stored in column-major order~\cite{Gilbert:Sparse}. The SpGEMM in the CSC format is almost the same as in the CSR format except rows are changed to columns and vice versa.

The above CSR-based SpGEMM algorithm can be performed by pseudocode in Algorithm~\ref{spgemm.jpdc.alg.spgemm}. An early description of this algorithm was given by Gustavson~\cite{Gustavson:Two}.

\begin{algorithm}[h!t]
  \caption{Pseudocode for the SpGEMM.}\label{spgemm.jpdc.alg.spgemm}
  \begin{algorithmic}[1]
      \For {\textbf{each} $a_{i*}$ in the matrix $A$}
        \State \texttt{set} $c_{i*}$ \texttt{to} $\emptyset$
        \For {\textbf{each} nonzero entry $a_{ij}$ in $a_{i*}$}
          \State \texttt{load} $b_{j*}$
          \For {\textbf{each} nonzero entry $b_{jk}$ in $b_{j*}$}
            \State $value\gets a_{ij}b_{jk}$
            \If {$c_{ik}\not\in c_{i*}$}
              \State \texttt{insert} $c_{ik}$ \texttt{to} $c_{i*}$
              \State $c_{ik}\gets value$
            \Else 
              \State $c_{ik}\gets c_{ik}+value$
            \EndIf
          \EndFor
        \EndFor
      \EndFor
  \end{algorithmic}
\end{algorithm}

\subsection{Prior SpGEMM Algorithms}

A classic CPU SpGEMM algorithm, also known as Matlab algorithm, was proposed by Gilbert et al.~\cite{Gilbert:Sparse}. This approach uses a dense vector-based sparse accumulator (or SPA) and takes $O(flops+nnz(B)+n)$ time to complete the SpGEMM, where $flops$ is defined as the number of necessary arithmetic operations on the nonzero entries, $nnz(B)$ is defined as the number of nonzero entries in the matrix $B$, and $n$ is the number of rows/columns of the input square matrices. Matam et al.~\cite{Matam:Sparse} developed a similar Matlab algorithm implementation for GPUs. 
Sulatycke and Ghose~\cite{Sulatycke:Caching} proposed a cache hits-oriented algorithm runs in relatively longer time $O(flops+n^2)$. 
A fast serial SpGEMM algorithm with time complexity $O(nnz^{0.7}n^{1.2}+n^{2+o(1)})$ was developed by Yuster and Zwick~\cite{Yuster:Fast}. Bulu\c{c} and Gilbert~\cite{Buluc:On} presented an SpGEMM algorithm with time complexity independent to the size of the input matrices under assumptions that the algorithm is used as a sub-routine of 2D distributed memory SpGEMM and the input matrices are hypersparse ($nnz < n$).

Recent GPU-based SpGEMM algorithms showed better time complexity. The SpGEMM algorithm in the cuSPARSE library~\cite{Demouth:Sparse, NVIDIA:CUSPARSE} utilized GPU hash table for the insert operations (lines 7--11 in Algorithm~\ref{spgemm.jpdc.alg.spgemm}). So time complexity of this approach is $O(flops)$ on average and $O(flops\; nnzr(C))$ in the worst case, where $nnzr(C)$ is defined as the average number of nonzero entries in the rows of the matrix $C$. Because the algorithm allocates one hash table of fixed size for each row of $C$, the space complexity is $O(nnz(A)+nnz(B)+n+nnz(C))$.

The CUSP library~\cite{Bell:Exposing, Dalton:CUSP} developed an SpGEMM method called expansion, sorting and compression (ESC) that expands all candidate nonzero entries generated by the necessary arithmetic operations (line 6 in Algorithm~\ref{spgemm.jpdc.alg.spgemm}) into an intermediate sparse matrix $\widehat{C}$, sorts the matrix by rows and columns and compresses it into the resulting matrix $C$ by eliminating entries in duplicate positions. By using GPU radix sort algorithm (with linear time complexity while size of the index data type of the matrices is fixed) and prefix-sum scan algorithm (with linear time complexity) as building blocks, time complexity of the ESC algorithm is $O(flops+nnz(\widehat{C})+nnz(\widehat{C}))$. Since $nnz(\widehat{C})$ equals half of $flops$, the ESC algorithm takes the optimal $O(flops)$ time. Dalton et al.~\cite{Dalton:Optimizing} improved the ESC algorithm by executing sorting and compression on the rows of $\widehat{C}$, but not on the entire matrix. Therefore fast on-chip memory has a chance to be utilized more efficiently. The improved method sorts the very short rows (of size no more than 32) by using sorting network algorithm (with time complexity $O(nnzr(\widehat{C})\log^2 (nnzr(\widehat{C})))$) instead of the radix sort algorithm which is mainly efficient for long lists. So the newer method is more efficient in practice, even though its time complexity is not lower than the original ESC algorithm. Because both of the ESC algorithms allocate an intermediate matrix $\widehat{C}$, they have the same space complexity $O(nnz(A)+nnz(B)+nnz(\widehat{C})+nnz(C)$).

RMerge algorithm, recently proposed by Gremse et al.~\cite{Gremse:GPU}, iteratively merges rows in the matrix $B$ into the resulting matrix $C$. Because this approach underutilizes thread interaction and generates one intermediate sparse matrix for each iteration step, it works best for input matrices with evenly distributed short rows. For irregular input matrices, load imbalance and large memory allocation make this method inefficient.

\subsection{Terminology Definition for GPU Programming}
Because CUDA and OpenCL are both widely used in GPU programming and they actually deliver comparable performance~\cite{Fang:A}, our SpGEMM algorithm support both of them. We use CUDA implementation on nVidia GPUs and OpenCL implementation on AMD GPU in our SpGEMM evaluation.
 
For simplicity, we define the following unified terminologies: (1) \textit{thread} denotes \textit{thread} in CUDA and \textit{work item} in OpenCL, (2) \textit{thread bunch} denotes \textit{warp} in nVidia GPU and \textit{wavefront} in AMD GPU, (3) \textit{thread group} denotes \textit{thread block} or \textit{cooperative thread array (CTA)} in CUDA and \textit{work group} in OpenCL, (4) \textit{core} denotes \textit{streaming multiprocessor (SMX)} or \textit{Maxwell streaming multiprocessor (SMM)} in nVidia GPU and \textit{compute unit} in AMD GPU, and (5) \textit{scratchpad memory} denotes \textit{shared memory} in CUDA and \textit{local memory} in OpenCL.

\section{Performance Considerations}

\subsection{Memory Pre-allocation For the Resulting Matrix}

Compared to SpGEMM, other sparse matrix multiplication operations (e.g., multiplication of sparse matrix and dense matrix~\cite{Ortega:FastSpMM, Vazquez:Fast, Saule:Performance} and its special case sparse matrix-vector multiplication~\cite{Bell:Implementing, Liu:CSR5, Liu:Speculative, Williams:Optimization, Buluc:Reduced}) pre-allocate a dense resulting matrix or vector. Thus the size of the result of the multiplication is trivially predictable, and the corresponding entries are stored to predictable memory addresses. However, because the number of nonzero entries in the resulting sparse matrix $C$ is unknown in advance, precise memory allocation of the SpGEMM is impossible before real computation. Moreover, physical address of each new entry is unknown either (consider line 7 in Algorithm~\ref{spgemm.jpdc.alg.spgemm}, the position $k$ is only a column index that cannot trivially map to a physical address on memory space).

To solve this problem, the previous SpGEMM algorithms proposed four different solutions: (1) precise method, (2) probabilistic method, (3) upper bound method, and (4) progressive method.

The first method, \textit{precise method}, pre-computes a simplified SpGEMM in the same computational pattern. We can imagine that multiplication of sparse boolean matrices is more efficient than multiplication of sparse floating-point matrices. RMerge algorithm and the SpGEMM methods in cuSPARSE and MKL are representatives of this approach. Even though the pre-computation generates precise size of $nnz(C)$, this method is relatively expensive since the SpGEMM operation in the same pattern is executed twice.

The second method, \textit{probabilistic method}, estimates an imprecise $nnz(C)$. This group of approaches~\cite{Amossen:Better, Cohen:On, Pagh:The} are based on random sampling and probability analysis on the input matrices. Since they do not guarantee a safe lower bound for the resulting matrix $C$ and extra memory has to be allocated while the estimation fails, they were mostly used for estimating the shortest execution time of multiplication of multiple sparse matrices.

The third method, \textit{upper bound method}, computes an upper bound of the number of nonzero entries in the resulting matrix $C$ and allocates corresponding memory space. Numerically, the upper bound size equals $nnz(\widehat{C})$, or half of $flops$, the number of necessary arithmetic operations. The ESC algorithms use this method for memory pre-allocation. Even though this approach saves cost of the pre-computation in the precise method, it brings another problem that the intermediate matrix $\widehat{C}$ may be too large to fit in the device global memory. Since the SpGEMM algorithm does not take into consideration cancellation that eliminates zero entries generated by arithmetic operations, the resulting matrix is normally larger than the input matrices. Table~\ref{spgemm.jpdc.tab.benchmarksuite} shows that $nnz(\widehat{C})$ is much larger than $nnz(C)$ while squaring some matrices. For example, the sparse matrix \textit{Wind Tunnel} generates 626.1 million nonzero entries (or 7.5 GB memory space for 32-bit index and 64-bit value) for the intermediate matrix $\widehat{C}$ while the real product $C$ (i.e., $A^2$) only contains 32.8 million nonzero entries. Although the upper bound method can partition the intermediate matrix $\widehat{C}$ into multiple sub-matrices, higher global memory pressure may reduce overall performance.

The last method, \textit{progressive method}, first allocates memory of a proper size, starts sparse matrix computation and re-allocates the buffer if larger space is required. Some CPU sparse matrix libraries use this method. For instance, sparse matrix computation in the Matlab~\cite{Gilbert:Sparse} increases the buffer by a ratio of 50\% if the current memory space is exhausted.

Since the upper bound method sacrifices space efficiency for the sake of improved performance and the progressive method is good at saving space, we use a hybrid method composed of the both approaches. However, compared to the relatively convenient upper bound method, it is hard to directly implement a progressive method for discrete GPUs. The reason is that although modern GPU devices have the ability of allocating device global memory while kernels are running, they still cannot re-allocate device memory on the fly. We will describe our hybrid method designed for discrete GPUs in the next section. 

On the other hand, emerging heterogeneous processors, composed of multiple CPU cores and GPU cores in one chip, supply both flexibility and efficiency. AMD Accelerated Processing Units (APUs)~\cite{Branover:Llano, AMD:Compute}, Intel multi-CPU and GPU system-on-a-chips (SoC) devices~\cite{Damaraju:22nm}, nVidia Echelon heterogeneous GPU architecture~\cite{Keckler:GPUs}, and many mobile processors (e.g., nVidia Tegra~\cite{nVidia:Tegra} and Qualcomm Snapdragon~\cite{Qualcomm:Snapdragon}) are representatives of the heterogeneous processor. Heterogeneous system architecture (HSA)~\cite{HSA:Manual} and OpenCL 2.0~\cite{Munshi:The} deliver programming tools for some heterogeneous processors. In this architecture, integrated GPU cores can directly use system memory allocated by the CPU part. Then data transfer through connection interfaces such as PCIe link can be avoided to obtain higher performance~\cite{Gregg:Where}. This gives our SpGEMM algorithm a chance to let integrated GPUs use re-allocatable system memory for a better overall performance. Later on, we will show the corresponding performance gain by using an AMD APU.

\subsection{Parallel Insert Operations}
As shown in Algorithm~\ref{spgemm.jpdc.alg.spgemm}, for each trivial arithmetic computation (line 6), one much more expensive insert operation (lines 7--11) is required. To the best of our knowledge, none of the previous GPU SpGEMM methods takes into account that the input sequence (line 4) is ordered because of the CSR format\footnote{Actually according to the CSR format standard, the column indices in each row do not necessarily have to be sorted. But most implementations choose to do so, thus our method reasonably makes this assumption.}. One of our algorithm design objectives is to efficiently utilize this property. Based on experiments by Kim et al.~\cite{Kim:Sort}, as the SIMD units are getting wider and wider, merge sort methods will outperform hash table methods on the join-merge problem, which is a similar problem in the SpGEMM. Then our problem converts to finding a fast GPU method for merging sorted sequences. Later on we will describe our strategy in detail.

\subsection{Load Balancing}
Because distribution patterns of nonzero entries in both input sparse matrices can be very diverse (consider plots of the matrices in Table~\ref{spgemm.jpdc.tab.benchmarksuite}), input space-based data decomposition~\cite{Demouth:Sparse, Sulatycke:Caching} normally does not bring efficient load balancing. One exception is that computing SpGEMM for huge sparse matrices on large scale distributed memory systems, 2D and 3D decomposition on input space methods demonstrated good load balancing and scalability by utilizing efficient communication strategies~\cite{Gilbert:High, Ballard:Communication, Buluc:Parallel}. However, in this paper we mainly consider load balancing for fine-grained parallelism in GPU and CPU-GPU shared memory architectures.

Therefore we use the other group of load balancing methods based on output space decomposition. Dalton et al.~\cite{Dalton:Optimizing} presented a method that sorts rows of the intermediate matrix $\widehat{C}$, divides it into 3 sub-matrices that include the rows in different size ranges, and uses differentiated ESC methods for the sub-matrices. We have a similar consideration, but our implementation is completely different. We do not strictly sort rows of the intermediate matrix $\widehat{C}$ but just assign rows to a fixed number of bins through a much faster linear time traverse on CPU. Moreover, we decompose the output space in a more detailed way that guarantees much more efficient load balancing. We will demonstrate that our method is always load balanced in all stages for maximizing resource utilization of GPUs.

\section{Methodology}

\subsection{Framework and Algorithm Design}

\begin{figure*}[h!t]
\centering
\includegraphics[trim=0in 0in 0in 0in,clip, width=4.5in]{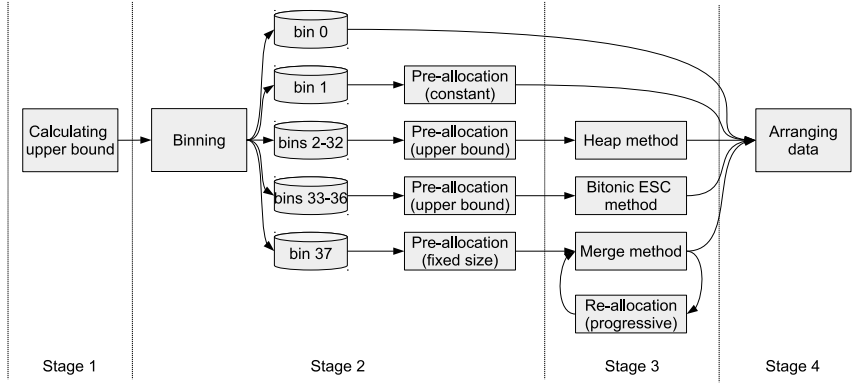}
\caption{The SpGEMM framework composed of four stages.}
\label{spgemm.jpdc.fig.framework}
\end{figure*}

Our SpGEMM framework includes four stages: (1) calculating upper bound, (2) binning, (3) computing the resulting matrix, and (4) arranging data. Figure~\ref{spgemm.jpdc.fig.framework} plots this framework.

\textbf{The first stage}, calculating upper bound, generates the upper bound number of nonzero entries in each row of the resulting matrix $C$. We create an array $U$ of size $m$, where $m$ is the number of rows of $C$, for the upper bound sizes of the rows. We use one GPU thread for computing each entry of the array $U$. Algorithm~\ref{spgemm.jpdc.alg.stage1} describes this procedure.

\begin{algorithm}[h!t]
  \caption{Pseudocode for the first stage on GPUs.}\label{spgemm.jpdc.alg.stage1}
  \begin{algorithmic}[1]
      \For {\textbf{each} entry $u_i$ in $U$ \textbf{in parallel}}
        \State $u_i\gets 0$
        \For {\textbf{each} nonzero entry $a_{ij}$ in $a_{i*}$}
          \State $u_i\gets u_i+nnz(b_{j*})$
        \EndFor
      \EndFor
  \end{algorithmic}
\end{algorithm}

\textbf{The second stage}, binning, deals with load balancing and memory pre-allocation. We first allocate 38 bins and put them into five bin groups. The bins contain the indices of the entries in the array $U$ and present as one array of size $m$ with 38 segments. Then all rows are assigned to corresponding bins according to the number of nonzero entries. Finally, based on the sizes of the bins, we allocate a temporary matrix for nonzero entries in the resulting matrix $C$.

The first bin group includes one bin that contains the indices of the rows of size 0. The second bin group also only has one bin that contains the indices of the rows of size 1. Because the rows in the first two bins only require trivial operations, they are excluded from subsequent more complex computation on GPUs. Thus a better load balancing can be expected.

The third bin group is composed of 31 bins that contain the indices of the rows of sizes 2--32, respectively. Since the sizes of these rows are no more than the size of a single thread bunch (32 in current nVidia GPUs or 64 in current AMD GPUs) and these rows require non-trivial computation, using one thread bunch or one thread group for each row cannot bring efficient instruction throughput on GPUs. Therefore, we use one thread for each row. Further, because each bin only contains the rows of the same upper bound size, the bins can be executed separately on GPUs with different kernel programs for efficient load balancing. In other words, 31 GPU kernel programs will be executed for the 31 bins, if not empty. 

The fourth bin group consists of 4 bins that contain the indices of the rows located in size ranges 33--64, 65--128, 129--256 and 257--512, respectively. The rows of these sizes are grouped because of three reasons: (1) each of them is large enough to be efficiently executed by a thread group, (2) each of them is small enough for scratchpad memory (48 kB per core in current nVidia Kepler GPUs, 96 kB per core in current nVidia Maxwell GPUs and 64 kB per core in current AMD Graphics Core Next, or GCN, GPUs), and (3) the final sizes of these rows in the resulting matrix $C$ are predictable in a reasonable small range (no less than the lower bound of size 1 and no more than the corresponding upper bound sizes). Even though the rows in each bin do not have exactly the same upper bound size, a good load balancing still can be expected because each row is executed by using one thread group and inter-thread group load balancing is naturally guaranteed by the GPU low-level scheduling sub-systems.

The fifth bin group includes the last bin that contains the indices of the rest of the rows of size larger than 512. These rows have two common features: (1) their sizes can be too large (recall $nnzr(\widehat{C})$ in Table~\ref{spgemm.jpdc.tab.benchmarksuite}) to fit in the scratchpad memory, and (2) predicting the final sizes of these rows to a small range (scratchpad memory level) is not possible in advance. Therefore, we execute them in a unified progressive method described later. Again because we use one thread group for each row, load balancing is naturally guaranteed.

Since we do not use precise method for memory pre-allocation, a temporary memory space for the resulting matrix $C$ is required. We design a hybrid method that allocates a CSR format sparse matrix $\widetilde{C}$ of the same size of the resulting matrix $C$ as temporary matrix. We set $nnz(\widetilde{c}_{i*})$ to $u_i$ while the row index $i$ is located in the bin groups 1--4 because compared with modern GPU global memory capacity, the total space requirement of these rows is relatively small. For the rows in the bin group 5, we set $nnz(\widetilde{c}_{i*})$ to a fixed size 256 since normally this is an efficient working size for the scratchpad memory. Therefore, we can see that if all of the indices of the rows are in the bin groups 1--4, our hybrid method converts to the upper bound method, on the other extreme end, our method converts to the progressive method. But generally, we obtain benefits from the both individual methods. The stage 2 is executed on CPU since it only requires a few simple linear time traverses, which are more efficient for the CPU cache sub-systems. The pseudocode is shown in Algorithm~\ref{spgemm.jpdc.alg.stage2}.

\begin{algorithm}[h!t]
  \caption{Pseudocode for the second stage on a CPU core.}\label{spgemm.jpdc.alg.stage2}
  \begin{algorithmic}[1]
      \For {\textbf{each} entry $u_i$ in $U$}
        \If {$u_i = 0$} \Comment {The 1st bin group}
          \State \texttt{insert} $i$ \texttt{to} $bin_0$
          \State $nnz(\widetilde{c}_{i*}) \gets 0$
        \ElsIf {$u_i = 1$} \Comment {The 2nd bin group}
          \State \texttt{insert} $i$ \texttt{to} $bin_1$
          \State $nnz(\widetilde{c}_{i*}) \gets 1$
        \ElsIf {$u_i \geq 2 \; \&\& \; u_i \leq 32$} \Comment {The 3rd bin group}
          \State \texttt{insert} $i$ \texttt{to} $bin_{u_i}$
          \State $nnz(\widetilde{c}_{i*}) \gets u_i$
        \ElsIf {$u_i \geq 33 \; \&\& \; u_i \leq 64$} \Comment {The 4th bin group}
          \State \texttt{insert} $i$ \texttt{to} $bin_{33}$
          \State $nnz(\widetilde{c}_{i*}) \gets u_i$
        \ElsIf {$u_i \geq 65 \; \&\& \; u_i \leq 128$} \Comment {The 4th bin group}
          \State \texttt{insert} $i$ \texttt{to} $bin_{34}$
          \State $nnz(\widetilde{c}_{i*}) \gets u_i$
        \ElsIf {$u_i \geq 129 \; \&\& \; u_i \leq 256$} \Comment {The 4th bin group}
          \State \texttt{insert} $i$ \texttt{to} $bin_{35}$
          \State $nnz(\widetilde{c}_{i*}) \gets u_i$
        \ElsIf {$u_i \geq 257 \; \&\& \; u_i \leq 512$} \Comment {The 4th bin group}
          \State \texttt{insert} $i$ \texttt{to} $bin_{36}$
          \State $nnz(\widetilde{c}_{i*}) \gets u_i$
        \ElsIf {$u_i > 512$} \Comment {The 5th bin group}
          \State \texttt{insert} $i$ \texttt{to} $bin_{37}$
          \State $nnz(\widetilde{c}_{i*}) \gets 256$
        \EndIf
      \EndFor
      \State $nnz(\widetilde{C}) \gets \sum nnz(\widetilde{c}_{i*})$
  \end{algorithmic}
\end{algorithm}

\textbf{The third stage}, computing the resulting matrix, generates $nnz(c_{i*})$ and the final nonzero entries stored in the temporary matrix $\widetilde{C}$.

For the rows in the bin groups 1--2, we simply update the numbers of corresponding nonzero entries. For the rows in the bin groups 3--5, we use three totally different methods: (1) heap method, (2) bitonic ESC method, and (3) merge method, respectively. Note that each bin has a counter (at the host side) that records the number of rows included. So the host can easily decide if a GPU kernel will be issued for a certain bin. In other words, our approach only issue kernels for non-empty bins.

\begin{figure}[h!t]
\centering
\subfloat[]{\includegraphics[trim=0in 6in 6.62in 0.1in,clip,width=1.1in]{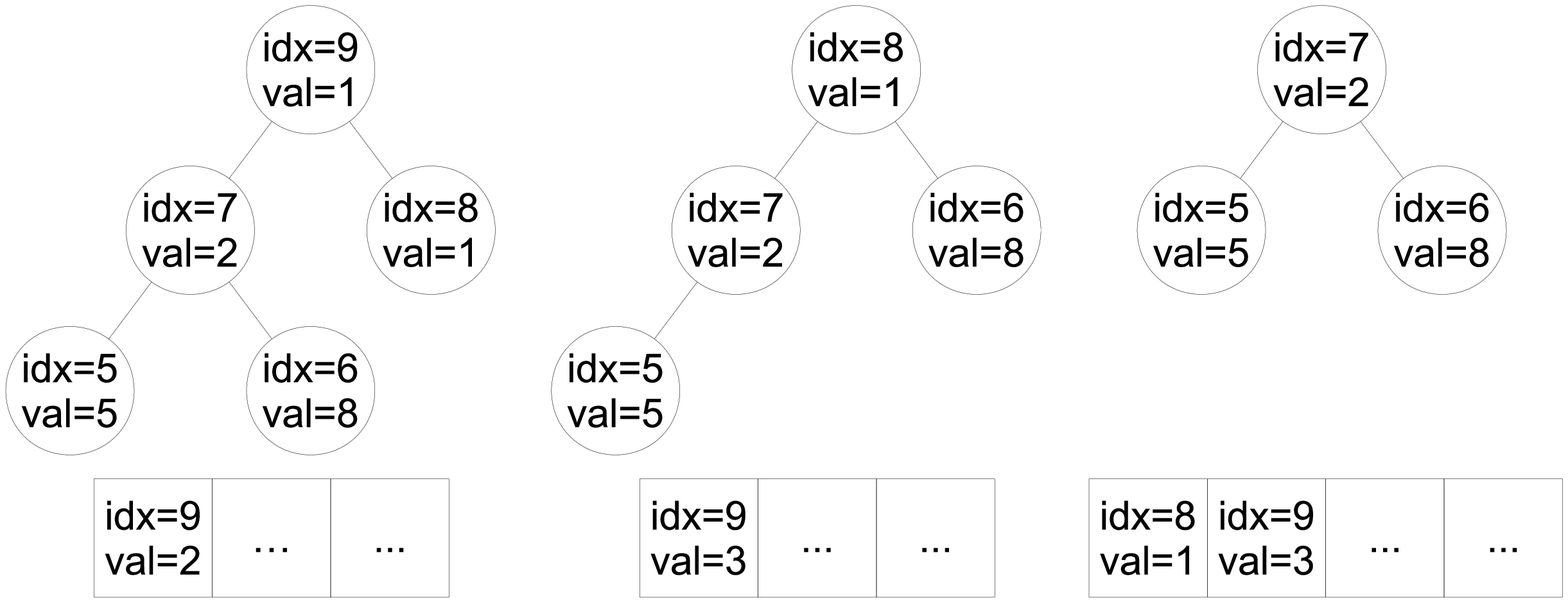}}
\subfloat[]{\includegraphics[trim=3.31in 6in 3.31in 0.1in,clip,width=1.1in]{methodheap}}
\subfloat[]{\includegraphics[trim=6.62in 6in 0in 0.1in,clip,width=1.1in]{methodheap}}
\caption{Two steps of an example of the heap method. From (a) to (b), the root entry is fused to the first entry in resulting sequence since they share the same index. From (b) to (c), the root entry is inserted to the sequence since they have different indices. After each step, the heap property is reconstructed.}
\label{spgemm.jpdc.fig.2heap}
\end{figure}

The heap method first creates an empty implicit index-value pair heap (or priority queue) of the upper bound size for each row in the bin group 3. The heaps are located in the scratchpad memory and collect all candidate nonzero entries for corresponding rows. Then each heap executes a heapsort-like operation to generate an ordered sequence located in the tail part of the heap. The difference between this operation and the classic heapsort operation is that the entries in the resulting sequence are duplicate-free while the initial heap includes duplicate entries. In each delete-max step in our variant heapsort, the root node and the first entry of the resulting sequence are fused if they share the same index; otherwise the root node is inserted to the head part of the sequence. Our method is also distinguished from a heap-based sparse accumulator given by Gilbert et al.~\cite{Gilbert:Ordered} by the mechanism of eliminating duplicate entries.  Figure~\ref{spgemm.jpdc.fig.2heap} gives two steps of an example of our heap method. Finally, the sorted sequence without duplicate indices is generated in the scratchpad memory and saved to the matrix $\widetilde{C}$ in the global memory. In addition, the numbers of nonzero entries in the rows of the resulting matrix $C$ are updated to the sizes of the corresponding resulting sequences.

\begin{figure}[h!t]
\centering
\includegraphics[trim=0in 4.75in 0in 0in,clip,width=3.4in]{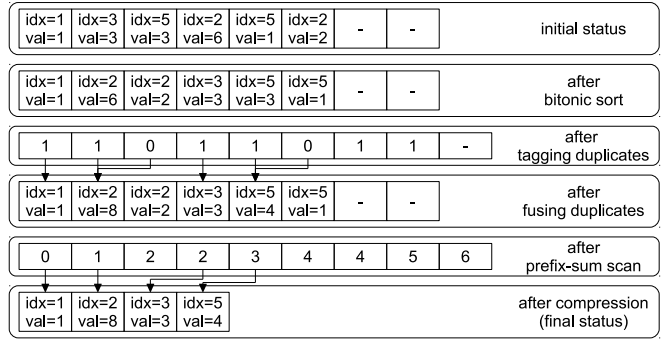}
\caption{An example of the bitonic ESC method.}
\label{spgemm.jpdc.fig.bitonic}
\end{figure}

For the rows in each bin of the bin group 4, a typical ESC algorithm is used. The method first collects all candidate nonzero entries to an array in the scratchpad memory, then sorts the array by using basic bitonic sort and compresses duplicate indices in the sequence by using prefix-sum scan. Figure~\ref{spgemm.jpdc.fig.bitonic} shows an example of this procedure. Finally, a sorted sequence without duplicate indices is generated in the scratchpad memory and saved to the matrix $\widetilde{C}$, and the numbers of nonzero entries in the rows are updated.

For the rows in the bin group 5, our method inserts each input nonzero entry to the corresponding row of the resulting matrix $C$ (lines 7--11 in Algorithm~\ref{spgemm.jpdc.alg.spgemm}) in parallel. We can see that the input sequence (the candidate nonzero entries) and the resulting sequence (the selected nonzero entries in the current row of $C$) should always be kept ordered and duplicate-free because of the CSR format. Therefore, we can convert the parallel insert operations to parallel merge operations that merge ordered sequences and the final resulting sequence is ordered and duplicate-free.

\begin{figure}[!t]
\centering
\includegraphics[trim=0in 2.35in 0in 0in,clip,width=3.4in]{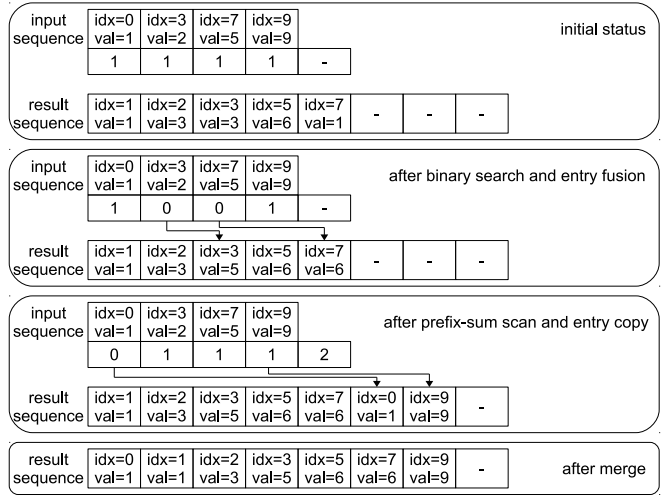}
\caption{An example of the merge method. The input sequence is in the register file. Its mask sequence and the resulting sequence are in the scratchpad memory.}
\label{spgemm.jpdc.fig.merge}
\end{figure}

Each parallel merge operation can be split into multiple sub-steps: (1) a binary search operation on the resulting sequence for fusing entries with the same indices and tagging them, (2) a prefix-sum scan operation on the input sequence for getting continuous positions in the incremental part of the resulting sequence, (3) copying non-duplicate entries from the input sequence to the resulting sequence, and (4) merging the two sequences in one continuous memory space. Figure~\ref{spgemm.jpdc.fig.merge} shows an example of this procedure. After all input sequences are merged into one resulting sequence, it is saved to the matrix $\widetilde{C}$. Then the numbers of nonzero entries in the rows are updated.

As we allocate a limited scratchpad memory space for the resulting sequence, a potential overflow may happen. In this case, we first compare total size of the two sequences (note that the input sequence is in the thread registers, but not in the scratchpad memory yet) with the allocated size of the resulting sequence in the scratchpad memory. If a merge operation is not allowed, our method records current computation position as a checkpoint and dumps the resulting sequence from the scratchpad memory to the global memory. Then the host allocates more global memory (we use 2x each time) and re-launches kernel with a 2x large scratchpad memory setting. The relaunched kernels obtain checkpoint information, and load existing results to the scratchpad memory and continue the computation. The global memory dumping and reloading bring an extra overhead, but actually it does not affect the total execution time too much because of three reasons: (1) the global memory access is almost completely coalesced, (2) the latency could be hidden by subsequent computation, and (3) this overhead is only a small factor of large computation (short rows normally do not face this problem). For very long rows that exceed the scratchpad memory capacity, our method still allocates a space in the scratchpad memory as a level-1 merge sequence, executes the same merge operations on it and merges the level-1 sequence in the scratchpad memory and the resulting sequence in the global memory only once before the kernel is ready to return.

It is worth noting that the parameters of the binning depends on specifications (e.g., thread bunch size and scratchpad memory capacity) of GPU architectures. In this paper, we use the abovementioned fixed-size parameters for assigning the rows into the bins since the current nVidia GPUs and AMD GPUs have comparable hardware specifications. However, the strategies in stages 2 and 3 can be easily extended for future GPUs with changed architecture designs.

\textbf{The fourth stage}, arranging data, first sums the numbers of nonzero entries in all rows of the resulting matrix $C$ and allocates its final memory space. Then our method copies existing nonzero entries from the temporary matrix $\widetilde{C}$ to the resulting matrix $C$. For the rows in the bin group 1, the copy operation is not required. For the rows in the bin group 2, we use one thread for each row. For the rest of the rows in the bin groups 3--5, we use one thread group for each row. After all copy operations, the SpGEMM computation is done.

\subsection{Evaluating GPU Merge algorithms}

\begin{figure*}[h!t]
\centering
\subfloat[]{\includegraphics[width=1.75in]{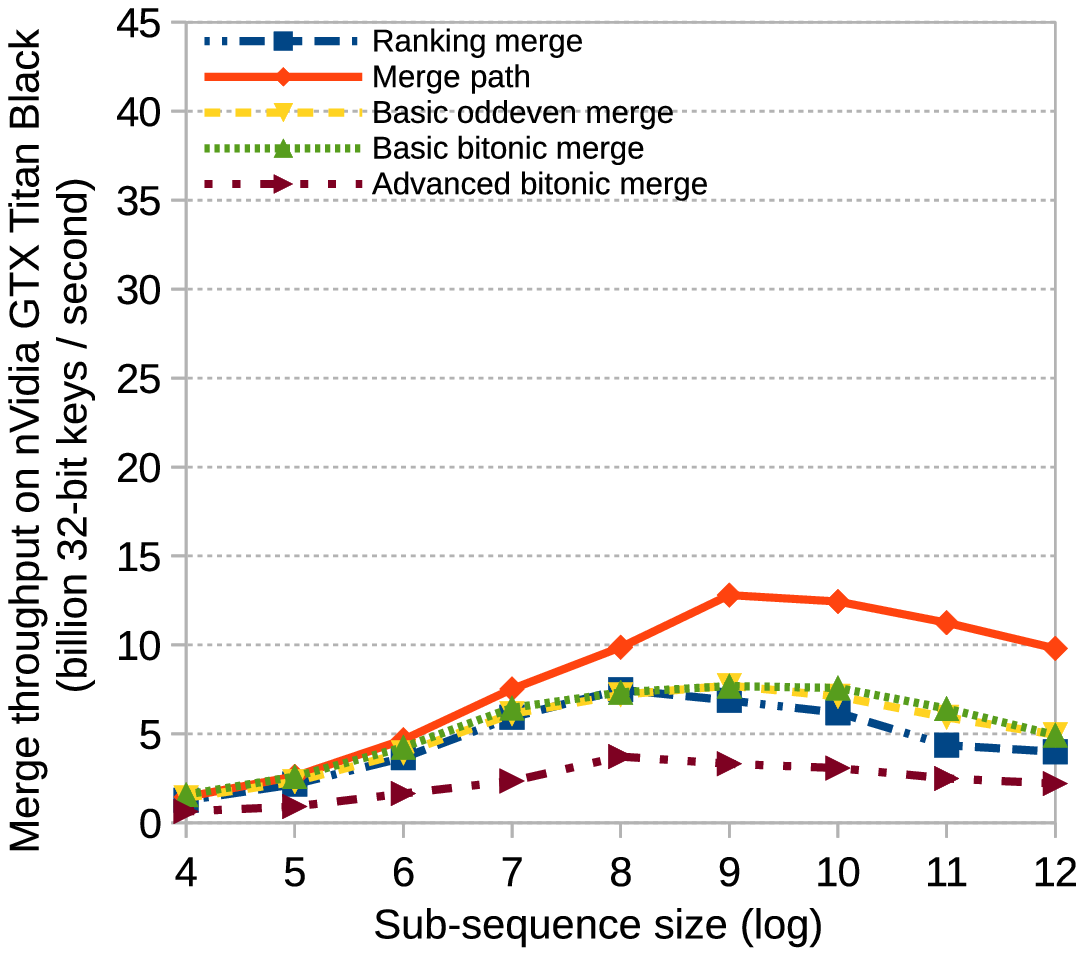}}
\subfloat[]{\includegraphics[width=1.75in]{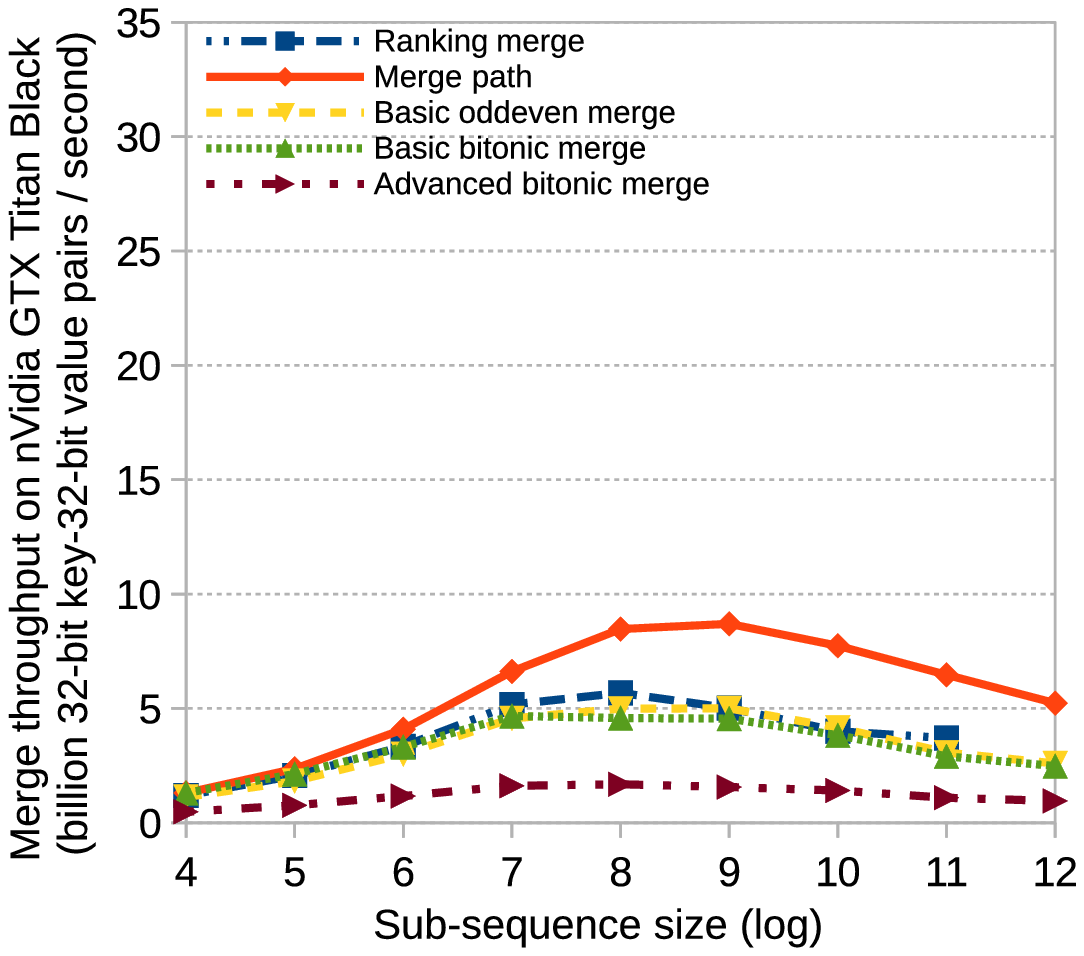}}
\subfloat[]{\includegraphics[width=1.75in]{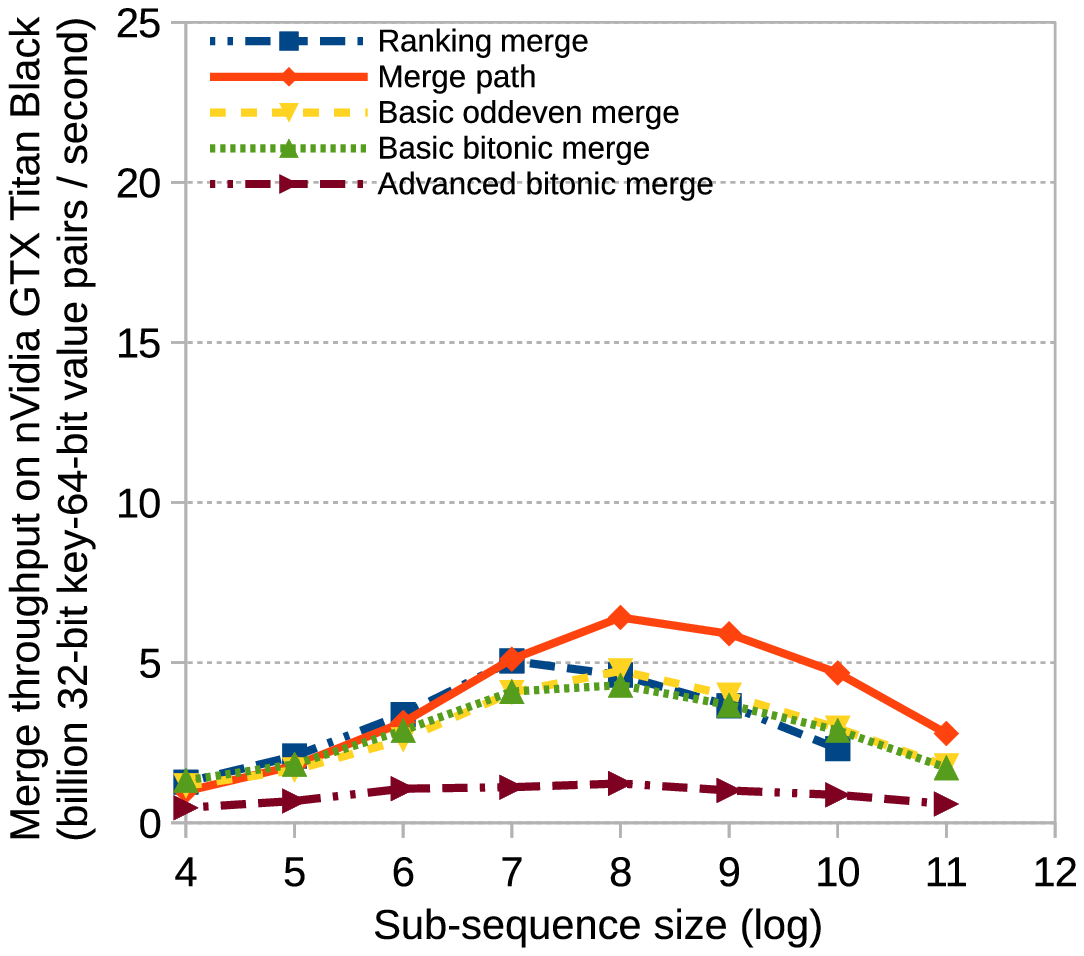}}
\linebreak
\subfloat[]{\includegraphics[width=1.75in]{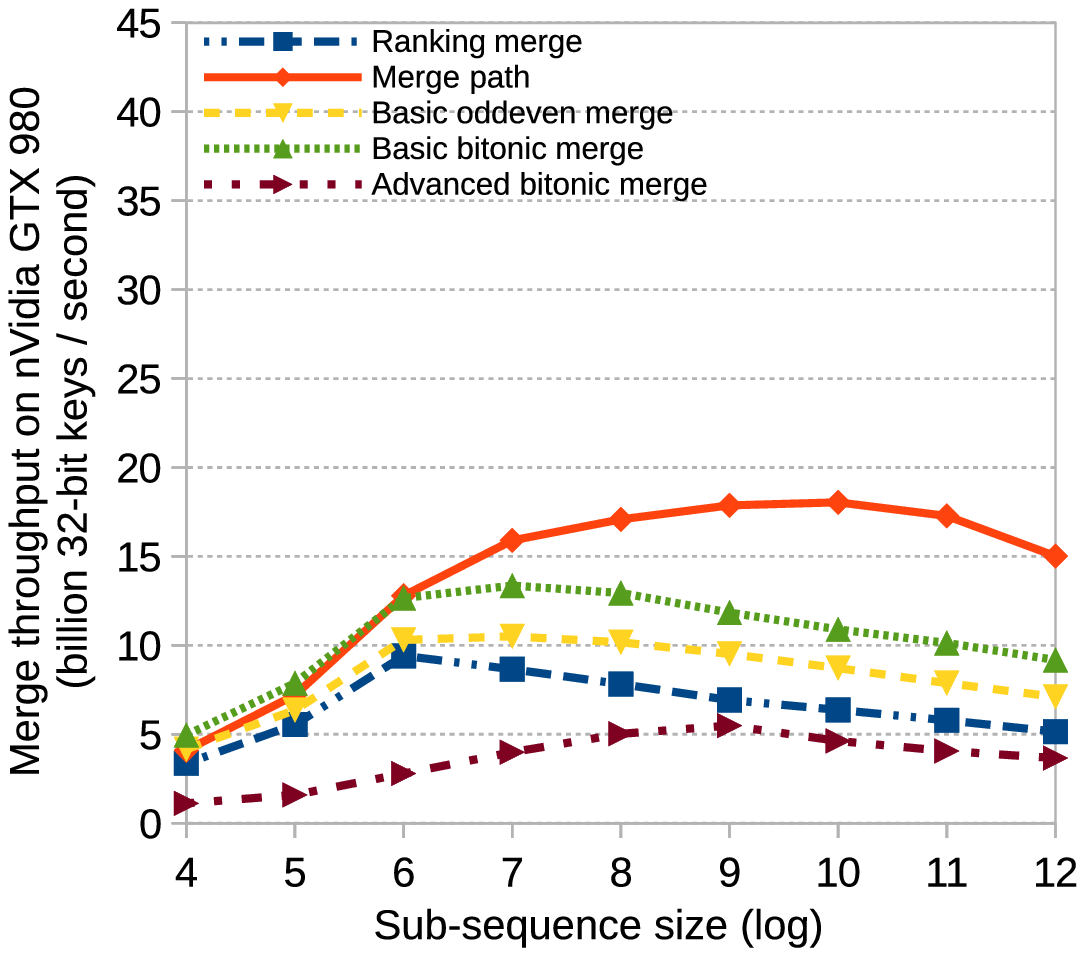}}
\subfloat[]{\includegraphics[width=1.75in]{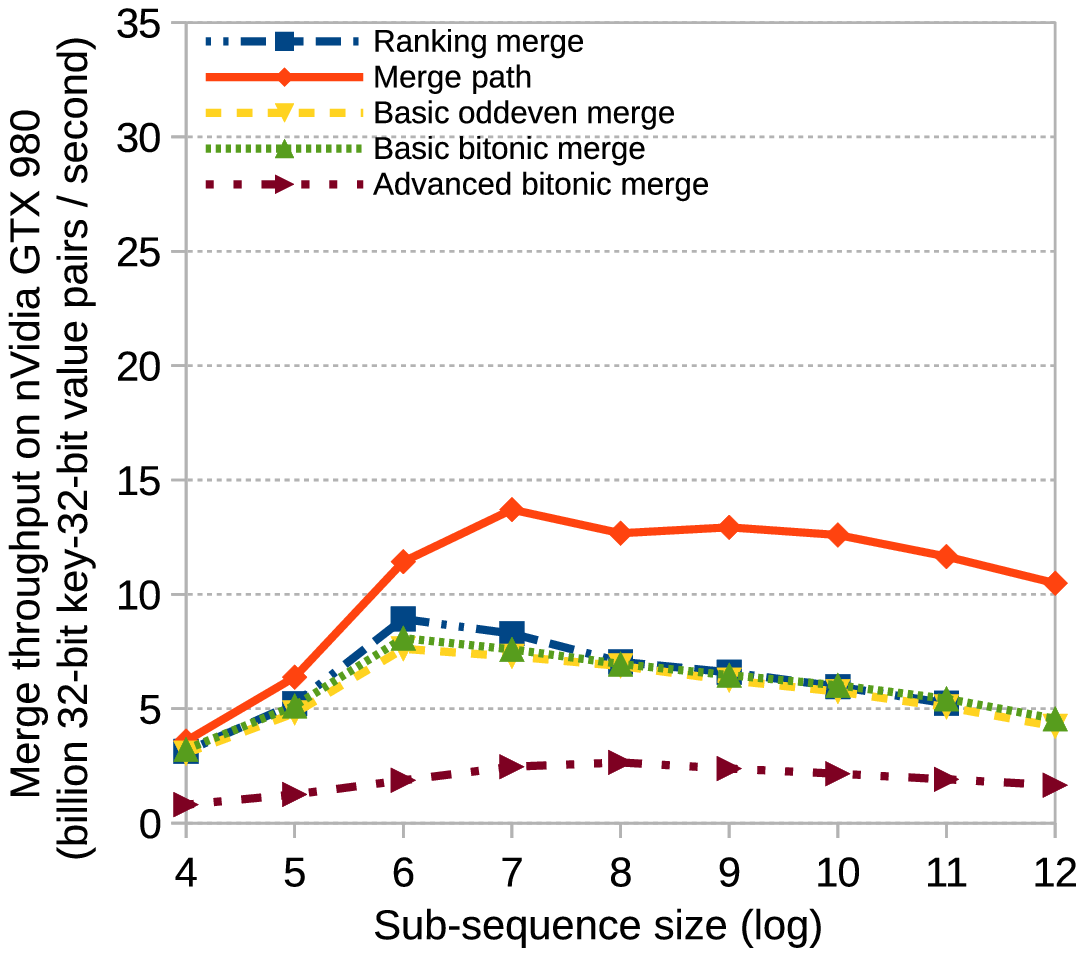}}
\subfloat[]{\includegraphics[width=1.75in]{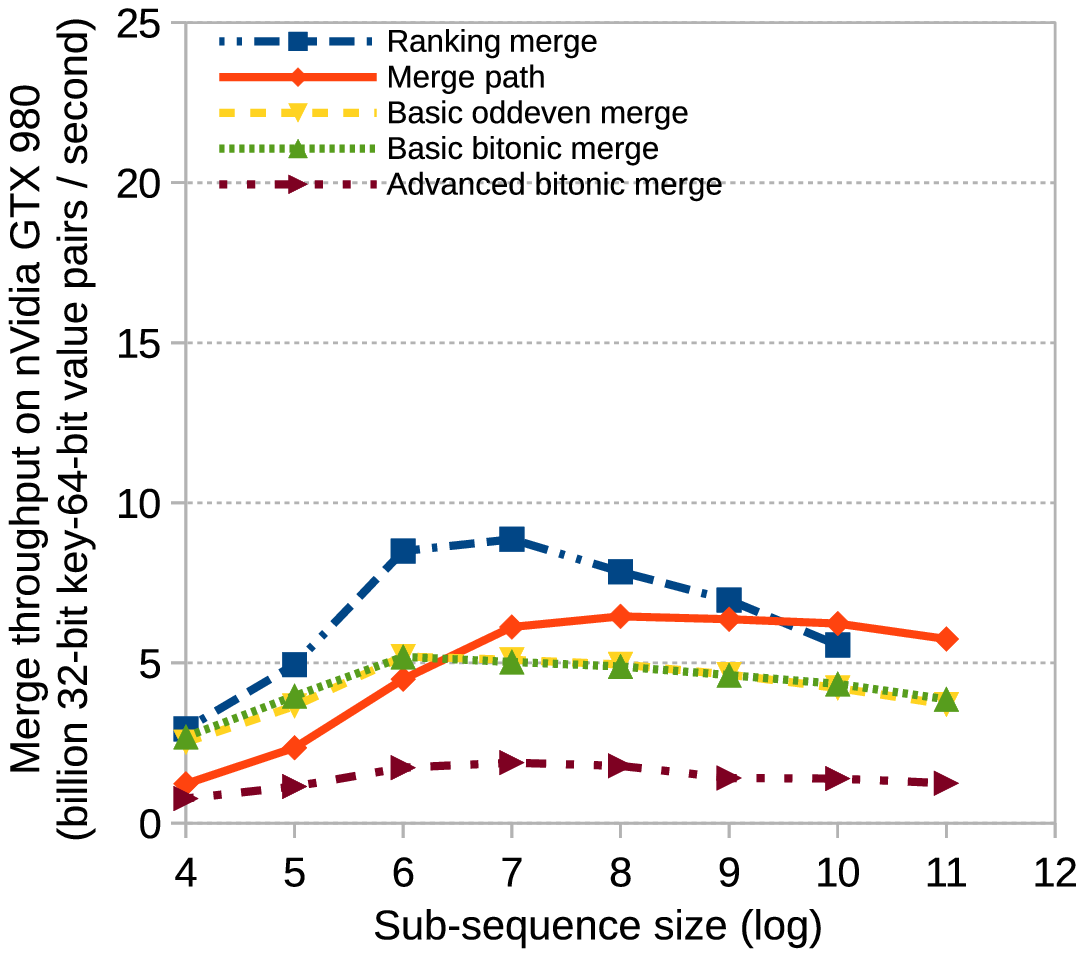}}
\linebreak
\subfloat[]{\includegraphics[width=1.75in]{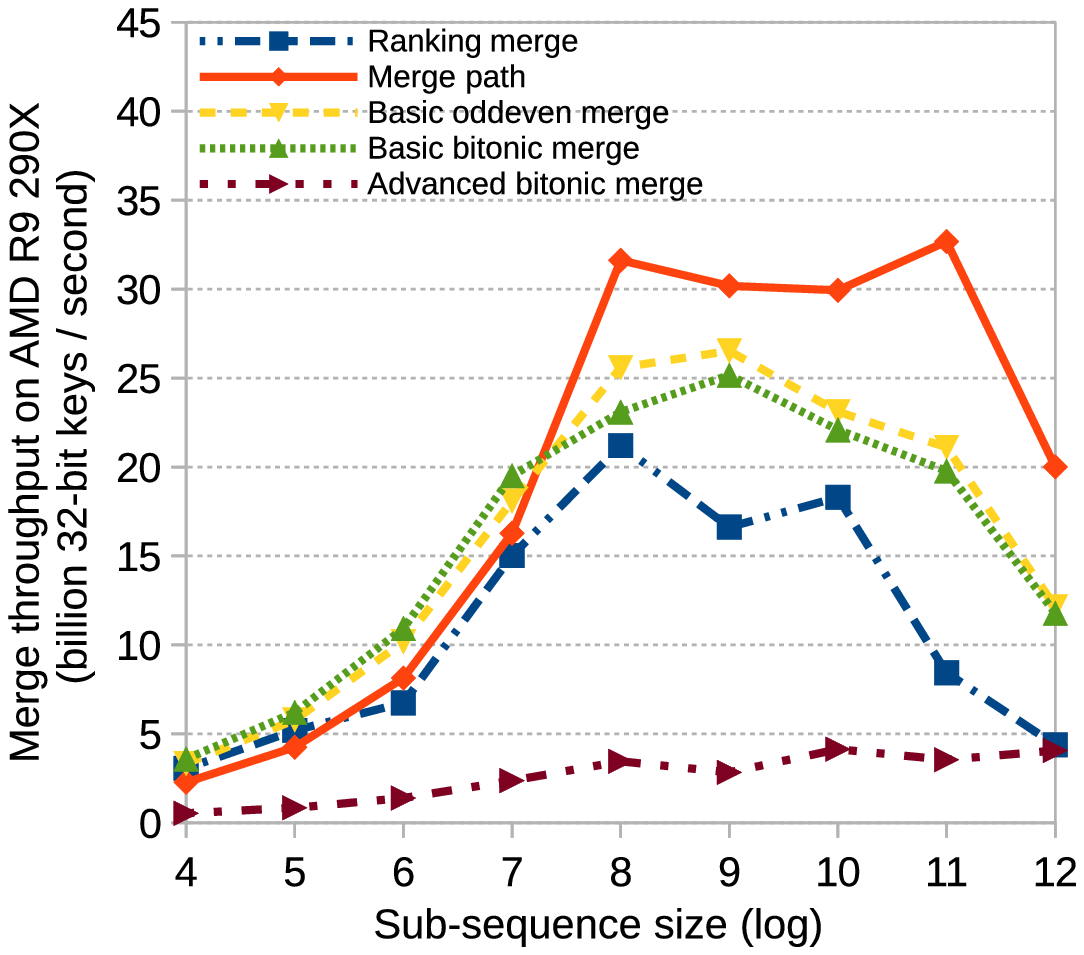}}
\subfloat[]{\includegraphics[width=1.75in]{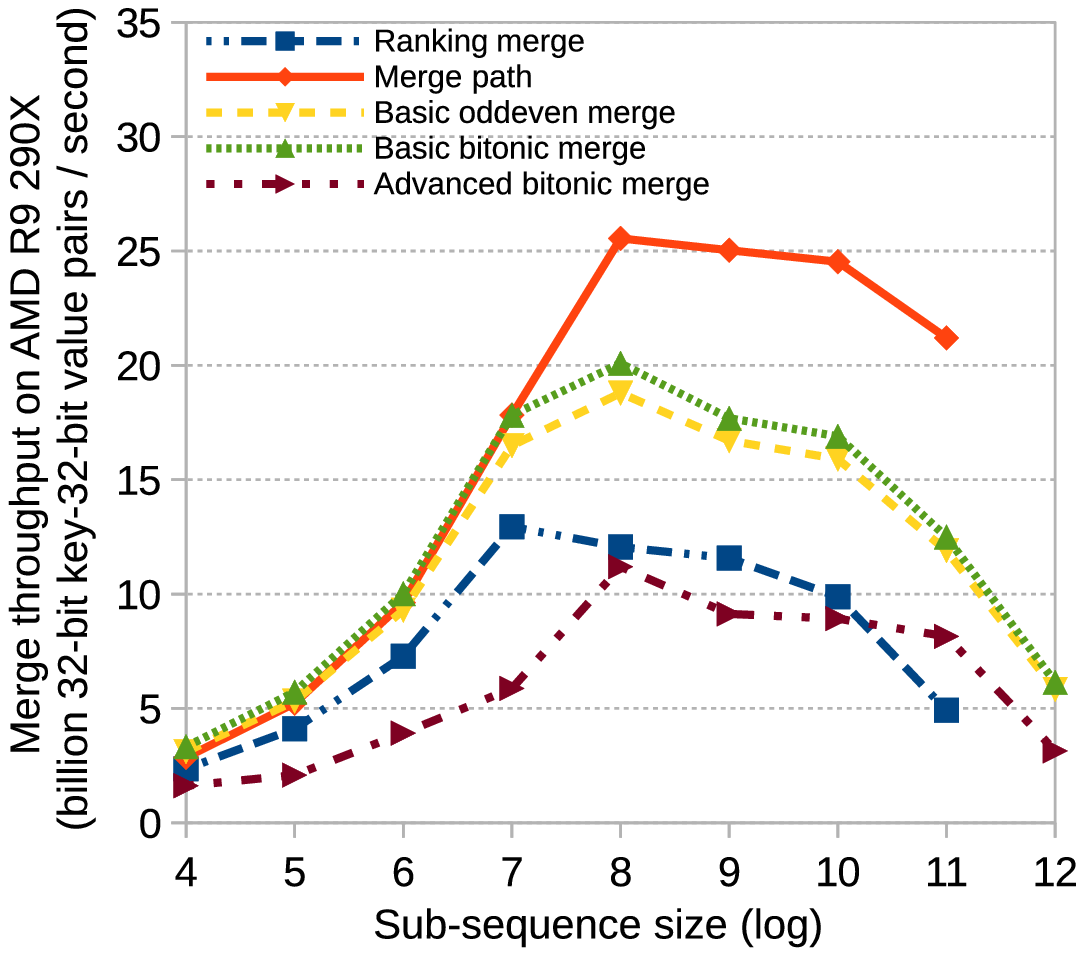}}
\subfloat[]{\includegraphics[width=1.75in]{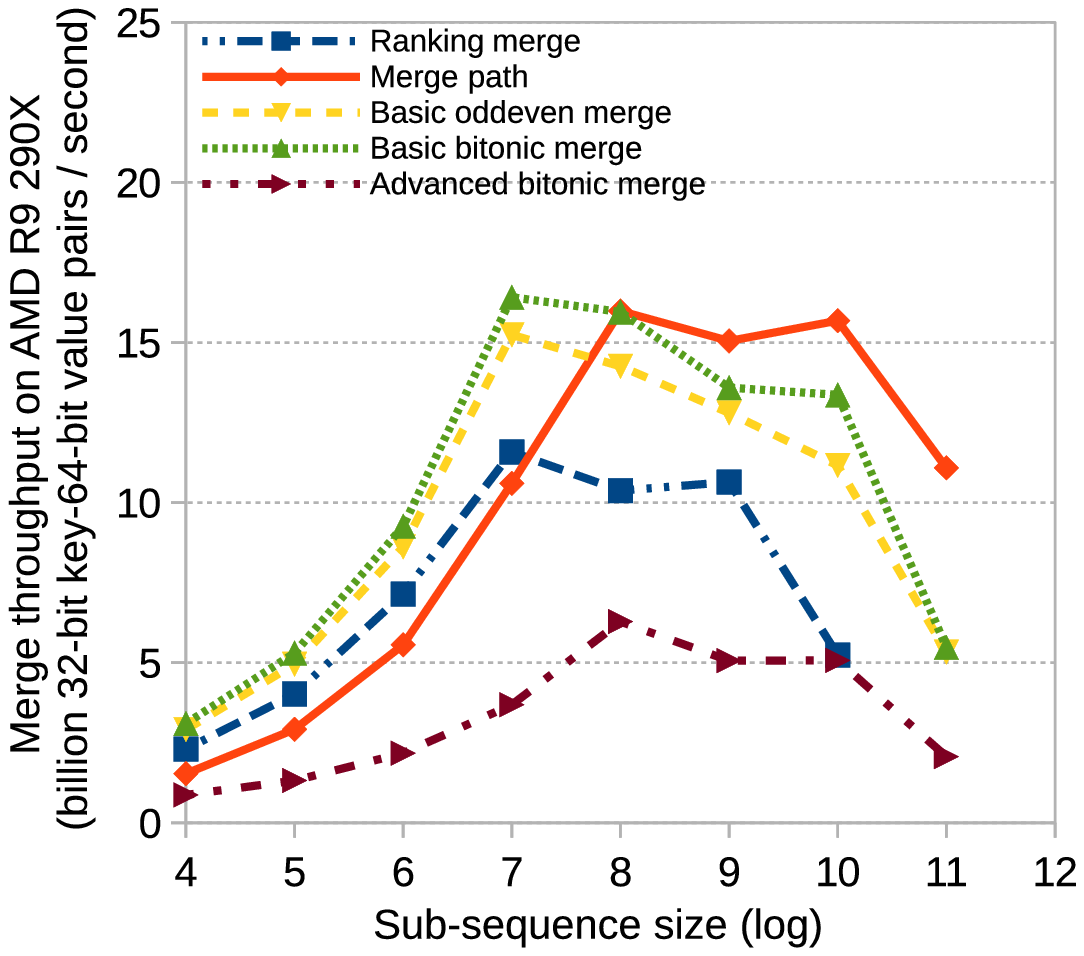}}
\caption{Performance comparison of merging 32-bit keys, 32-bit key-32-bit value pairs and 32-bit key-64-bit value pairs through 5 GPU merge algorithms: ranking merge, merge path, basic oddeven merge, basic bitonic merge and advanced bitonic merge on three different GPUs: nVidia GeForce GTX Titan Black, nVidia GeForce GTX 980 and AMD Radeon R9 290X.}
\label{spgemm.jpdc.fig.testmergemethods}
\end{figure*}

Because both the binary search and the prefix-sum scan take fast logarithmic time for each entry in the input sequence, these operations have relatively good efficiency and performance stability on modern GPUs. Therefore, a fast merge algorithm is very crucial for the performance of the merge method in our SpGEMM framework.

Recently some new merge algorithms~\cite{Satish:Designing, Green:GPU, Kipfer:Improved, Peters:Comparison, Peters:A, Inoue:AA, Davidson:Efficient} have been proposed for GPUs. But which one is the fastest in practice is still an open question. Because the main objective of the research~\cite{Peters:A, Inoue:AA, Davidson:Efficient} is efficiently merging large data in the global memory, they still use basic methods, such as bitonic sort and ranking-based merge, as building blocks for small data in the scratchpad memory. Peters et al.~\cite{Peters:Comparison} proposed a locality-oriented advanced bitonic sort method that can reduce synchronization overhead by merging data in fast private memory instead of relatively slow shared memory. Therefore we evaluate 5 GPU merge algorithms: (1) ranking merge~\cite{Satish:Designing}, (2) merge path~\cite{Green:GPU}, (3) basic oddeven merge~\cite{Kipfer:Improved}, (4) basic bitonic merge~\cite{Kipfer:Improved}, and (5) advanced bitonic merge~\cite{Peters:Comparison}. The implementation of the algorithm (2) is extracted from the Modern GPU library~\cite{Baxter:Modern}. The implementations of the algorithm (3) and (4) are extracted from the nVidia CUDA SDK. We implement the algorithm (1) and (5). Additionally, another reason why we conduct the evaluation is that none of the above literature presented performance of merging short sequences of size less than $2^{12}$, which is the most important length (consider the $nnzr(C)$ in Table~\ref{spgemm.jpdc.tab.benchmarksuite}) for our SpGEMM.

Our evaluation results of merging 32-bit keys, 32-bit key-32-bit value pairs and 32-bit key-64-bit value pairs are shown in Figure~\ref{spgemm.jpdc.fig.testmergemethods}. The experimental platforms are described in Table~\ref{spgemm.jpdc.tab.testbeds}. Each of the five algorithms merges two short ordered sequences of size $l$ into one ordered output sequence of size $2l$. The sorting network methods in our evaluation only execute the last stage, since both inputs are sorted. To saturate throughput of GPUs, the whole problem size is set to size $2^{25}$. For example, $2^{14}$ thread groups are launched while each of them merges two sub-sequences of size $l=2^{10}$. We execute each problem set through multiple thread groups of different sizes and record the best performance for the evaluation.

In Figure~\ref{spgemm.jpdc.fig.testmergemethods}, we can see that the GPU merge path algorithm almost always outperforms other methods while sub-sequence size is no less than $2^8$. Since our merge method starts from size 256, the merge path method is chosen for our SpGEMM implementation. The extra advantages of the merge path method are that it can evenly assign work load to threads and can easily deal with the input sequences of arbitrary sizes. Detailed description and complexity analysis of the GPU merge path algorithm can be found in~\cite{Green:GPU}.

Other algorithms are not chosen because of various reasons. We can see that the ranking merge is faster than the merge path method in Figure~\ref{spgemm.jpdc.fig.testmergemethods}(f). But it is not chosen in our implementation, since this algorithm is an out-of-place method that requires more scratchpad memory and thus cannot scale to longer sequences. Because the basic bitonic merge and the basic oddeven merge in general do not show better performance and cannot simply deal with data of arbitrary sizes, none of them is chosen. The advanced bitonic sort method is always the slowest because it loads data from the scratchpad memory to thread private memory (register file or an off-chip memory space) for data locality. However, due to the small or negatively large latency gap between the scratchpad memory and the thread private memory, the load operations actually reduce the overall performance. Thus this method should only be used for migrating global memory access to scratchpad memory access.

We can also see that the AMD Radeon R9 290X GPU is almost always much faster than the two nVidia GPUs in all tests. The reason is that the capacity of the scratchpad memory (2816 kB, 64 kB/core $\times$ 44 cores, in the AMD GPU, 1536 kB, 96 kB/core $\times$ 16 cores, in the nVidia Maxwell-based GTX 980 GPU and 720 kB, 48 kB/core $\times$ 15 cores, in the nVidia Kepler-based GTX Titan Black GPU) heavily influence the performance of merging small sequences. For the same reason, the GTX 980 GPU delivers better overall performance than the GTX Titan Black GPU. On the other hand, even though the AMD GPU has 64 kB scratchpad memory per core, each instance of the kernel program can only use up to 32 kB. Thus the AMD GPU cannot scale to longer sub-sequences (e.g., $2^{12}$ with 32-bit key-32-bit value pairs) that can be executed by using the nVidia GPUs.

\section{Experimental Results}

\subsection{Testbeds}
We use four platforms (one CPU and three GPUs) shown in Table~\ref{spgemm.jpdc.tab.testbeds} for evaluating the SpGEMM algorithms. The host side of all GPUs is a quad-core 3.7GHz CPU in an AMD A10-7850K APU with 8 GB DDR3-1600 dual-channel system memory and 64-bit Ubuntu Linux 14.04.

\begin{table*}[h!t]
\scriptsize
\renewcommand{\arraystretch}{1.3}
\caption{One CPU and three GPUs used for benchmarking}
\label{spgemm.jpdc.tab.testbeds}
\centering
\begin{tabular}{p{2cm} p{2.3cm} p{2.3cm} p{2.3cm} p{2.55cm}} 
\hline
Vendor & Intel  & nVidia  & nVidia & AMD\\ 
\hline
Family & Xeon CPU & GeForce GPU & GeForce GPU & Radeon GPU\\ 
\hline
Device & E5-2630 & GTX Titan Black & GTX 980 & R9 290X\\ 
\hline
Codename & Sandy Bridge & Kepler GK110 & Maxwell GM204 & GCN Hawaii\\ 
\hline
\hline
\#Cores & 6 & 15 & 16 & 44\\ 
\hline
\#SIMD units & 6$\times$256-bit wide & 2880 CUDA cores & 2048 CUDA cores & 2816 Radeon cores\\ 
\hline
Clock & 2.3 GHz & 889 MHz & 1126 MHz & 1050 MHz\\ 
\hline
SP flop/cycle & 96 & 5760 & 4096 & 5632\\ 
\hline
SP Peak & 220.8 GFlop/s & 5120.6 GFlop/s & 4612.1 GFlop/s & 5913.6 GFlop/s\\ 
\hline
DP flop/cycle & 48 & 1920 & 128 & 704\\ 
\hline
DP Peak & 110.4 GFlop/s & 1706.9 GFlop/s & 144.1 GFlop/s & 739.2 GFlop/s\\ 
\hline
On-chip scratchpad & N/A & 720 kB & 1536 kB & 2816 kB\\ 
\hline
\hline
Memory & 32 GB DDR3-1333 (4 channels) & 6 GB GDDR5  & 4 GB GDDR5 & 4 GB GDDR5\\ 
\hline
Bandwidth & 42.6 GB/s & 336 GB/s & 224 GB/s & 345.6 GB/s\\ 
\hline
\hline
OS (64-bit) & Ubuntu 12.04 & Ubuntu 14.04 & Ubuntu 14.04 & Ubuntu 14.04\\ 
\hline
Device driver & N/A & v344.16 & v344.16 & v14.41\\ 
\hline
Compiler & Intel C++ v14.0 & g++ v4.9, \newline nvcc v6.5.19 & g++ v4.9, \newline nvcc v6.5.19  & g++ v4.9, \newline OpenCL v1.2\\ 
\hline
Library & Intel MKL v11.0 & CUSP v0.4.0, \newline cuSPARSE v6.5, \newline RMerge, \newline bhSPARSE & CUSP v0.4.0, \newline cuSPARSE v6.5, \newline RMerge, \newline bhSPARSE & bhSPARSE\\ 
\hline
\end{tabular}
\end{table*}



\subsection{Performance Comparison for Galerkin Products}

Calculating Galerkin products plays an important role in AMG. We use smoothed aggregation preconditioner with Jacobi smoother (described in~\cite{Bell:Exposing} and implemented in the CUSP library~\cite{Dalton:CUSP}) as a test scenario for evaluating SpGEMM algorithms. In each level of an AMG hierarchy in this context, we multiply three sparse matrices $P^T$, $A$ and $P$, where rectangular matrix $P^T$ is a restriction operator, square matrix $A$ is initially the system matrix, and rectangular matrix $P$ is a prolongation operator.

\begin{figure*}[h!t]
\captionsetup[subfigure]{labelformat=empty}
\subfloat[(a) 2D 5-point]{\includegraphics[width=1.33in]{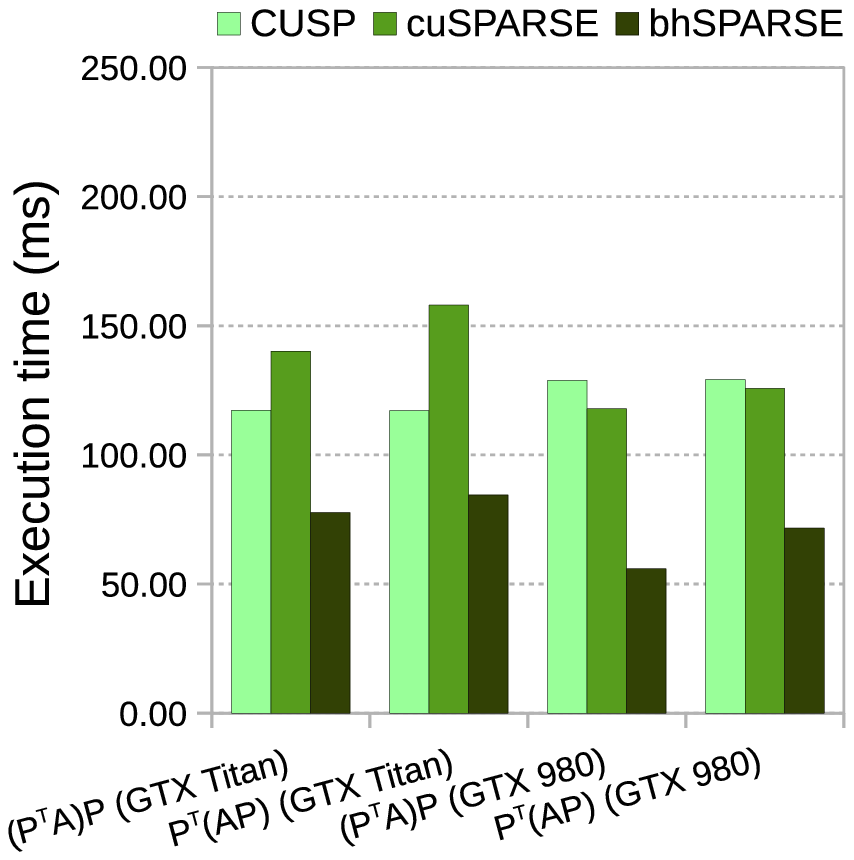}}
\subfloat[(b) 2D 9-point]{\includegraphics[width=1.33in]{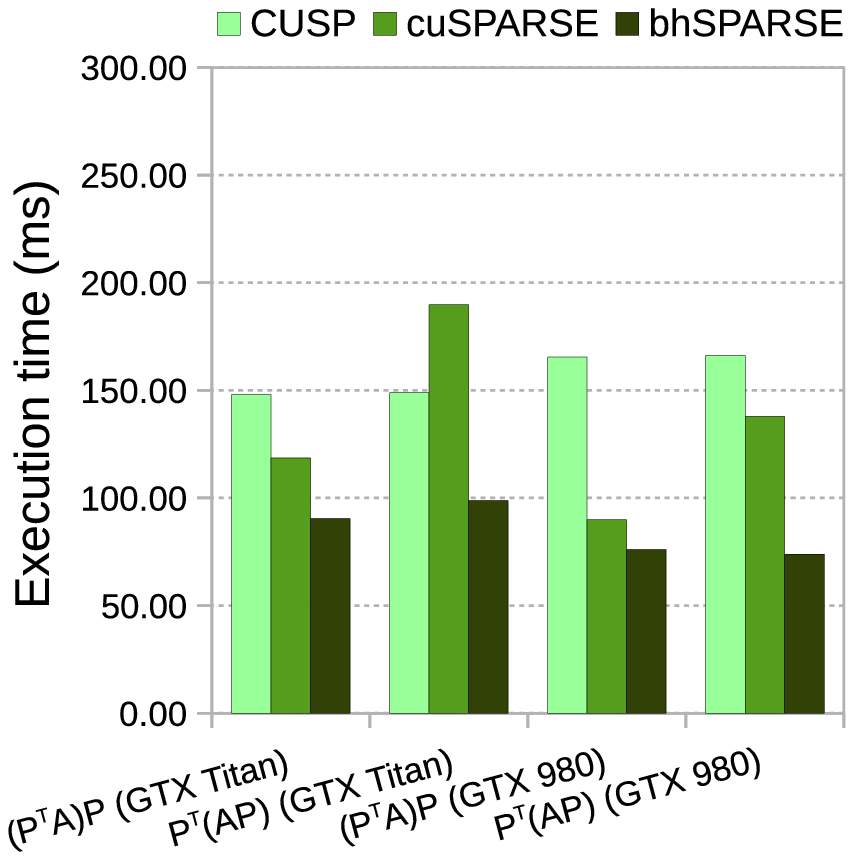}}
\subfloat[(c) 3D 7-point]{\includegraphics[width=1.33in]{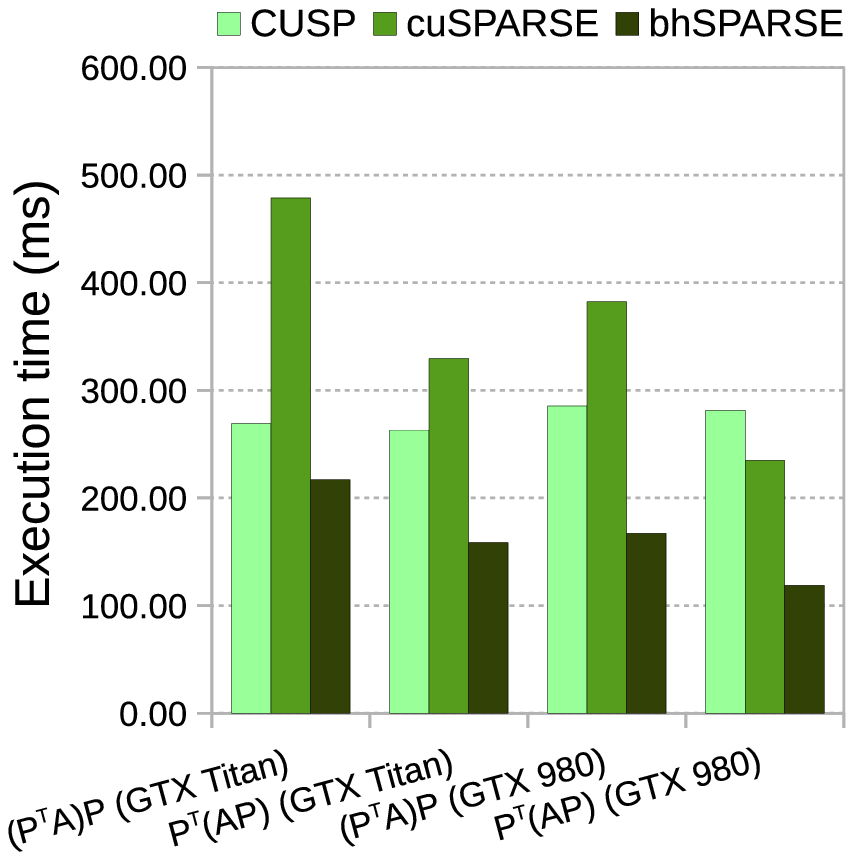}}
\subfloat[(d) 3D 27-point]{\includegraphics[width=1.33in]{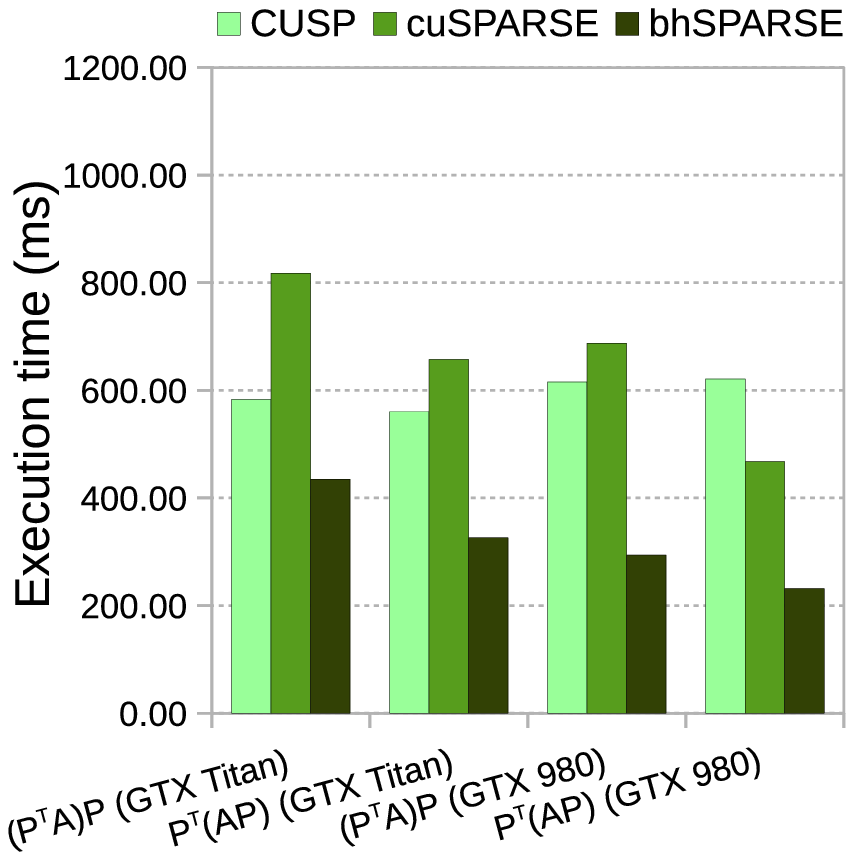}}
\caption{Execution time (in milliseconds) comparison of single precision SpGEMM (SpSGEMM) from three libraries CUSP, cuSPARSE and bhSPARSE in the context of  smoothed aggregation preconditioner with Jacobi smoother. The system matrices are from four Poisson problems. Both $(P^TA)P$ and $P^T(AP)$ are tested on two nVidia GPUs.}
\label{spgemm.jpdc.fig.amg-sp}
\end{figure*}

\begin{figure*}[h!t]
\captionsetup[subfigure]{labelformat=empty}
\subfloat[(a) 2D 5-point]{\includegraphics[width=1.33in]{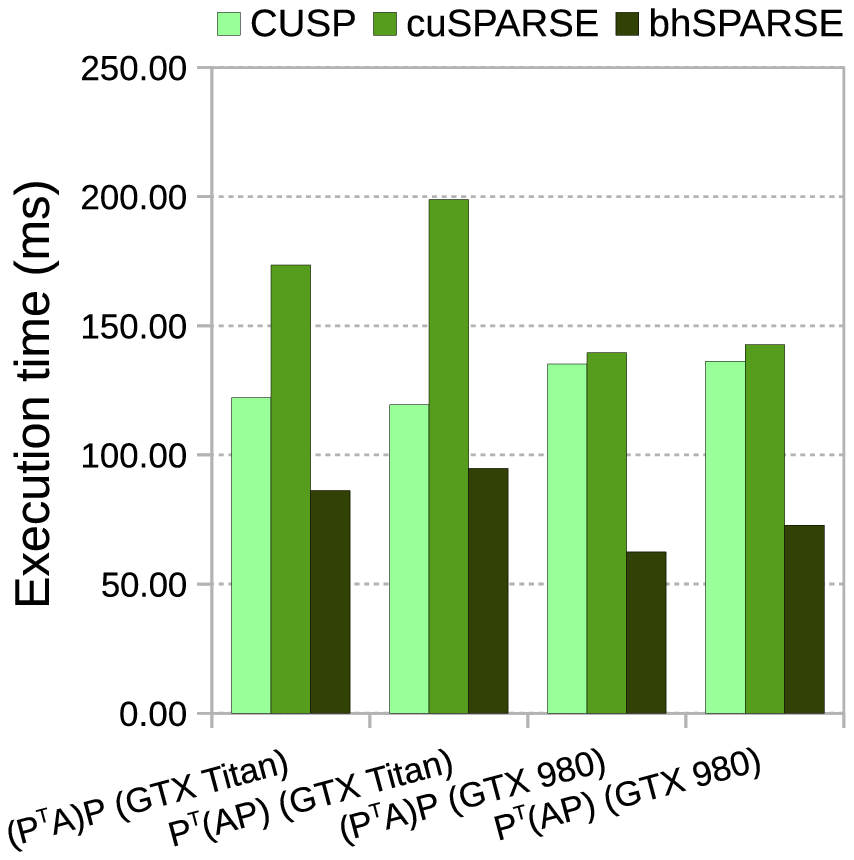}}
\subfloat[(b) 2D 9-point]{\includegraphics[width=1.33in]{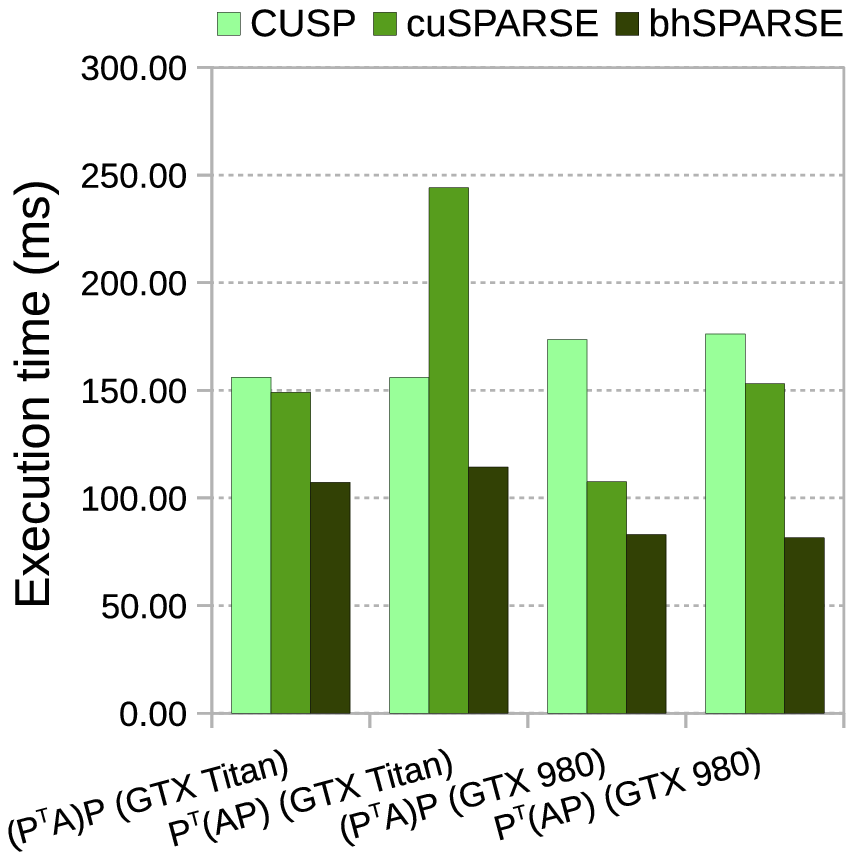}}
\subfloat[(c) 3D 7-point]{\includegraphics[width=1.33in]{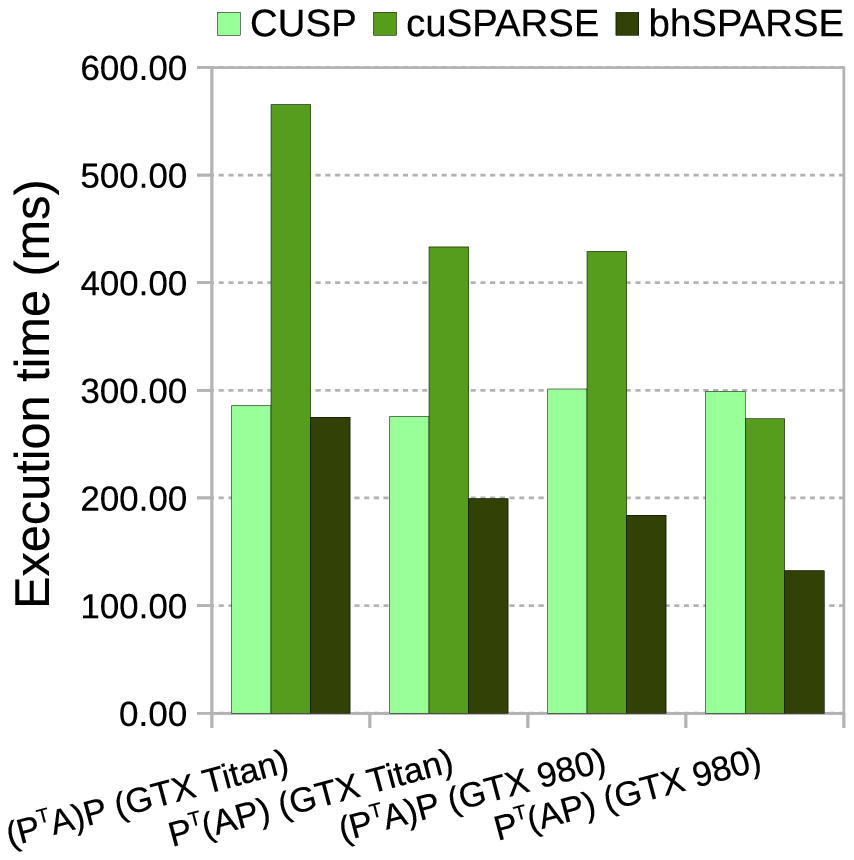}}
\subfloat[(d) 3D 27-point]{\includegraphics[width=1.33in]{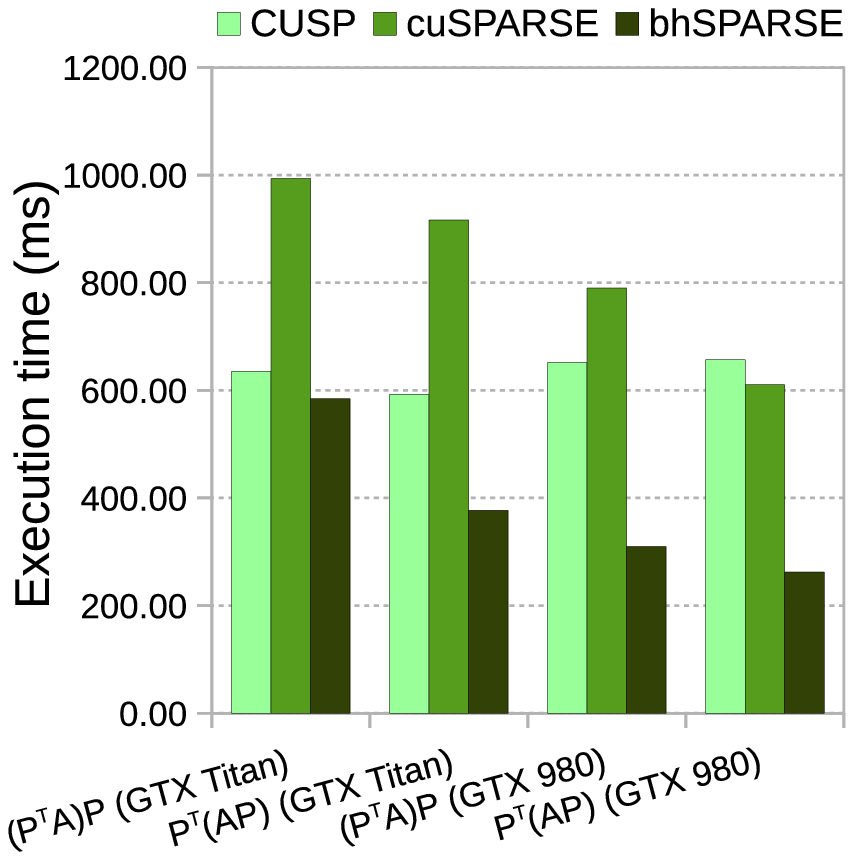}}
\caption{Execution time (in milliseconds) comparison of double precision SpGEMM (SpDGEMM) from three libraries CUSP, cuSPARSE and bhSPARSE in the context of  smoothed aggregation preconditioner with Jacobi smoother. The system matrices are from four Poisson problems. Both $(P^TA)P$ and $P^T(AP)$ are tested on two nVidia GPUs.}
\label{spgemm.jpdc.fig.amg-dp}
\end{figure*}

Figures~\ref{spgemm.jpdc.fig.amg-sp} and~\ref{spgemm.jpdc.fig.amg-dp} show execution time of Galerkin products $P^TAP$ in constructing an AMG hierarchy (typically including 3-5 levels) for a smoothed aggregation preconditioner in single precision and double precision, respectively. The input  system matrix $A$ is from 2D 5-point, 2D 9-point, 3D 7-point or 3D 27-point Poisson problem, respectively. The two 2D problems have dimensions $1024\times 1024$ and generate system matrices of size $1048576\times 1048576$. The two 3D problems have dimensions $101\times 101\times 101$ and generate system matrices of size $1030301\times 1030301$. The SpGEMM approaches in three libraries, CUSP v0.4.0, cuSPARSE v6.5 and bhSPARSE\footnote{We call our library bhSPARSE since this work is under the Project Bohrium~\cite{Kristensen:Bohrium}.}, are tested on nVidia GeForce GTX Titan Black and GeForce GTX 980 GPUs. To obtain the best SpGEMM performance, CUSP uses the coordinate (COO) format for its input matrices. The other two libraries use the CSR format. Because the operation multiplies three sparse matrices $P^T$, $A$ and $P$, the order of multiplication may influence overall performance. Here we test the two possible orders $(P^TA)P$ and $P^T(AP)$. In our experiments, matrix data transfer time between the host and the device is not included since the SpGEMM is normally one of the building blocks for more complex problem completely running on GPUs.

In Figures~\ref{spgemm.jpdc.fig.amg-sp} and~\ref{spgemm.jpdc.fig.amg-dp}, we can see that our method is constantly faster than SpGEMM algorithms in the other two libraries. When using system matrix from 3D 27-point Poisson problem, bhSPARSE delivers up to 2.6x and up to 2.7x speedups over cuSPARSE and CUSP, respectively. On average, speedups of 1.9x and 1.7x are achieved when compared with the above two libraries, respectively.

As for the order of multiplication, we can see that our method in general gives better performance while doing $P^T(AP)$, compared to running $(P^TA)P$. In contrast, the order of multiplication does not bring obvious performance difference for CUSP. When cuSPARSE is used, $(P^TA)P$ delivers better throughput for the two 2D problems, but degrades throughput for the two 3D problems.

\subsection{Benchmark Suite for Evaluating Matrix Squaring}

We also evaluate multiplication of sparse square matrix and itself (i.e., $C = A^2$) to avoid introducing another sparse matrix as a multiplier with different sparsity structure. 
We choose 23 sparse matrices as our benchmark suite. 16 of them were widely used for performance evaluations in previous sparse matrix computation research~\cite{Liu:CSR5, Liu:Speculative, Demouth:Sparse, Dalton:Optimizing, Gremse:GPU, Intel:MKL, Saule:Performance, Williams:Optimization, Buluc:Challenges}. The other 7 new matrices are chosen since they bring more diverse irregular sparsity structures that challenge the SpGEMM algorithm design. The variety of sparsity structures are from many application fields, such as finite element methods, macroeconomic model, protein data, circuit simulation, web connectivity and combinational problem. All of the 23 matrices are downloadable from the University of Florida Sparse Matrix Collection~\cite{Davis:The}. Note that symmetry in the sparse matrices is not used in our SpGEMM algorithm, although some matrices in the benchmark suite are symmetric. Also note that we use the standard CSR format that does not consider symmetric storage pattern.

Besides the input matrix $A$, the work complexities of the different SpGEMM algorithms also
depend on the intermediate matrix $\widehat{C}$ and the resulting matrix $C$. So we list characteristics of the three matrices in Table~\ref{spgemm.jpdc.tab.benchmarksuite}. The set of characteristics includes matrix dimension ($n$), the number of nonzero entries ($nnz$) and the average number of nonzero entries in rows ($nnzr$). The upper 9 matrices in the table have relatively regular nonzero entry distribution mostly on the diagonal. The other 14 matrices include various irregular sparsity structures.

\begin{table*}[h!t]
\scriptsize 
\renewcommand{\arraystretch}{1.8}
\caption{Overview of sparse matrices for benchmarking matrix squaring. Here $nnz(\widehat{C})$ is the upper bound size of $A^2$. Numerically, $nnz(\widehat{C})$ equals to half of $flops$, the number of necessary arithmetic operations while doing SpGEMM. $nnz(C)$ is the number of nonzero entries in the resulting matrix $C=A^2$.}
\label{spgemm.jpdc.tab.benchmarksuite}
\centering
\begin{tabular}{
>{\arraybackslash}m{0.17\columnwidth}
>{\arraybackslash}m{0.04\columnwidth}
>{\raggedleft\arraybackslash}m{0.07\columnwidth}
>{\centering\arraybackslash}m{0.17\columnwidth}
>{\centering\arraybackslash}m{0.17\columnwidth}
>{\centering\arraybackslash}m{0.17\columnwidth}
}
\hline
\textbf{Name} 
& \centering \textbf{Plot}
& \centering \boldmath{$n$}
& \begin{minipage}[c]{0.17\columnwidth} \centering \boldmath{$nnz(A)$, $nnzr(A)$} \end{minipage} 
& \begin{minipage}[c]{0.17\columnwidth} \centering \boldmath{$nnz(\widehat{C})$, $nnzr(\widehat{C})$} \end{minipage} 
& \begin{minipage}[c]{0.17\columnwidth} \centering \boldmath{$nnz(C)$, $nnzr(C)$} \end{minipage} \\
\hline
FEM/Cantilever
& \begin{minipage}[c]{0.045\columnwidth} \includegraphics[width=0.24in]{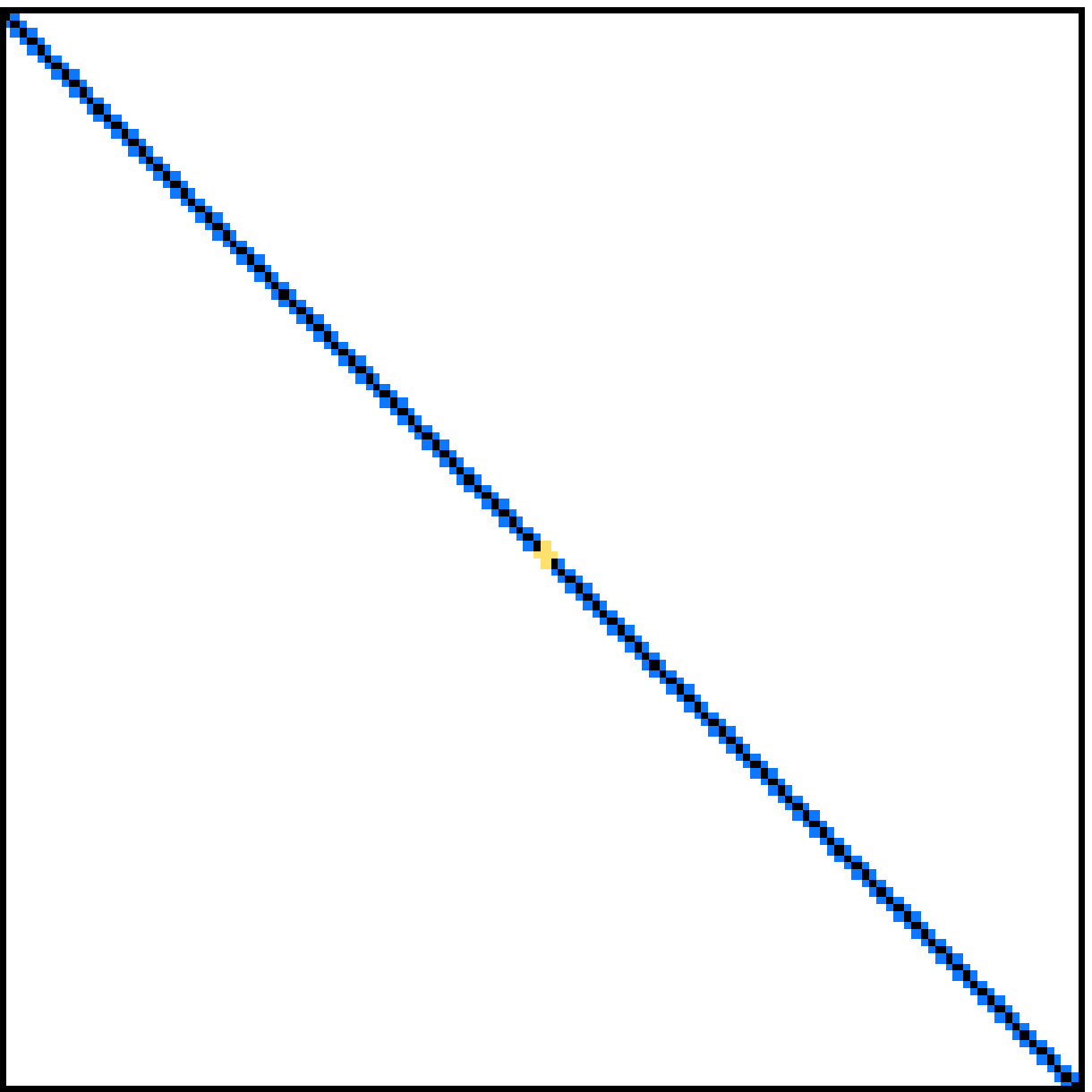} \end{minipage} 
& 63 K
& \begin{minipage}[c]{0.17\columnwidth} \centering  4 M, 64 \end{minipage} 
& \begin{minipage}[c]{0.17\columnwidth} \centering  269.5 M, 4315 \end{minipage} 
& \begin{minipage}[c]{0.17\columnwidth} \centering  17.4 M, 279 \end{minipage}  
\\ 
Economics 
& \begin{minipage}[c]{0.045\columnwidth} \centering \includegraphics[width=0.24in]{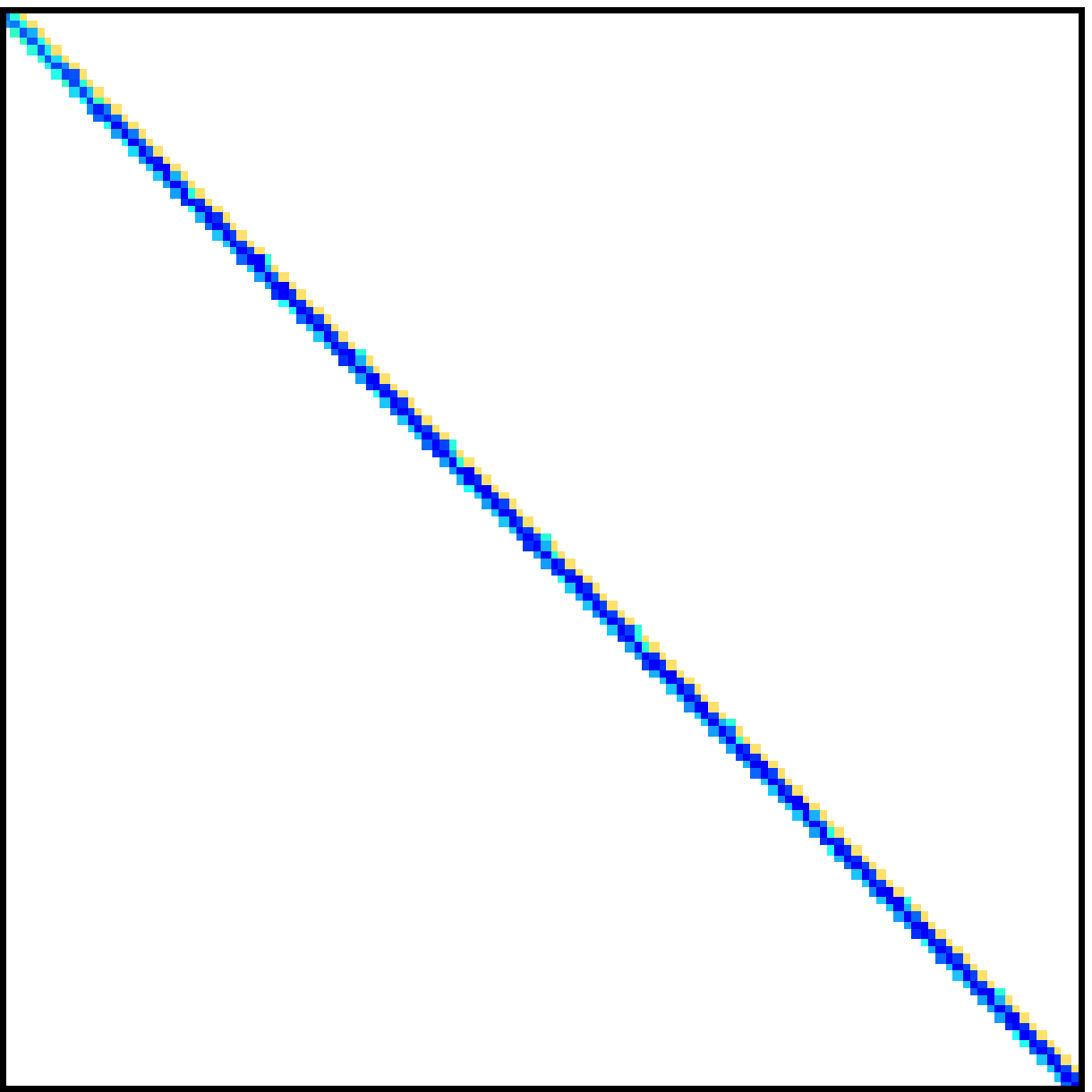} \end{minipage} 
& 207 K 
& \begin{minipage}[c]{0.17\columnwidth} \centering 1.3 M, 6 \end{minipage} 
& \begin{minipage}[c]{0.17\columnwidth} \centering 7.6 M, 37 \end{minipage} 
& \begin{minipage}[c]{0.17\columnwidth} \centering 6.7 M, 32 \end{minipage}
\\ 
Epidemiology 
& \begin{minipage}[c]{0.045\columnwidth} \centering \includegraphics[width=0.24in]{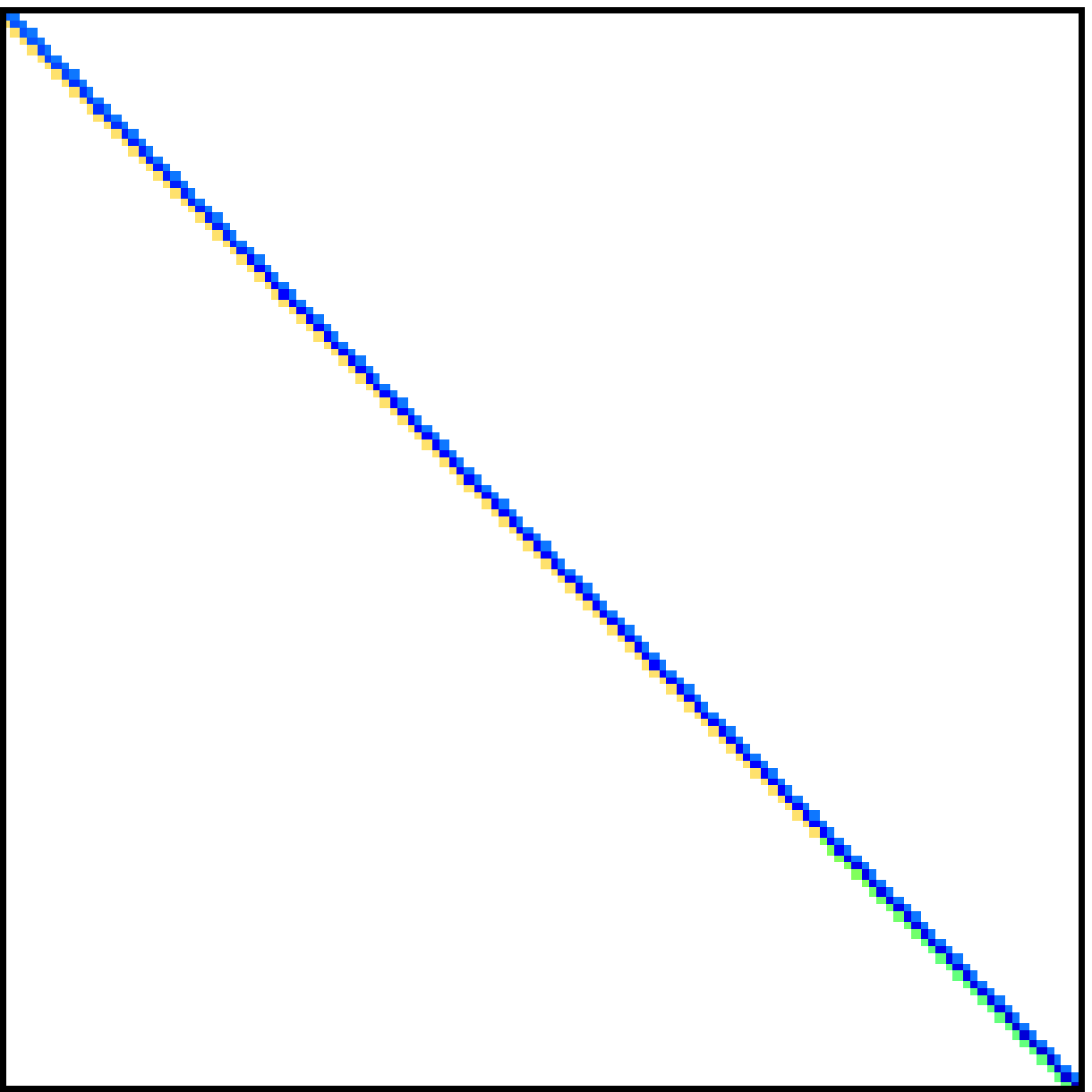} \end{minipage} 
& 526 K 
& \begin{minipage}[c]{0.17\columnwidth} \centering 2.1 M, 4 \end{minipage} 
& \begin{minipage}[c]{0.17\columnwidth} \centering 8.4 M, 16 \end{minipage} 
& \begin{minipage}[c]{0.17\columnwidth} \centering 5.2 M, 10 \end{minipage} 
\\ 
Filter3D 
& \begin{minipage}[c]{0.045\columnwidth} \centering \includegraphics[width=0.24in]{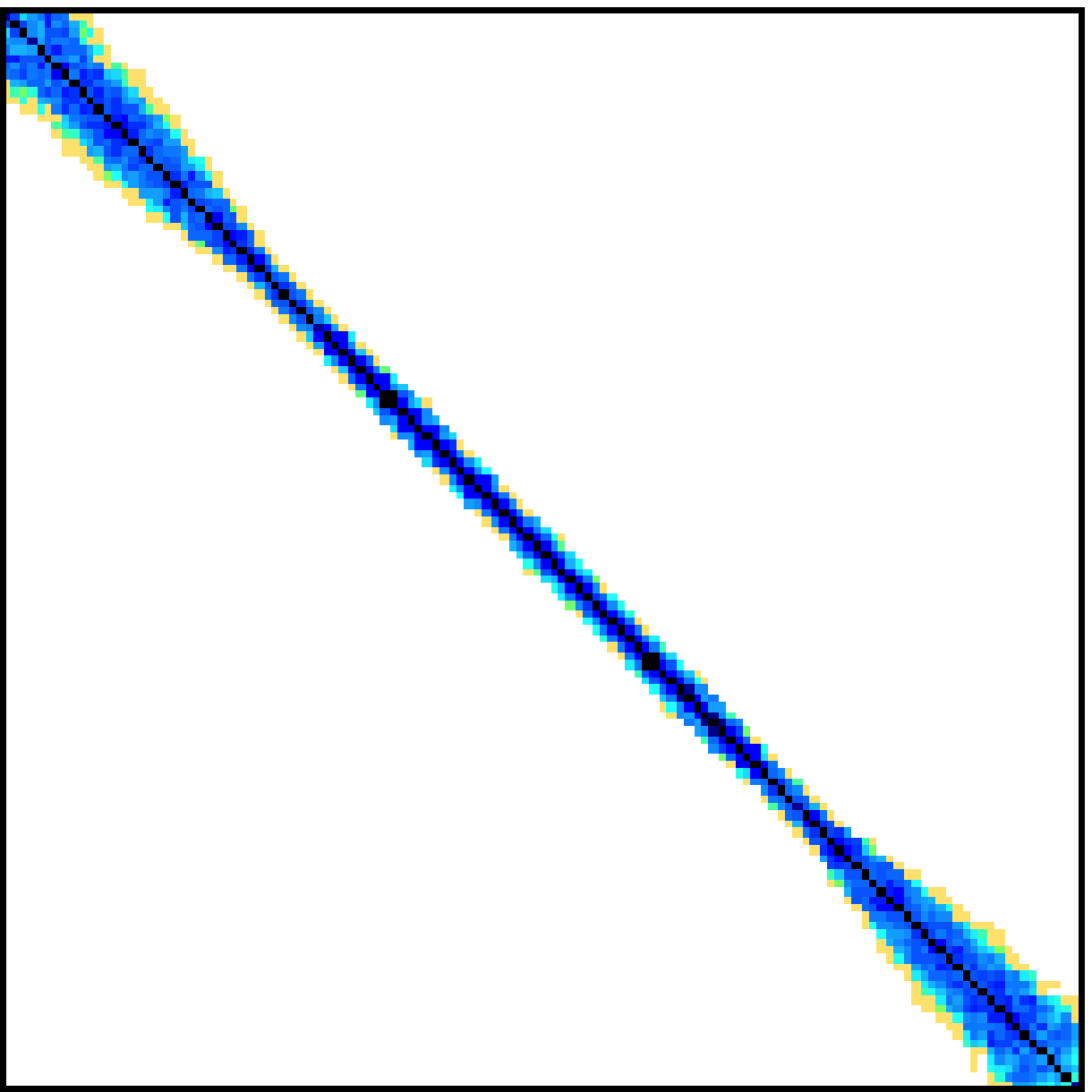} \end{minipage} 
& 106 K 
& \begin{minipage}[c]{0.17\columnwidth} \centering 2.7 M, 25 \end{minipage} 
& \begin{minipage}[c]{0.17\columnwidth} \centering 86 M, 808 \end{minipage} 
& \begin{minipage}[c]{0.17\columnwidth} \centering 20.2 M, 189 \end{minipage} 
\\ 
Wind Tunnel 
& \begin{minipage}[c]{0.045\columnwidth} \centering \includegraphics[width=0.24in]{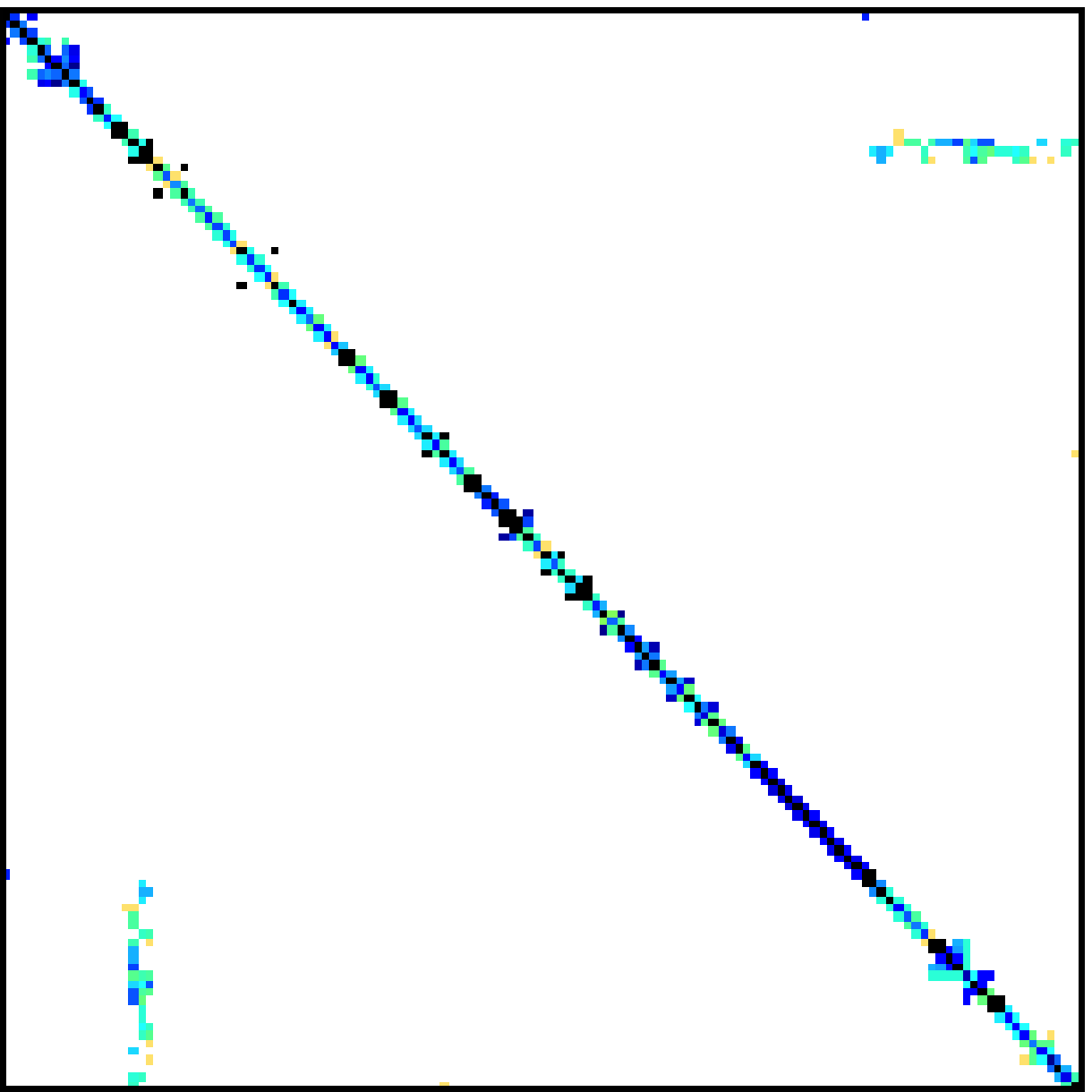} \end{minipage} 
& 218 K 
& \begin{minipage}[c]{0.17\columnwidth} \centering 11.6 M, 53 \end{minipage} 
& \begin{minipage}[c]{0.17\columnwidth} \centering 626.1 M, 2873 \end{minipage} 
& \begin{minipage}[c]{0.17\columnwidth} \centering 32.8 M, 150 \end{minipage} 
\\ 
FEM/Ship 
& \begin{minipage}[c]{0.045\columnwidth} \centering \includegraphics[width=0.24in]{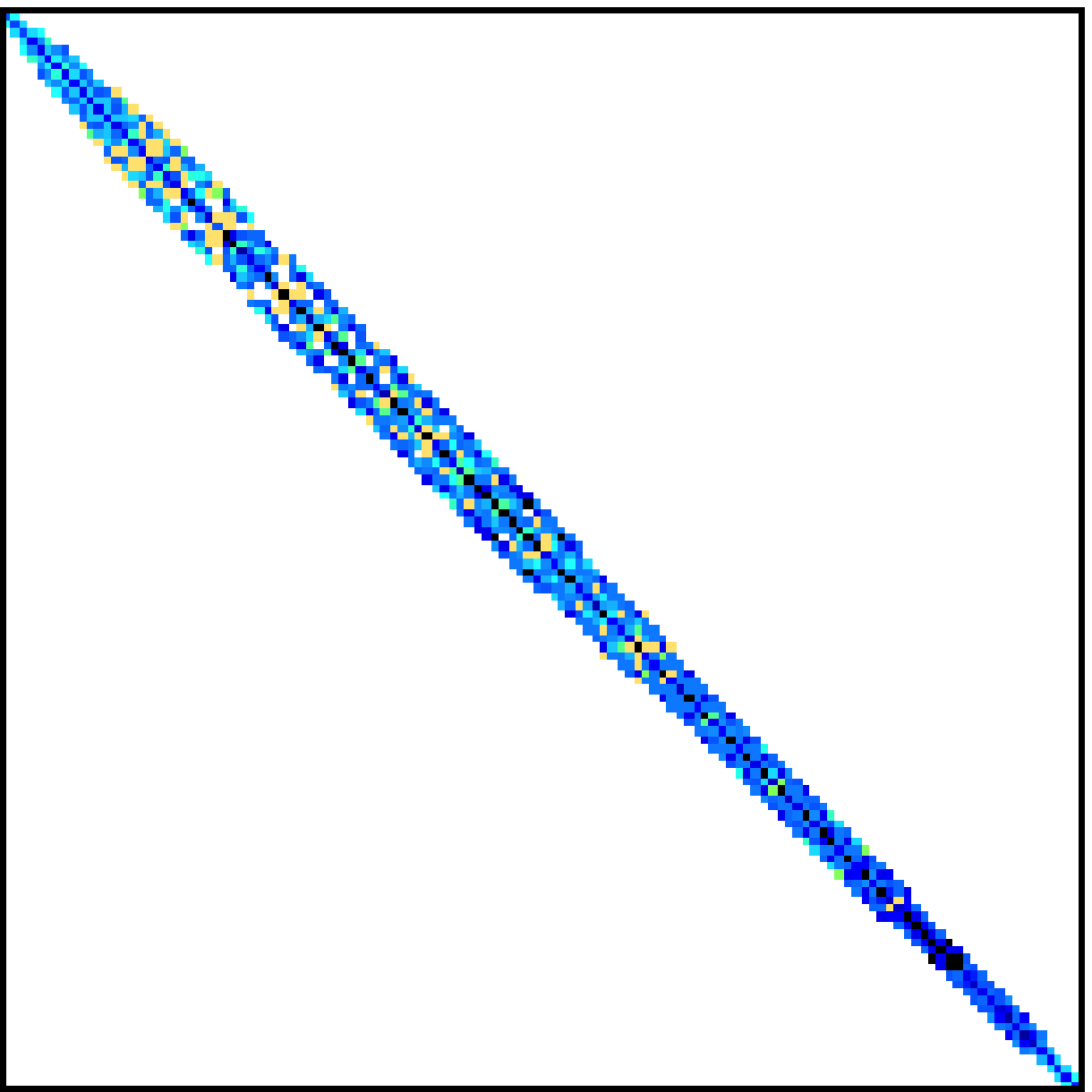} \end{minipage} 
& 141 K 
& \begin{minipage}[c]{0.17\columnwidth} \centering 7.8 M, 55 \end{minipage} 
& \begin{minipage}[c]{0.17\columnwidth} \centering 450.6 M, 3199 \end{minipage} 
& \begin{minipage}[c]{0.17\columnwidth} \centering 24.1 M, 171 \end{minipage} 
\\ 
FEM/Harbor
& \begin{minipage}[c]{0.045\columnwidth} \centering \includegraphics[width=0.24in]{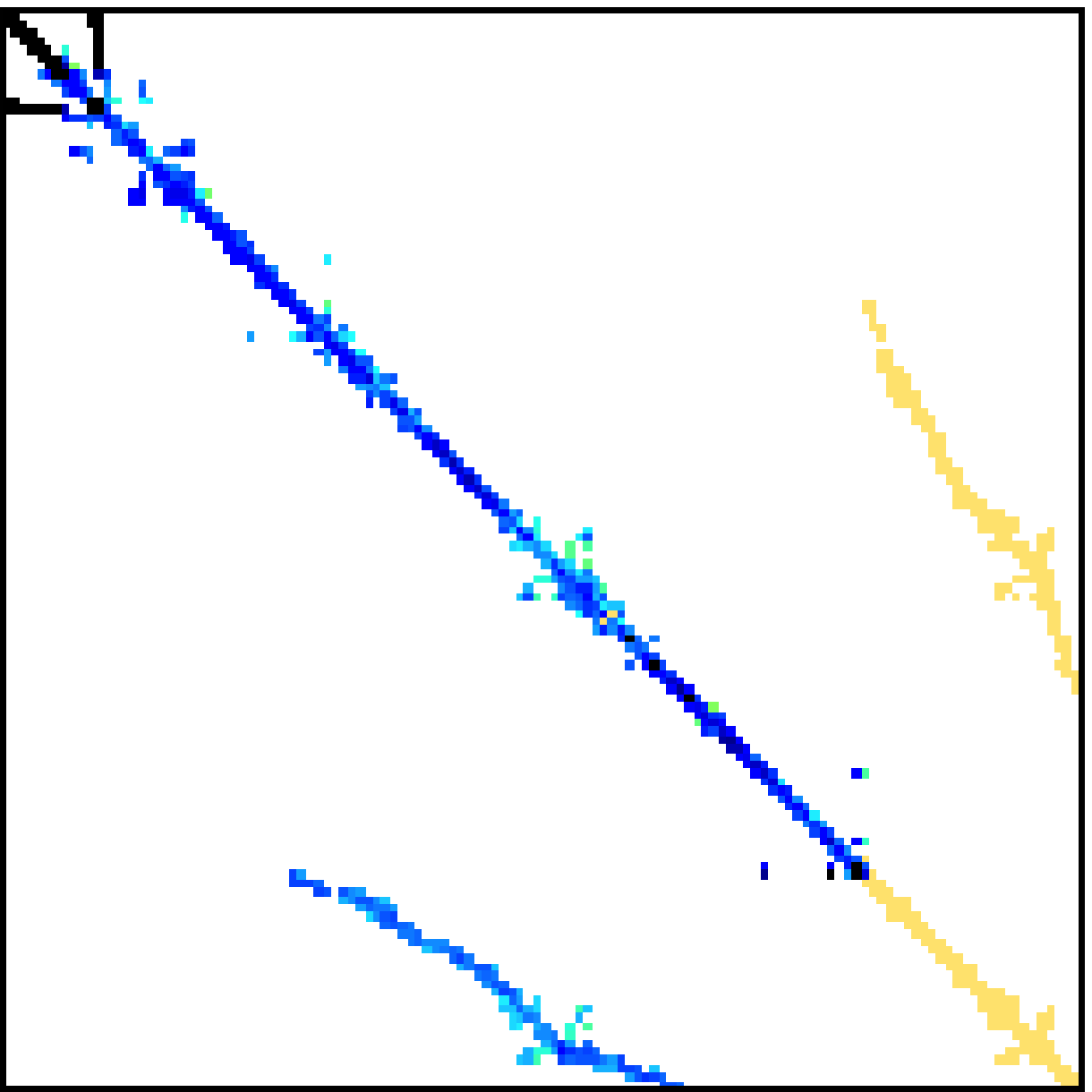} \end{minipage} 
& 47 K 
& \begin{minipage}[c]{0.17\columnwidth} \centering 2.4 M, 51 \end{minipage} 
& \begin{minipage}[c]{0.17\columnwidth} \centering 156.5 M, 3341 \end{minipage} 
& \begin{minipage}[c]{0.17\columnwidth} \centering 7.9 M, 169 \end{minipage}
\\ 
Protein 
& \begin{minipage}[c]{0.045\columnwidth} \centering \includegraphics[width=0.24in]{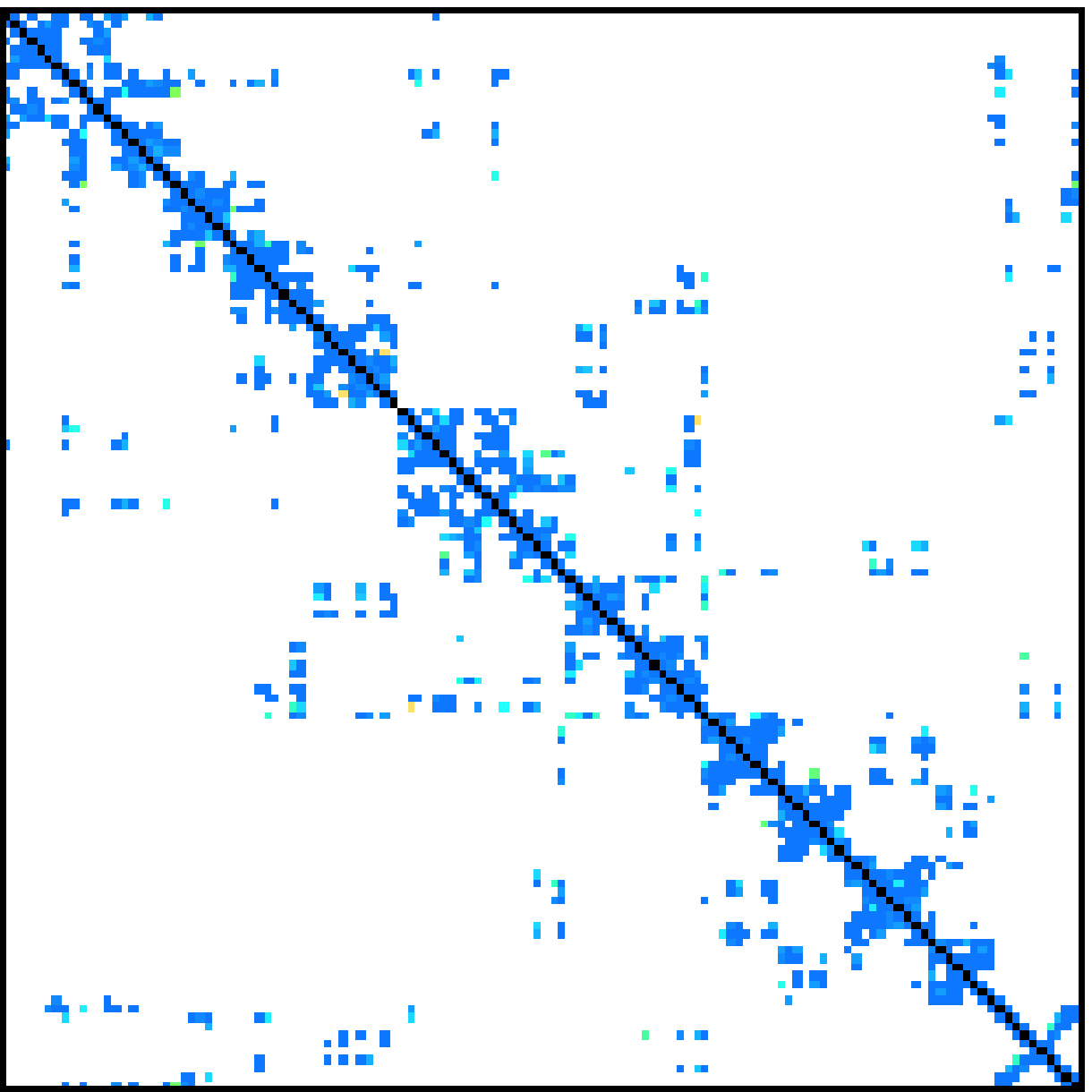} \end{minipage} 
& 36 K 
& \begin{minipage}[c]{0.17\columnwidth} \centering 4.3 M, 119 \end{minipage} 
& \begin{minipage}[c]{0.17\columnwidth} \centering 555.3 M, 15249 \end{minipage} 
& \begin{minipage}[c]{0.17\columnwidth} \centering 19.6 M, 538 \end{minipage} 
\\ 
FEM/Spheres 
& \begin{minipage}[c]{0.045\columnwidth} \centering \includegraphics[width=0.24in]{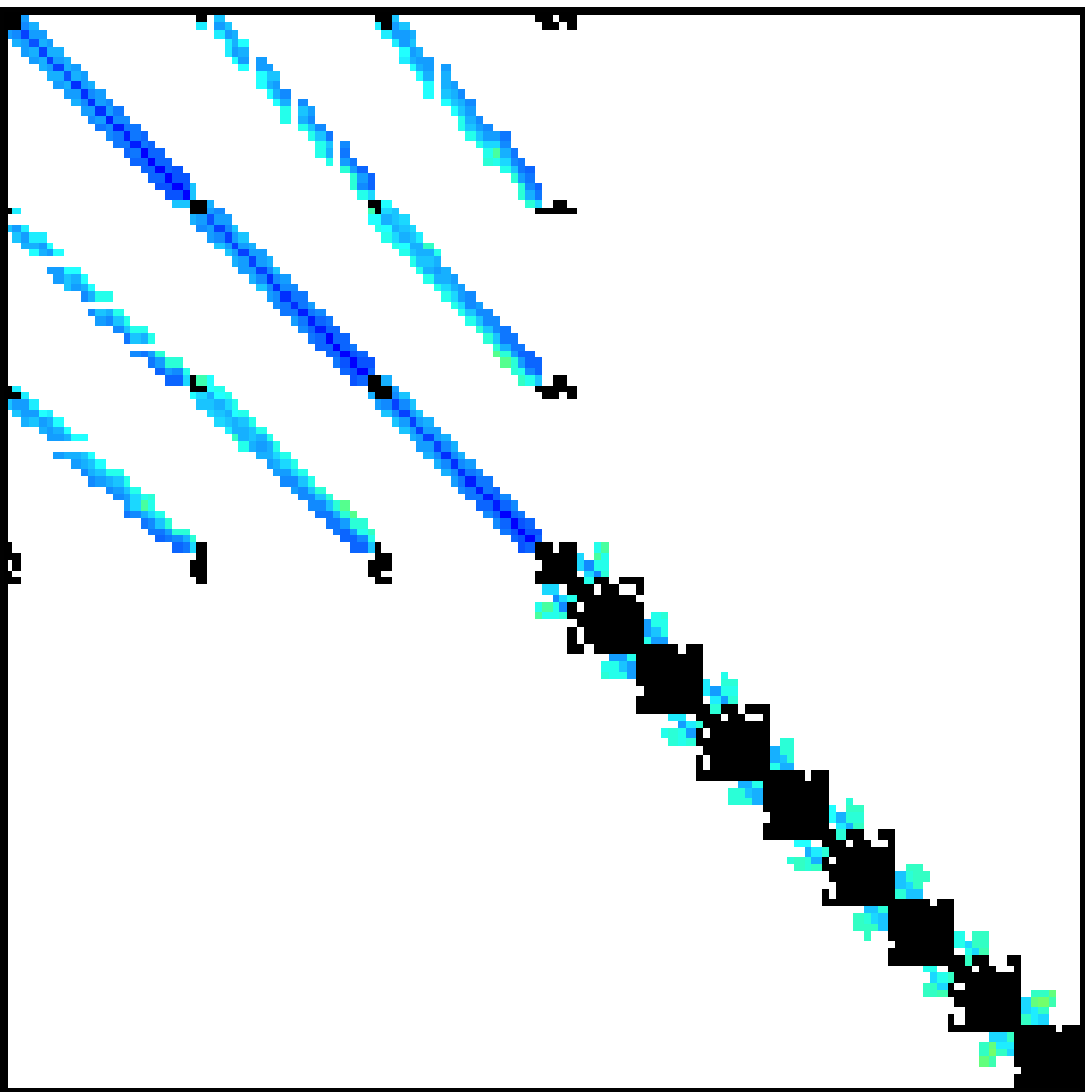} \end{minipage} 
& 83 K 
& \begin{minipage}[c]{0.17\columnwidth} \centering 6 M, 72 \end{minipage} 
& \begin{minipage}[c]{0.17\columnwidth} \centering 463.8 M, 5566 \end{minipage} 
& \begin{minipage}[c]{0.17\columnwidth} \centering 26.5 M, 318 \end{minipage}
\\ 
\hline
2cubes\_sphere
& \begin{minipage}[c]{0.045\columnwidth} \centering \includegraphics[width=0.24in]{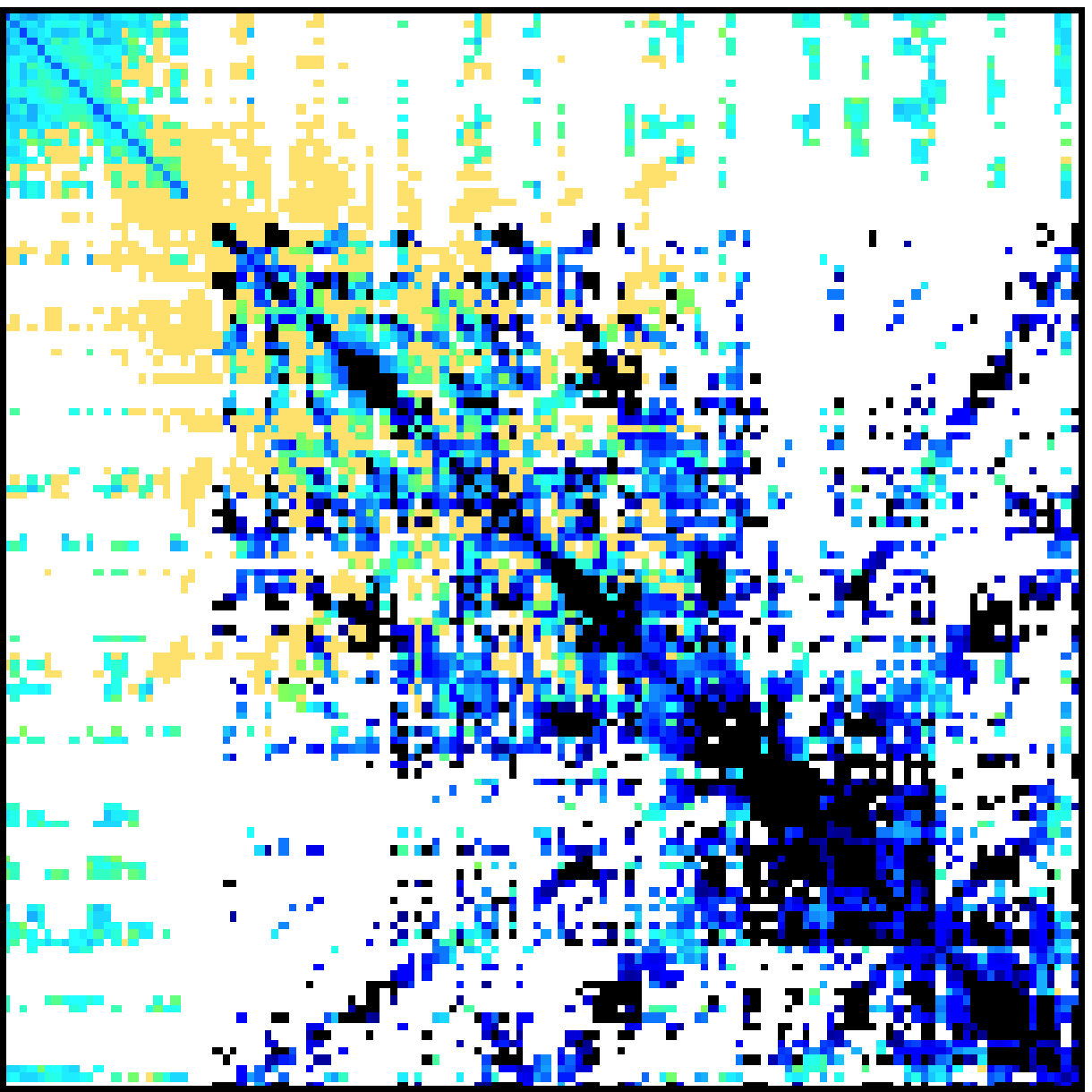} \end{minipage} 
& 102 K 
& \begin{minipage}[c]{0.17\columnwidth} \centering 1.6 M, 16 \end{minipage} 
& \begin{minipage}[c]{0.17\columnwidth} \centering 27.5 M, 270 \end{minipage} 
& \begin{minipage}[c]{0.17\columnwidth} \centering 9 M, 88 \end{minipage} 
\\ 
FEM/Accelerator 
& \begin{minipage}[c]{0.045\columnwidth} \centering \includegraphics[width=0.24in]{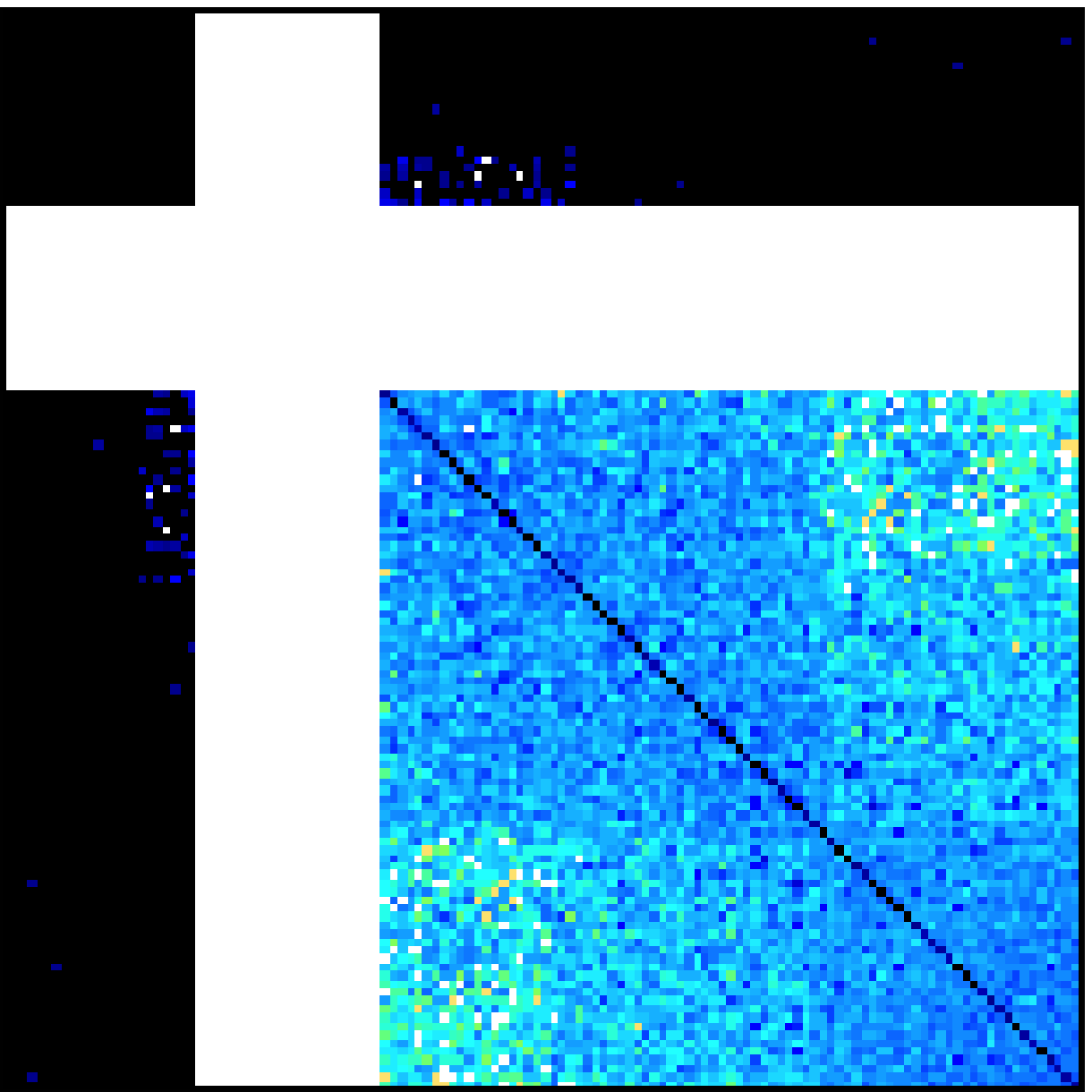} \end{minipage} 
& 121 K 
& \begin{minipage}[c]{0.17\columnwidth} \centering 2.6 M, 22 \end{minipage} 
& \begin{minipage}[c]{0.17\columnwidth} \centering 79.9 M, 659 \end{minipage} 
& \begin{minipage}[c]{0.17\columnwidth} \centering 18.7 M, 154 \end{minipage} 
\\ 
Cage12 
& \begin{minipage}[c]{0.045\columnwidth} \centering \includegraphics[width=0.24in]{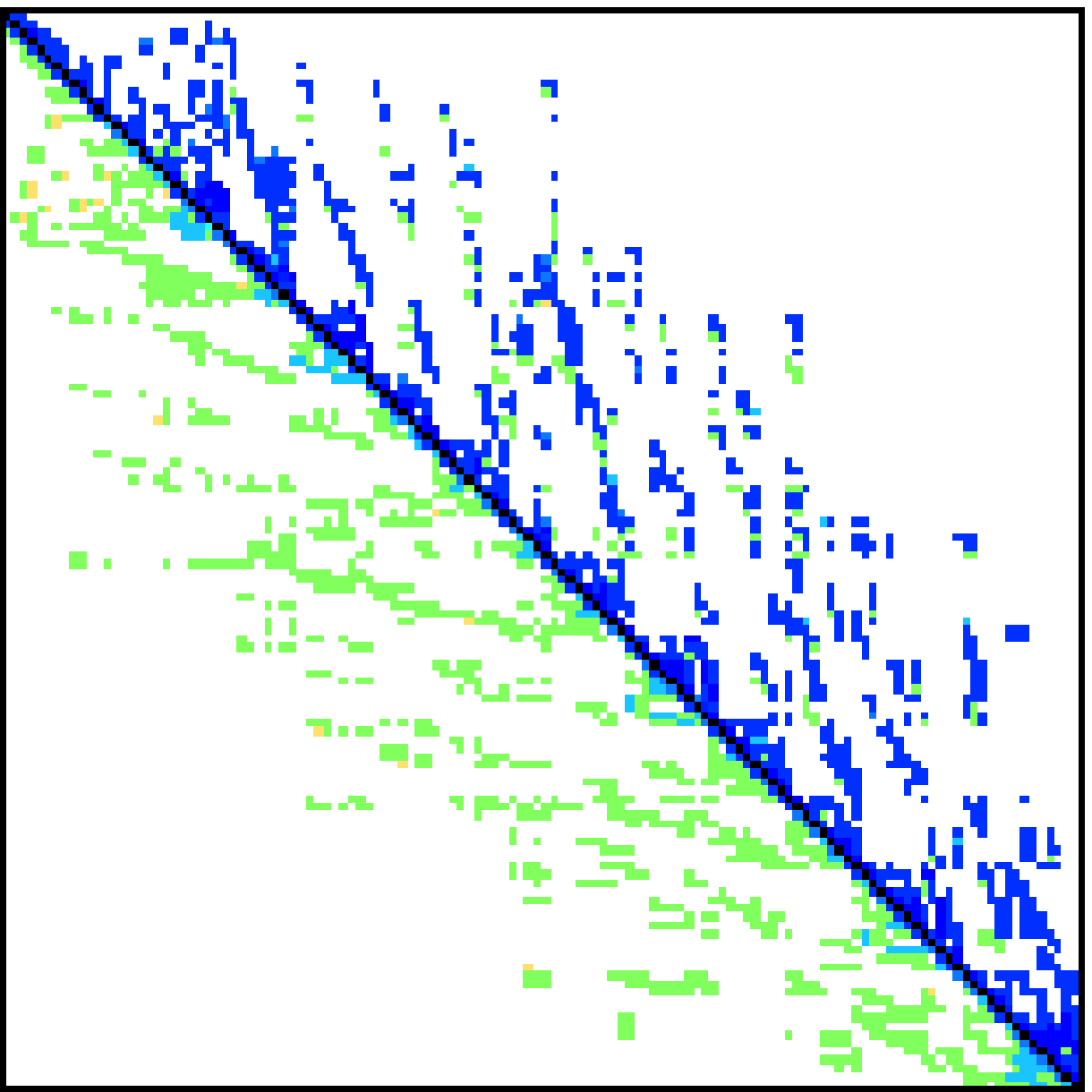} \end{minipage} 
& 130 K 
& \begin{minipage}[c]{0.17\columnwidth} \centering 2 M, 16 \end{minipage} 
& \begin{minipage}[c]{0.17\columnwidth} \centering 34.6 M, 266 \end{minipage} 
& \begin{minipage}[c]{0.17\columnwidth} \centering 15.2 M, 117 \end{minipage}
\\ 
Hood
& \begin{minipage}[c]{0.045\columnwidth} \centering \includegraphics[width=0.24in]{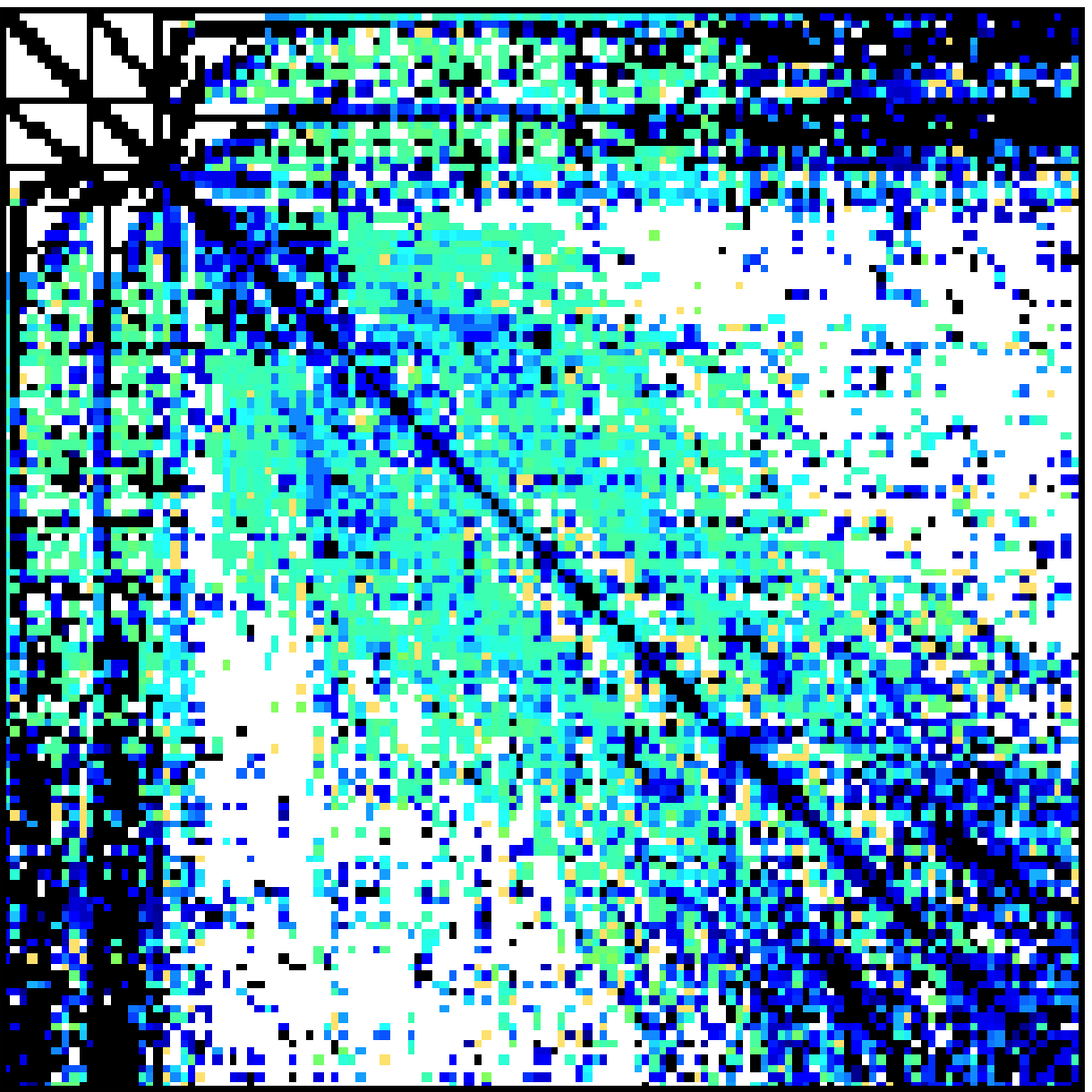} \end{minipage} 
& 221 K 
& \begin{minipage}[c]{0.17\columnwidth} \centering 10.8 M, 49 \end{minipage} 
& \begin{minipage}[c]{0.17\columnwidth} \centering 562 M, 2548 \end{minipage} 
& \begin{minipage}[c]{0.17\columnwidth} \centering 34.2 M, 155 \end{minipage}
\\  
M133-b3 
& \begin{minipage}[c]{0.045\columnwidth} \centering \includegraphics[width=0.24in]{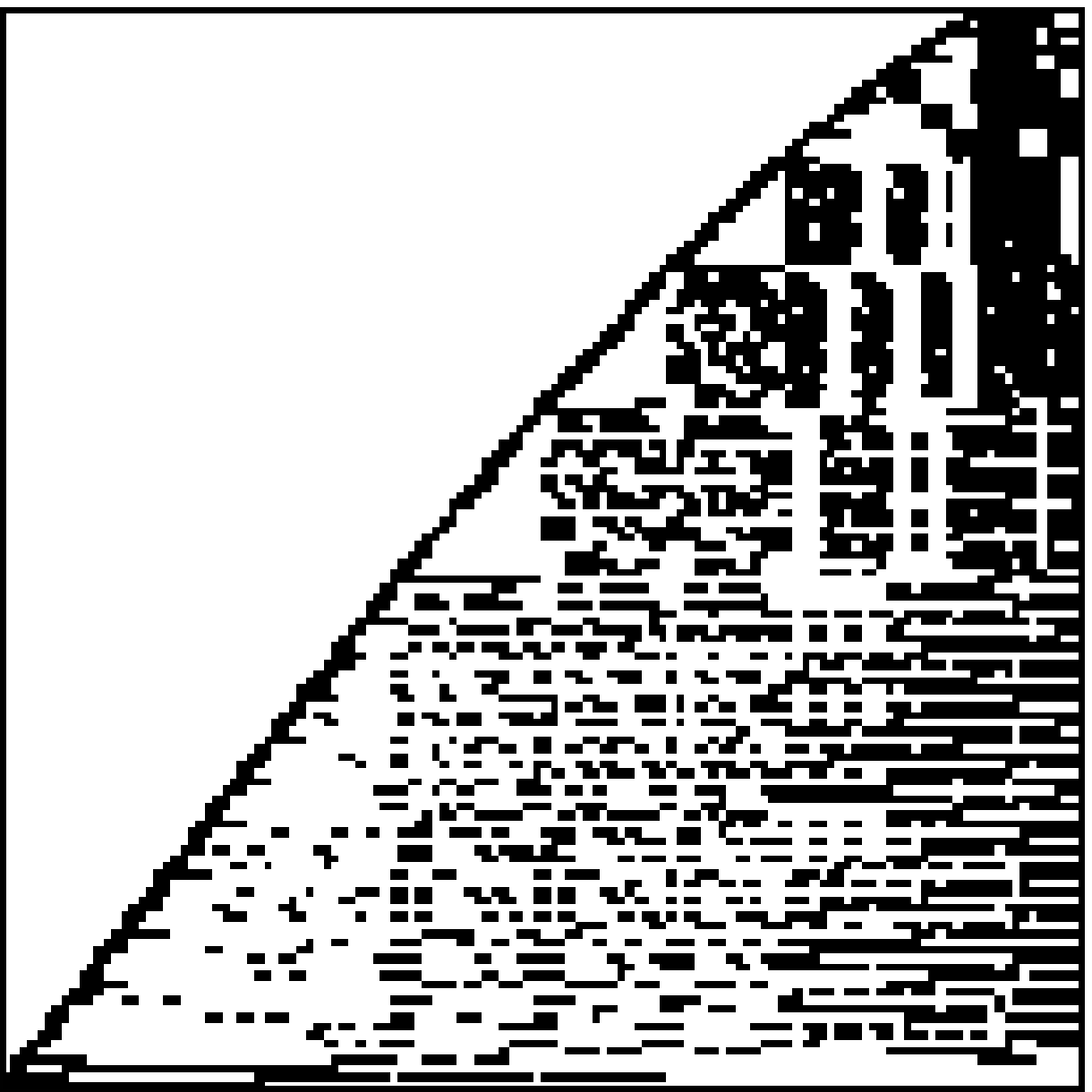} \end{minipage} 
& 200 K 
& \begin{minipage}[c]{0.17\columnwidth} \centering 0.8 M, 4 \end{minipage} 
& \begin{minipage}[c]{0.17\columnwidth} \centering 3.2 M, 16 \end{minipage} 
& \begin{minipage}[c]{0.17\columnwidth} \centering 3.2 M, 16 \end{minipage}
\\ 
Majorbasis 
& \begin{minipage}[c]{0.045\columnwidth} \centering \includegraphics[width=0.24in]{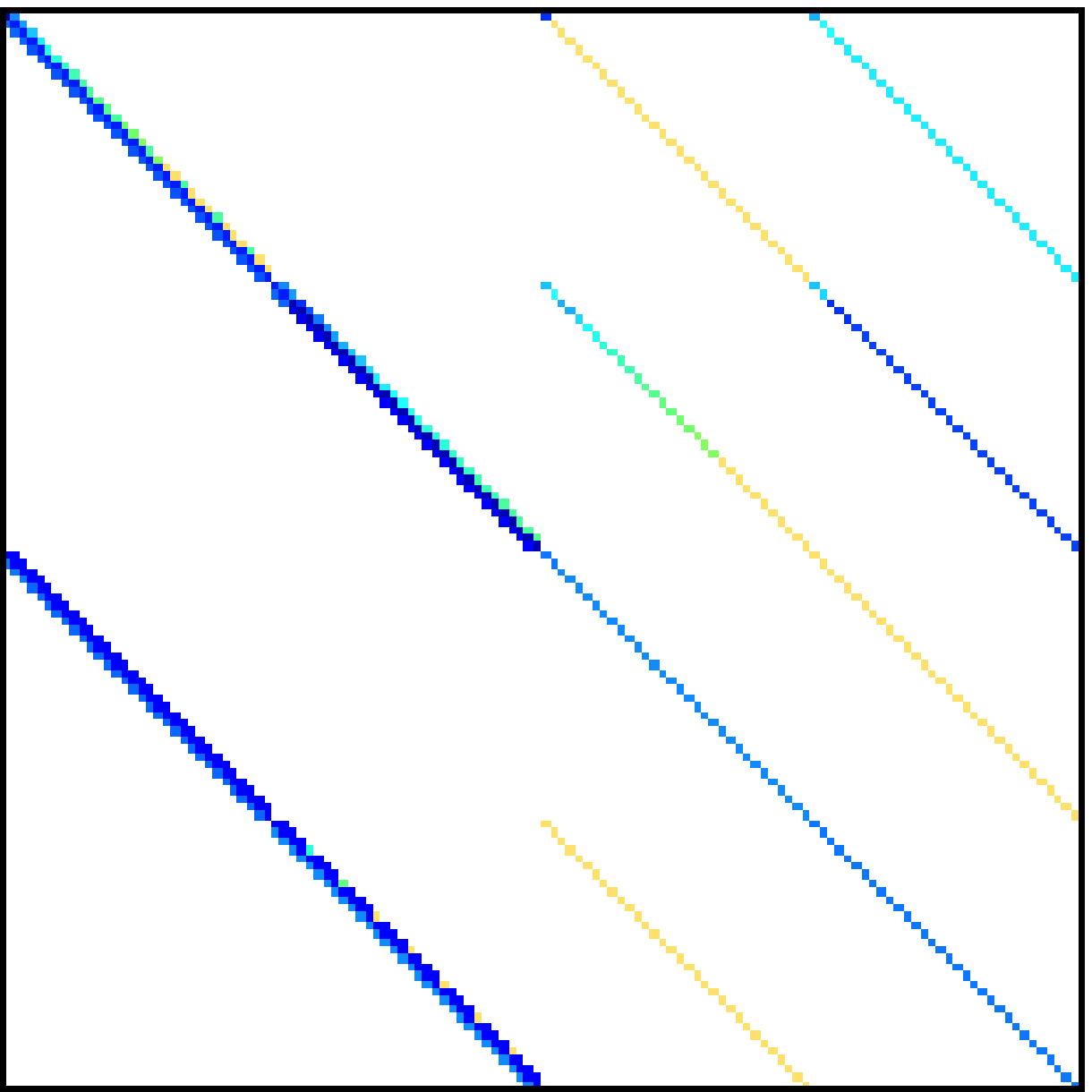} \end{minipage} 
& 160 K 
& \begin{minipage}[c]{0.17\columnwidth} \centering 1.8 M, 11 \end{minipage} 
& \begin{minipage}[c]{0.17\columnwidth} \centering 19.2 M, 120 \end{minipage} 
& \begin{minipage}[c]{0.17\columnwidth} \centering 8.2 M, 52 \end{minipage} 
\\ 
Mario002 
& \begin{minipage}[c]{0.045\columnwidth} \centering \includegraphics[width=0.24in]{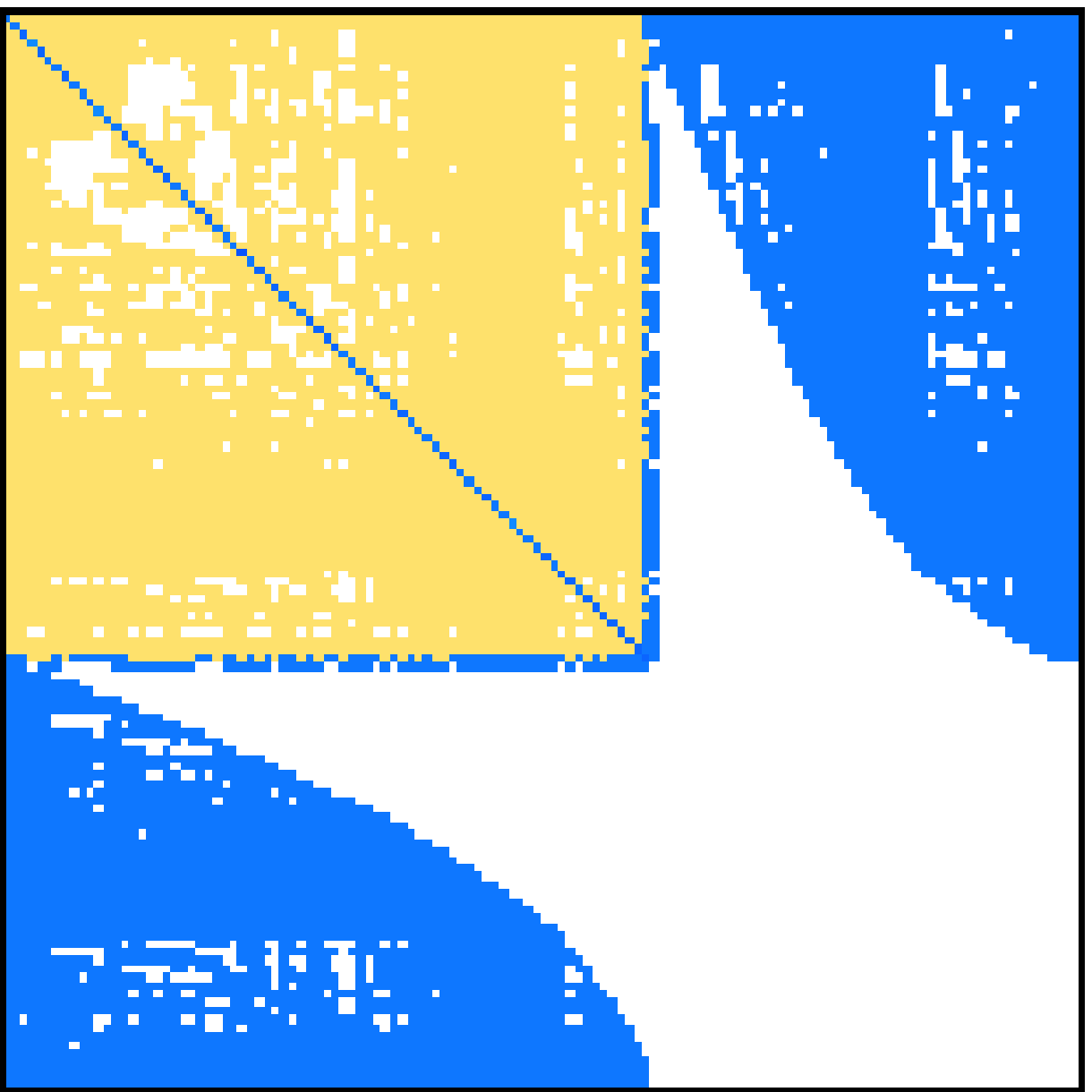} \end{minipage} 
& 390 K 
& \begin{minipage}[c]{0.17\columnwidth} \centering 2.1 M, 5 \end{minipage} 
& \begin{minipage}[c]{0.17\columnwidth} \centering 12.8 M, 33 \end{minipage} 
& \begin{minipage}[c]{0.17\columnwidth} \centering 6.4 M, 17 \end{minipage}
\\ 
Mono\_500Hz 
& \begin{minipage}[c]{0.045\columnwidth} \centering \includegraphics[width=0.24in]{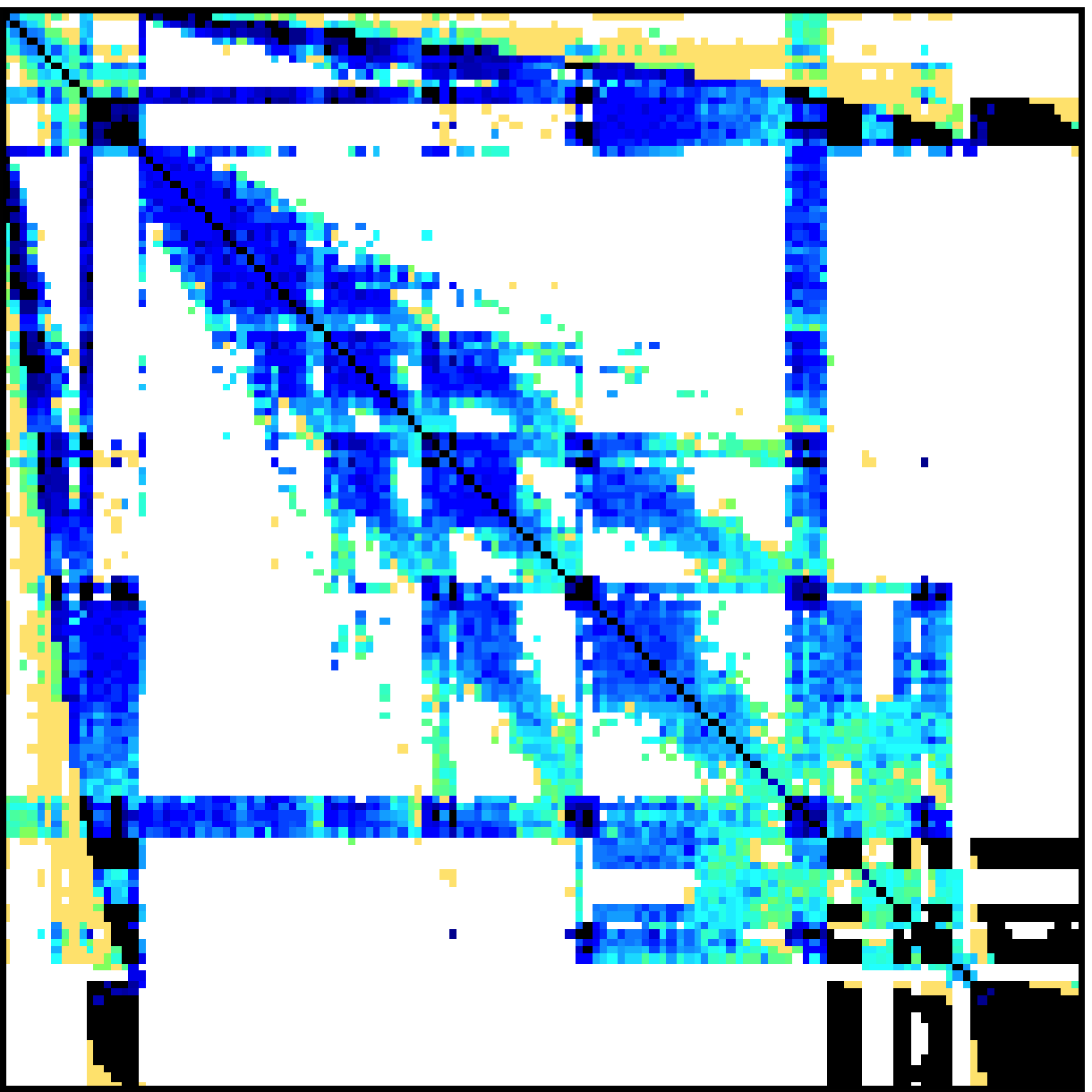} \end{minipage} 
& 169 K 
& \begin{minipage}[c]{0.17\columnwidth} \centering 5 M, 30 \end{minipage} 
& \begin{minipage}[c]{0.17\columnwidth} \centering 204 M, 1204 \end{minipage} 
& \begin{minipage}[c]{0.17\columnwidth} \centering 41.4 M, 244 \end{minipage}
\\ 
Offshore 
& \begin{minipage}[c]{0.045\columnwidth} \centering \includegraphics[width=0.24in]{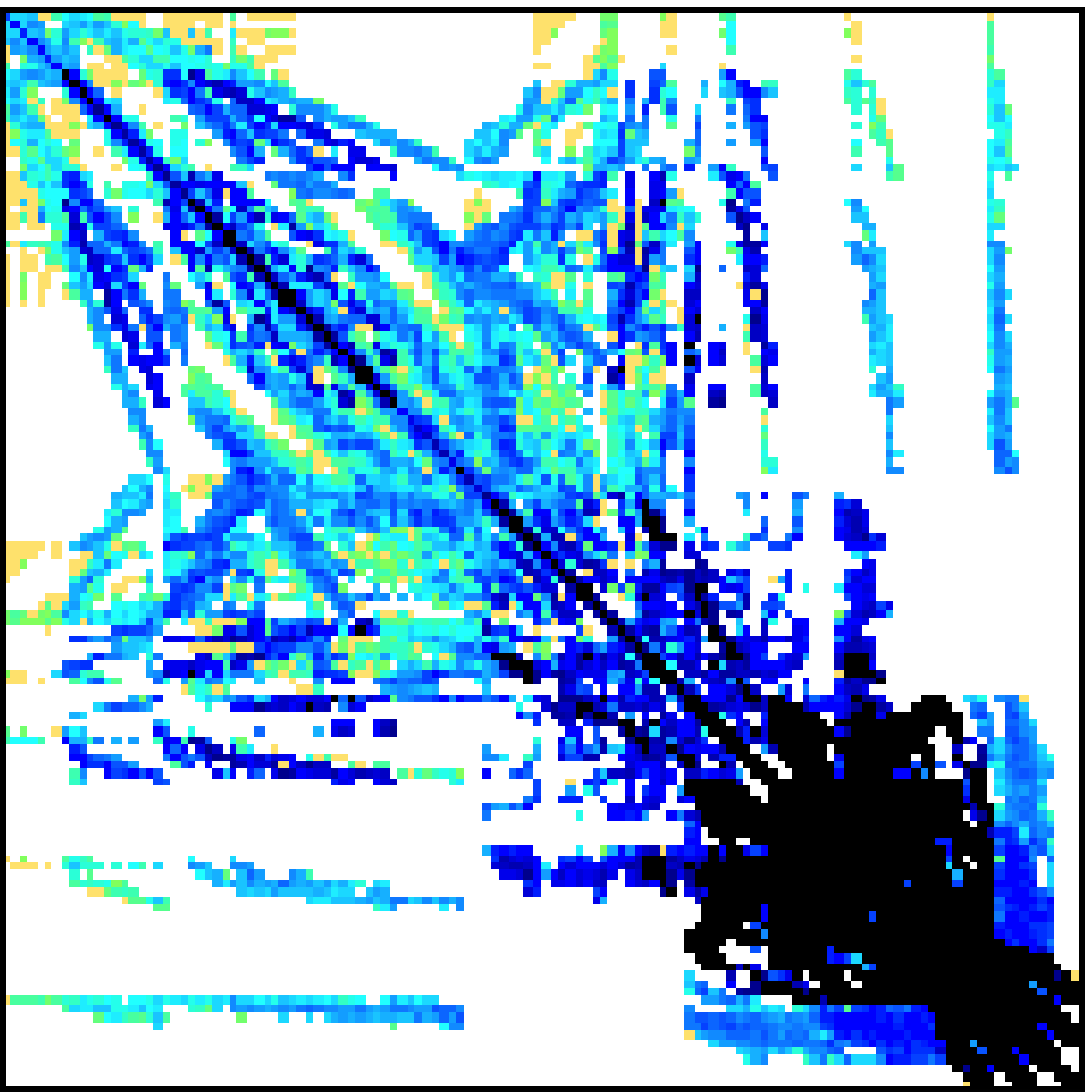} \end{minipage} 
& 260 K 
& \begin{minipage}[c]{0.17\columnwidth} \centering 4.2 M, 16 \end{minipage} 
& \begin{minipage}[c]{0.17\columnwidth} \centering 71.3 M, 275 \end{minipage} 
& \begin{minipage}[c]{0.17\columnwidth} \centering 23.4 M, 90 \end{minipage} 
\\ 
Patents\_main 
& \begin{minipage}[c]{0.045\columnwidth} \centering \includegraphics[width=0.24in]{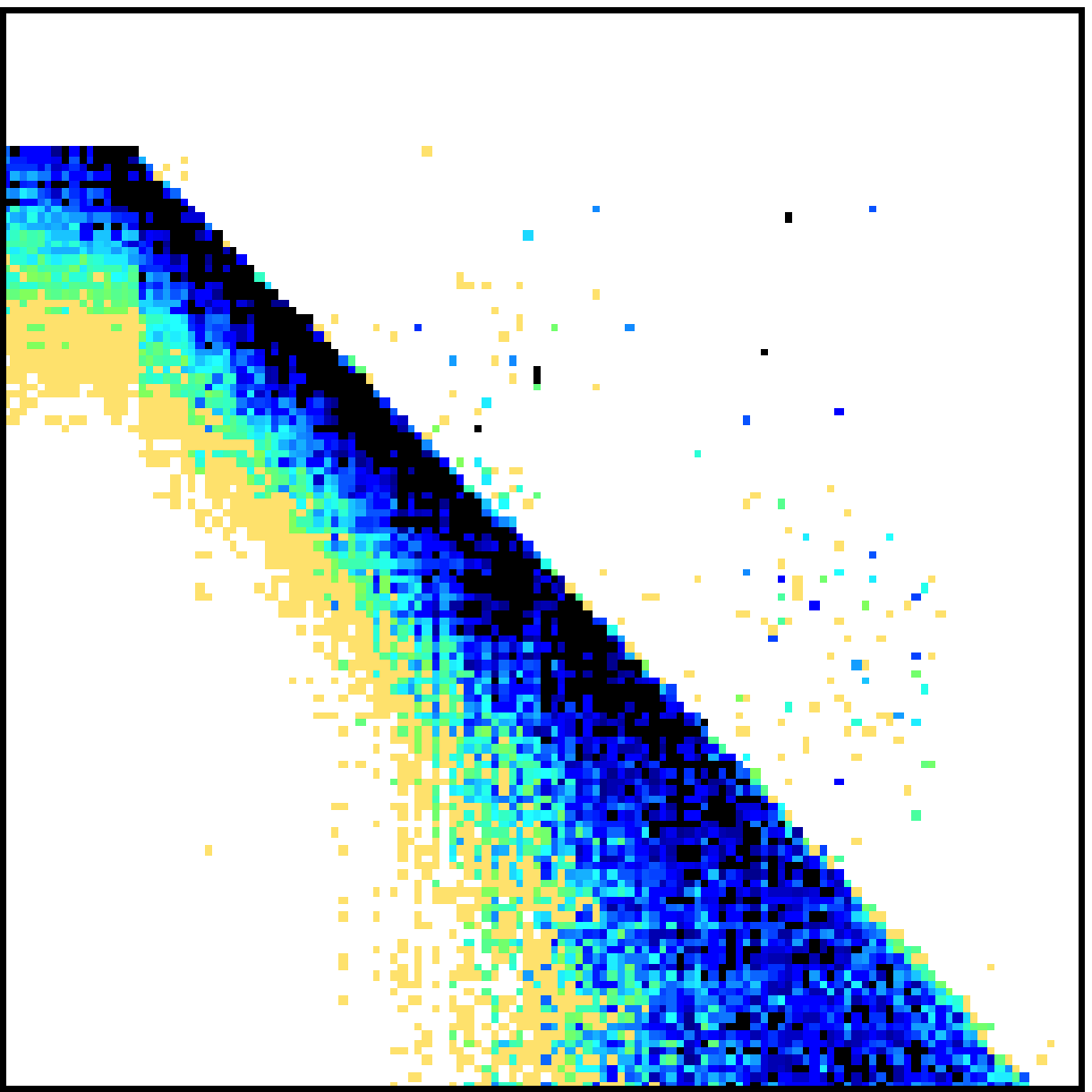} \end{minipage} 
& 241 K 
& \begin{minipage}[c]{0.17\columnwidth} \centering 0.6 M, 2 \end{minipage} 
& \begin{minipage}[c]{0.17\columnwidth} \centering 2.6 M, 11 \end{minipage} 
& \begin{minipage}[c]{0.17\columnwidth} \centering 2.3 M, 9 \end{minipage} 
\\ 
Poisson3Da 
& \begin{minipage}[c]{0.045\columnwidth} \centering \includegraphics[width=0.24in]{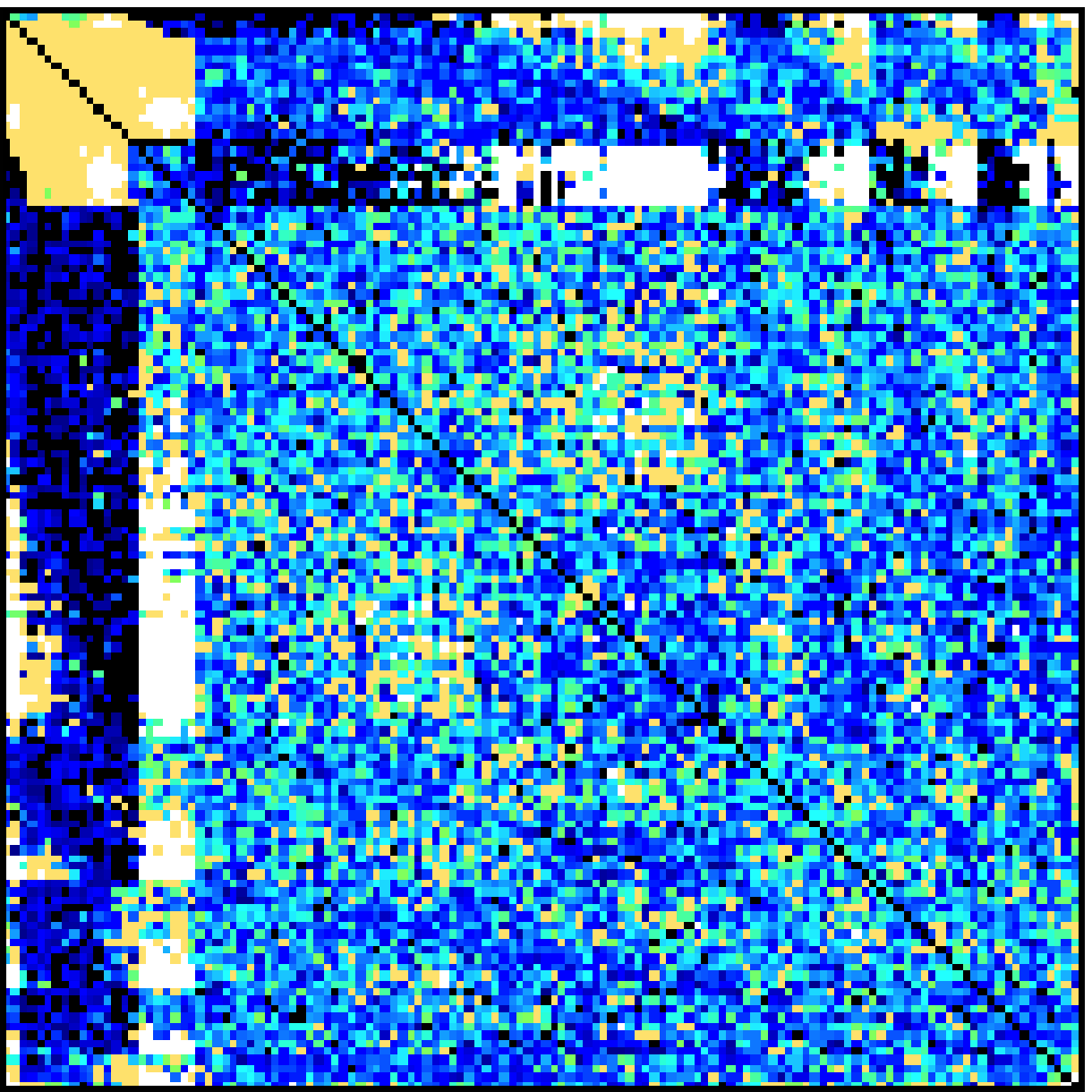} \end{minipage} 
& 14 K 
& \begin{minipage}[c]{0.17\columnwidth} \centering 0.4 M, 26 \end{minipage} 
& \begin{minipage}[c]{0.17\columnwidth} \centering 11.8 M, 871 \end{minipage} 
& \begin{minipage}[c]{0.17\columnwidth} \centering 3 M, 219 \end{minipage} 
\\ 
QCD 
& \begin{minipage}[c]{0.045\columnwidth} \centering \includegraphics[width=0.24in]{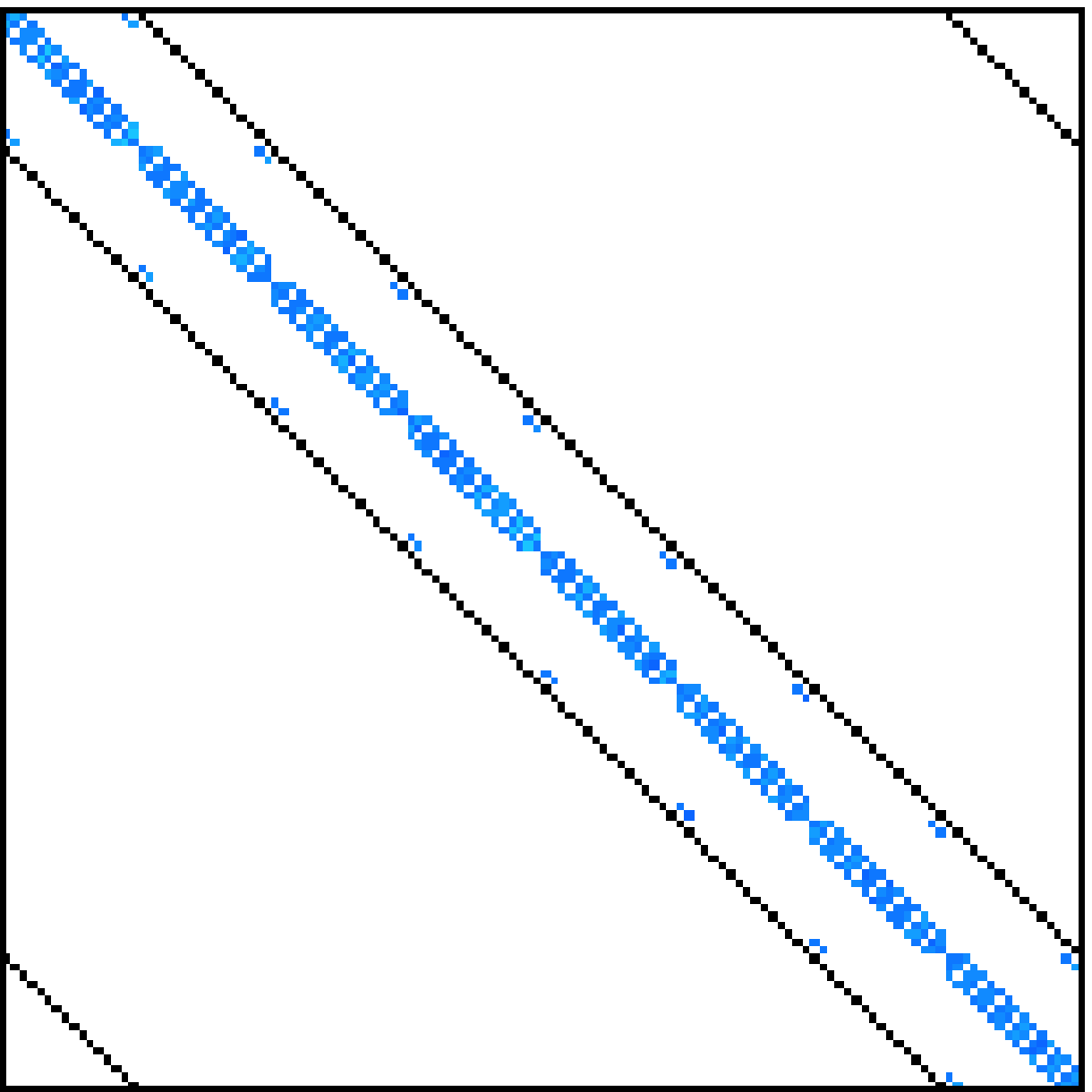} \end{minipage} 
& 49 K 
& \begin{minipage}[c]{0.17\columnwidth} \centering 1.9 M, 39 \end{minipage} 
& \begin{minipage}[c]{0.17\columnwidth} \centering 74.8 M, 1521 \end{minipage} 
& \begin{minipage}[c]{0.17\columnwidth} \centering 10.9 M, 222 \end{minipage} 
\\ 
Circuit 
& \begin{minipage}[c]{0.045\columnwidth} \centering \includegraphics[width=0.24in]{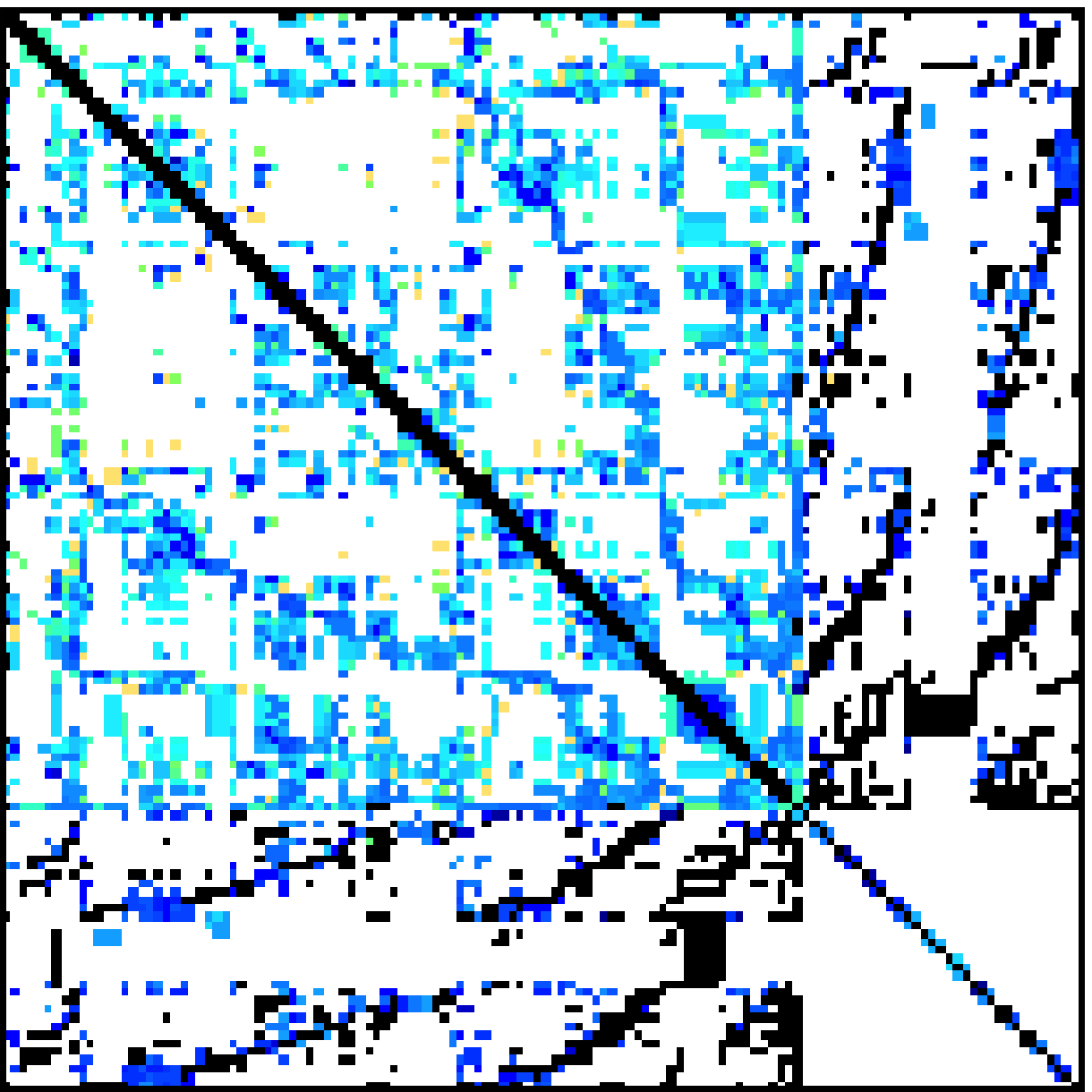} \end{minipage} 
& 171 K 
& \begin{minipage}[c]{0.17\columnwidth} \centering 1 M, 6 \end{minipage} 
& \begin{minipage}[c]{0.17\columnwidth} \centering 8.7 M, 51 \end{minipage} 
& \begin{minipage}[c]{0.17\columnwidth} \centering 5.2 M, 31 \end{minipage} 
\\ 
Webbase
& \begin{minipage}[c]{0.045\columnwidth} \centering \includegraphics[width=0.24in]{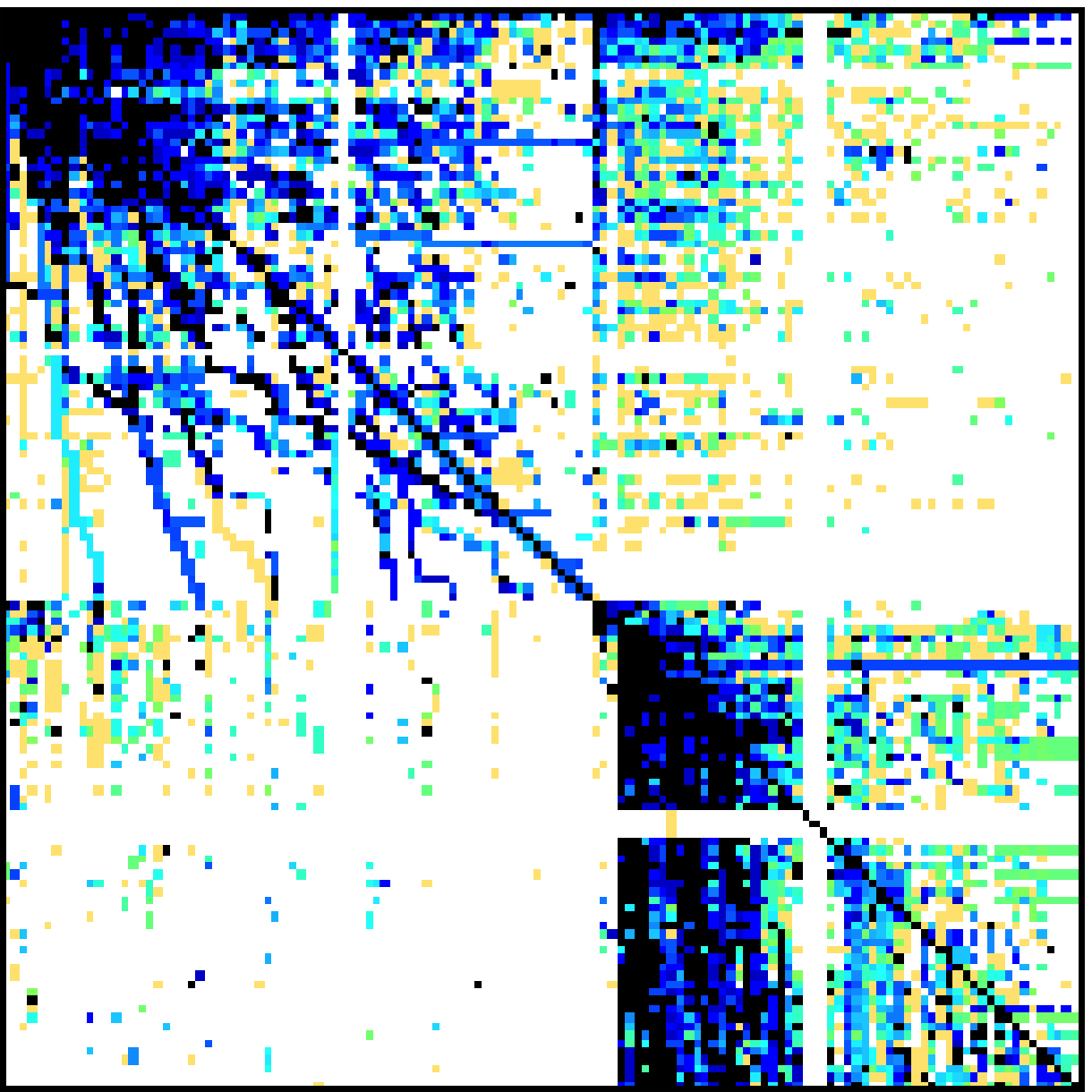} \end{minipage} 
& 1 M
& \begin{minipage}[c]{0.17\columnwidth} \centering 3.1 M, 3 \end{minipage} 
& \begin{minipage}[c]{0.17\columnwidth} \centering 69.5 M, 70 \end{minipage} 
& \begin{minipage}[c]{0.17\columnwidth} \centering 51.1 M, 51 \end{minipage} 
\\ 

\hline
\end{tabular}
\end{table*}

\subsection{Performance Comparison for Matrix Squaring}

\begin{figure*}[h!t]
\captionsetup[subfigure]{labelformat=empty}
\subfloat[]{\includegraphics[width=1.33in]{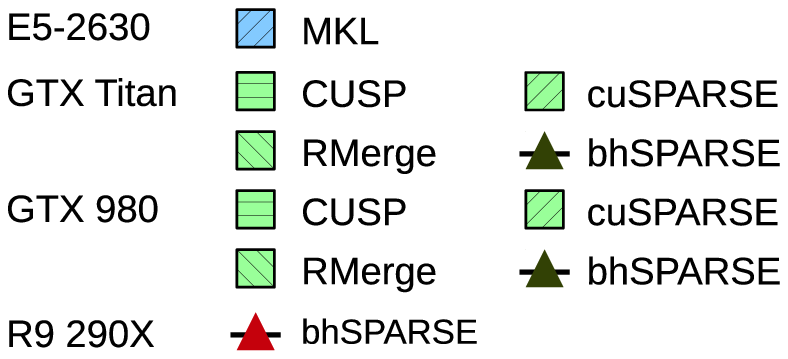}}
\subfloat[(a) FEM/Cantilever]{\includegraphics[width=1.33in]{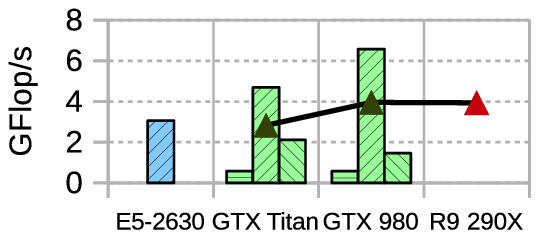}}
\subfloat[(b) Economics]{\includegraphics[width=1.33in]{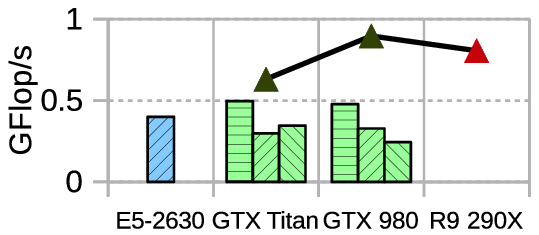}}
\subfloat[(c) Epidemiology]{\includegraphics[width=1.33in]{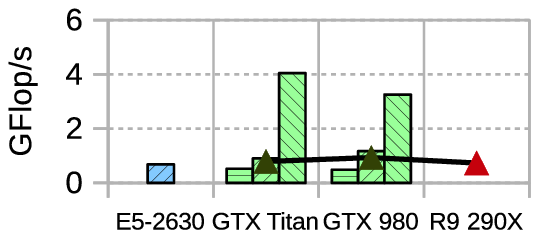}}
\linebreak
\vskip -6pt
\subfloat[(d) Filter3D]{\includegraphics[width=1.33in]{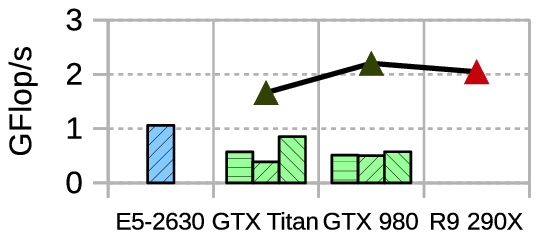}}
\subfloat[(e) Wind Tunnel]{\includegraphics[width=1.33in]{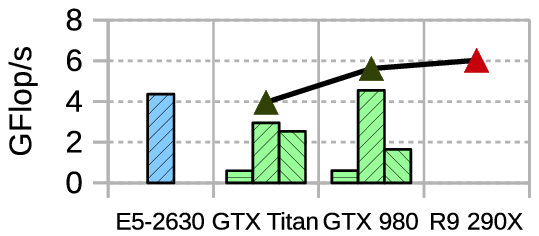}}
\subfloat[(f) FEM/Ship]{\includegraphics[width=1.33in]{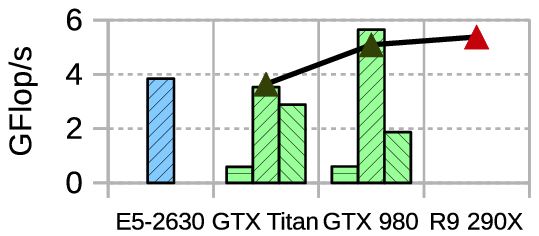}}
\subfloat[(g) FEM/Harbor]{\includegraphics[width=1.33in]{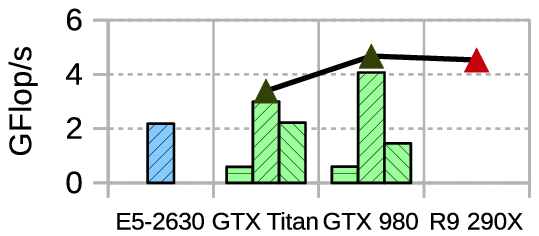}}
\linebreak
\vskip -6pt
\subfloat[(h) Protein]{\includegraphics[width=1.33in]{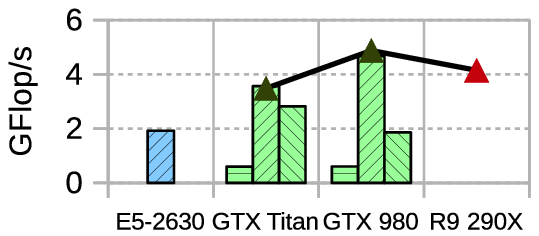}}
\subfloat[(i) FEM/Spheres]{\includegraphics[width=1.33in]{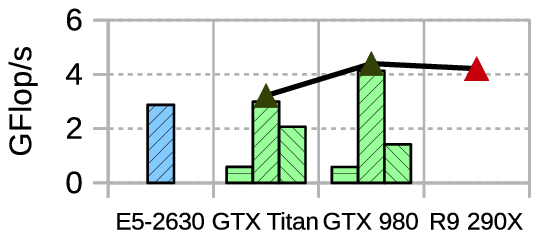}}
\subfloat[(j) 2cubes sphere]{\includegraphics[width=1.33in]{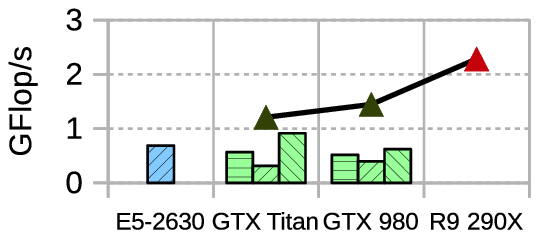}}
\subfloat[(k) FEM/Accelerator]{\includegraphics[width=1.33in]{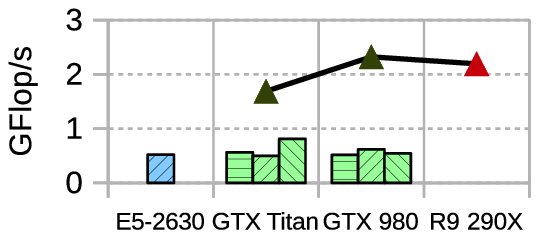}}
\linebreak
\vskip -6pt
\subfloat[(l) Cage12]{\includegraphics[width=1.33in]{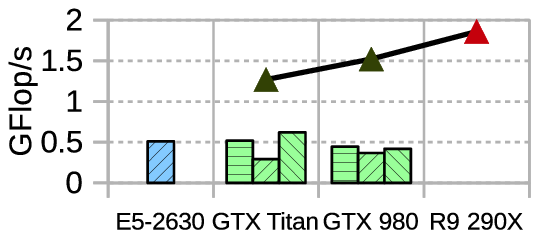}}
\subfloat[(m) Hood]{\includegraphics[width=1.33in]{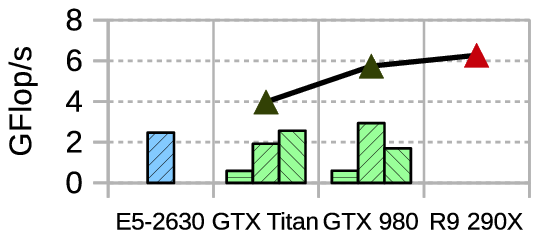}}
\subfloat[(n) M133-b3]{\includegraphics[width=1.33in]{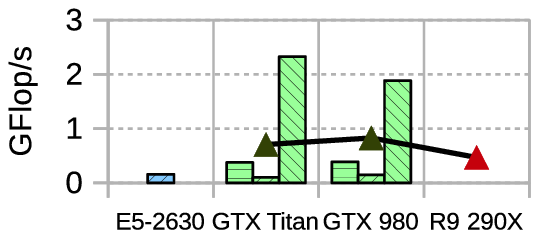}}
\subfloat[(o) Majorbasis]{\includegraphics[width=1.33in]{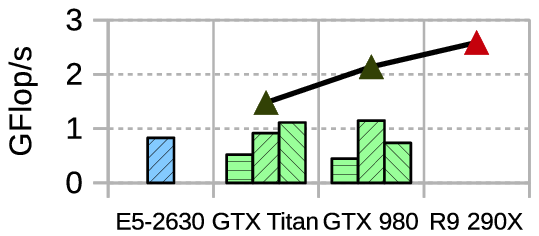}}
\linebreak
\vskip -6pt
\subfloat[(p) Mario002]{\includegraphics[width=1.33in]{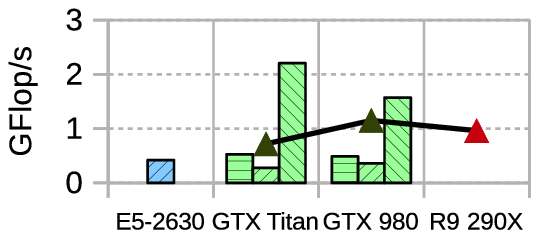}}
\subfloat[(q) Mono\_500Hz]{\includegraphics[width=1.33in]{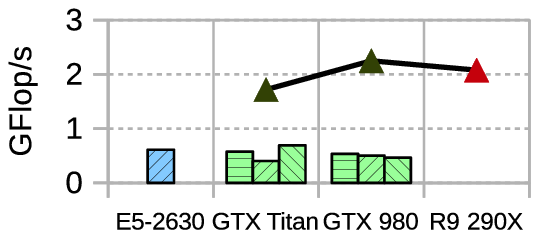}}
\subfloat[(r) Offshore]{\includegraphics[width=1.33in]{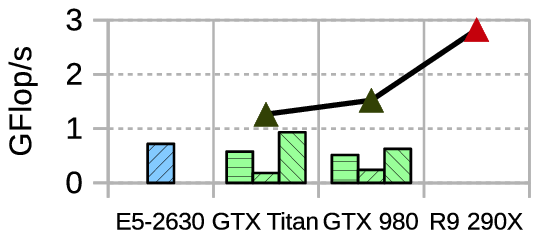}}
\subfloat[(s) Patents\_main]{\includegraphics[width=1.33in]{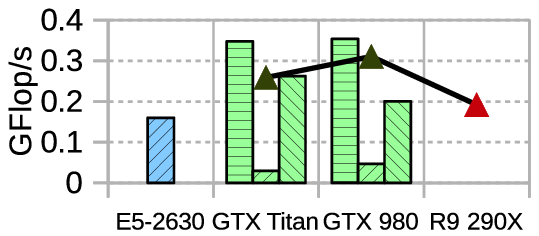}}
\linebreak
\vskip -6pt
\subfloat[(t) Poisson3Da]{\includegraphics[width=1.33in]{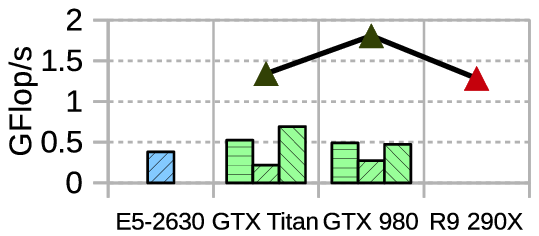}}
\subfloat[(u) QCD]{\includegraphics[width=1.33in]{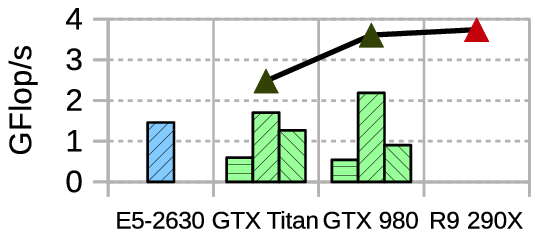}}
\subfloat[(v) Circuit]{\includegraphics[width=1.33in]{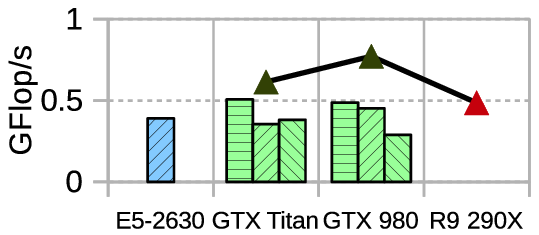}}
\subfloat[(w) Webbase]{\includegraphics[width=1.33in]{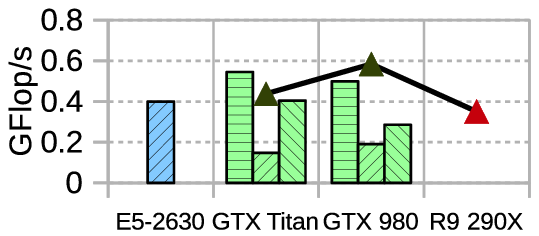}}
\caption{Single precision SpGEMM (SpSGEMM) GFlop/s comparison of 5 methods/libraries (MKL, CUSP, cuSPARSE, RMerge and bhSPARSE) on 4 platforms (Intel Xeon E5-2630, nVidia GeForce GTX Titan Black, nVidia GeForce GTX 980 and AMD Radeon R9 290X). The performance of bhSPARSE is shown by the points on the lines. The bars plot the throughout of the other tested approaches.}
\label{spgemm.jpdc.fig.spsgemm}
\end{figure*}

\begin{figure*}[h!t]
\captionsetup[subfigure]{labelformat=empty}
\subfloat[]{\includegraphics[width=1.33in]{legend}}
\subfloat[(a) FEM/Cantilever]{\includegraphics[width=1.33in]{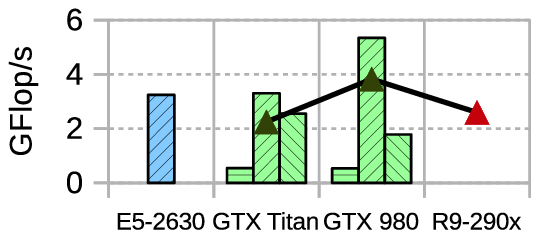}}
\subfloat[(b) Economics]{\includegraphics[width=1.33in]{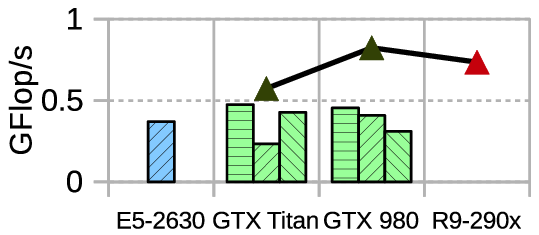}}
\subfloat[(c) Epidemiology]{\includegraphics[width=1.33in]{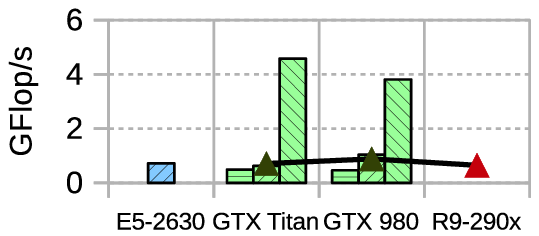}}
\linebreak
\vskip -6pt
\subfloat[(d) Filter3D]{\includegraphics[width=1.33in]{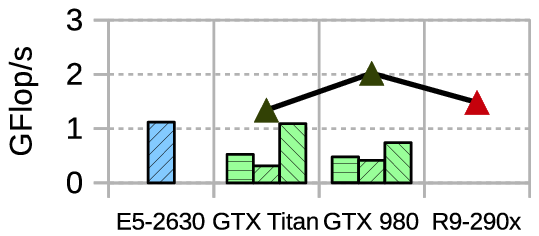}}
\subfloat[(e) Wind Tunnel]{\includegraphics[width=1.33in]{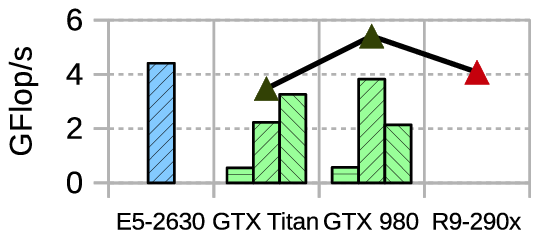}}
\subfloat[(f) FEM/Ship]{\includegraphics[width=1.33in]{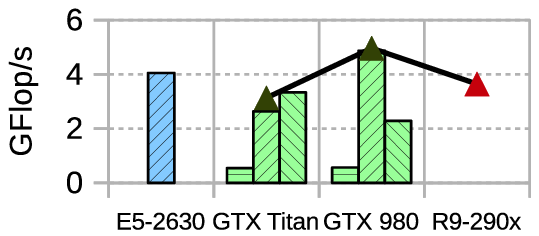}}
\subfloat[(g) FEM/Harbor]{\includegraphics[width=1.33in]{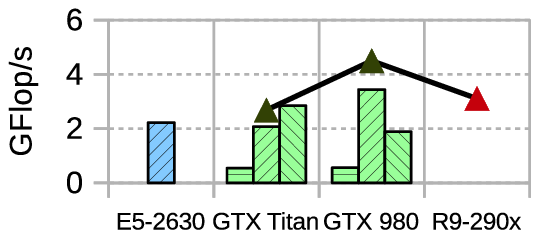}}
\linebreak
\vskip -6pt
\subfloat[(h) Protein]{\includegraphics[width=1.33in]{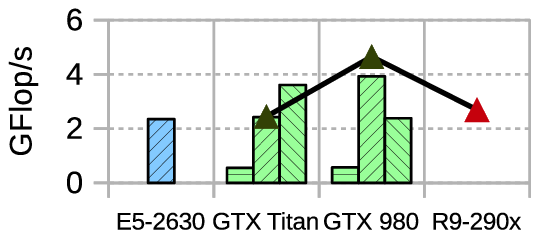}}
\subfloat[(i) FEM/Spheres]{\includegraphics[width=1.33in]{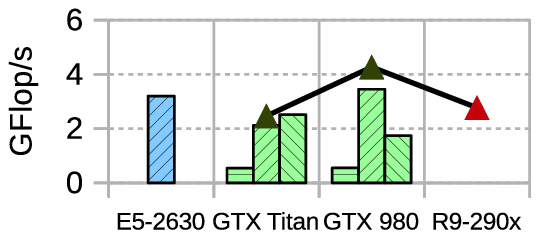}}
\subfloat[(j) 2cubes sphere]{\includegraphics[width=1.33in]{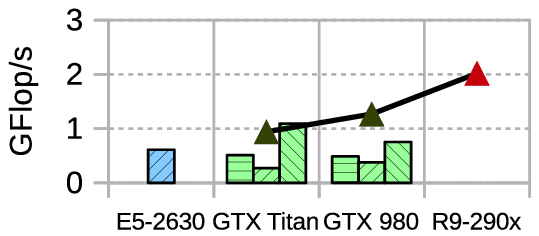}}
\subfloat[(k) FEM/Accelerator]{\includegraphics[width=1.33in]{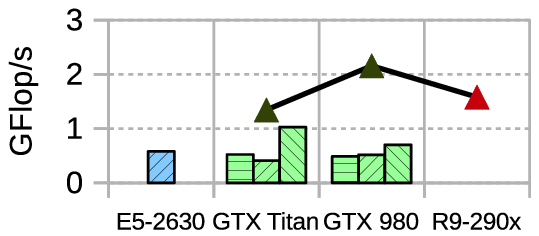}}
\linebreak
\vskip -6pt
\subfloat[(l) Cage12]{\includegraphics[width=1.33in]{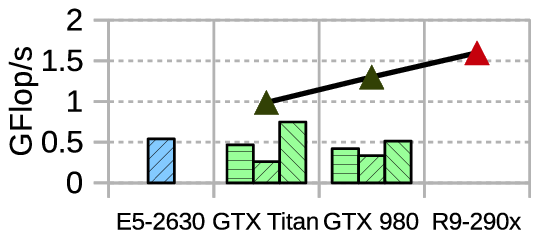}}
\subfloat[(m) Hood]{\includegraphics[width=1.33in]{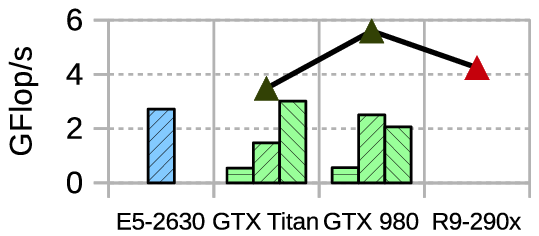}}
\subfloat[(n) M133-b3]{\includegraphics[width=1.33in]{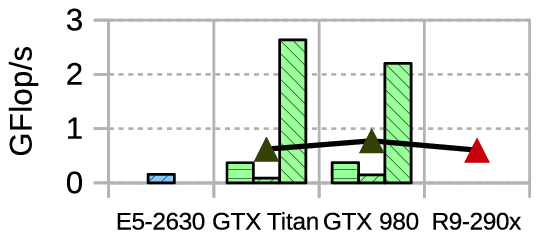}}
\subfloat[(o) Majorbasis]{\includegraphics[width=1.33in]{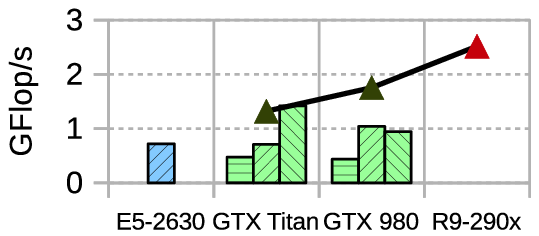}}
\linebreak
\vskip -6pt
\subfloat[(p) Mario002]{\includegraphics[width=1.33in]{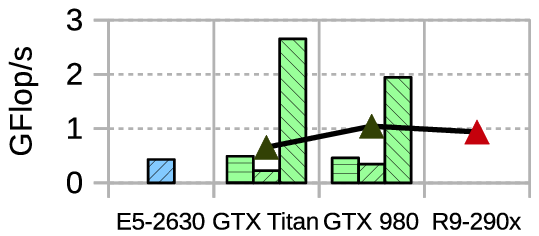}}
\subfloat[(q) Mono\_500Hz]{\includegraphics[width=1.33in]{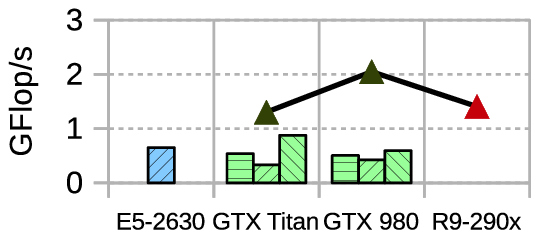}}
\subfloat[(r) Offshore]{\includegraphics[width=1.33in]{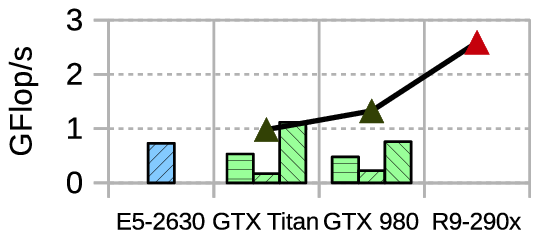}}
\subfloat[(s) Patents\_main]{\includegraphics[width=1.33in]{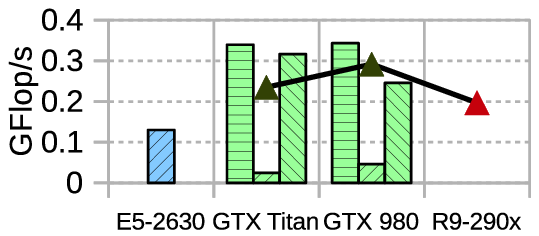}}
\linebreak
\vskip -6pt
\subfloat[(t) Poisson3Da]{\includegraphics[width=1.33in]{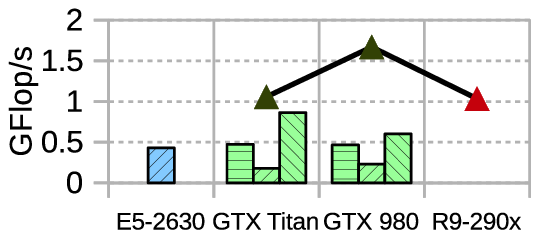}}
\subfloat[(u) QCD]{\includegraphics[width=1.33in]{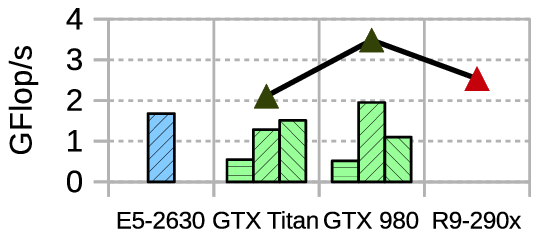}}
\subfloat[(v) Circuit]{\includegraphics[width=1.33in]{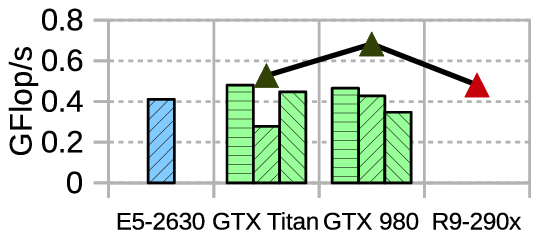}}
\subfloat[(w) Webbase]{\includegraphics[width=1.33in]{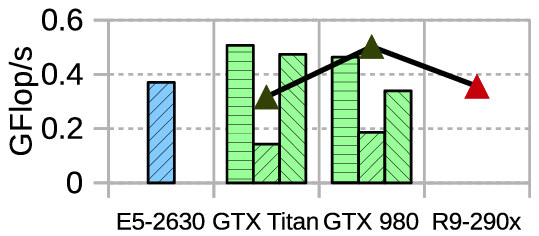}}
\caption{Double precision SpGEMM (SpDGEMM) GFlop/s comparison of 5 methods/libraries (MKL, CUSP, cuSPARSE, RMerge and bhSPARSE) on 4 platforms (Intel Xeon E5-2630, nVidia GeForce GTX Titan Black, nVidia GeForce GTX 980 and AMD Radeon R9 290X). The performance of bhSPARSE is shown by the points on the lines. The bars plot the throughout of the other tested approaches.}
\label{spgemm.jpdc.fig.spdgemm}
\end{figure*}

The single precision and double precision absolute performance of the SpGEMM algorithms that compute $C = A^2$ are shown in Figures~\ref{spgemm.jpdc.fig.spsgemm} and~\ref{spgemm.jpdc.fig.spdgemm}, respectively. Four GPU methods from CUSP v0.4.0, cuSPARSE v6.5, RMerge~\cite{Gremse:GPU} and bhSPARSE are evaluated on three GPUs: nVidia GeForce GTX Titan Black, nVidia GeForce GTX 980 and AMD Radeon R9 290X. One CPU method in Intel MKL v11.0 is evaluated on Intel Xeon E5-2630 CPU. The performance of another recent ESC-based GPU SpGEMM work~\cite{Dalton:Optimizing} is not included in the comparison because its source code is not available to us yet. The Intel MKL SpGEMM program is multithreaded and utilizes all six cores in the Intel Xeon CPU. For GPU algorithms, again, the host-device data transfer time is not included.

We first compare the performance of the four different GPU SpGEMM algorithms on the nVidia GPUs. We can see that bhSPARSE always outperforms CUSP, cuSPARSE and RMerge on most sparse matrices in the benchmark suite. Compared to the two vendor supplied libraries, our method obtains better SpSGEMM and SpDGEMM performance on 21 and 21 matrices out of the whole 23 matrices over CUSP, and on 19 and 21 matrices over cuSPARSE, respectively. Compared to RMerge, another CUDA-specific method, bhSPARSE achieves better SpSGEMM and SpDGEMM performance on 19 and 10 matrices on the GTX Titan Black GPU, and on 19 and 20 matrices on the GTX 980 GPU.

From the perspective of speedup, our method delivers on average 4.6x (up to 9.6x) and 3.1x (up to 8.8x) speedup on SpSGEMM performance over CUSP and cuSPARSE, and on average 4.6x (up to 9.9x) and 3.1x (up to 9.5x) speedup on SpDGEMM performance over them, respectively. Compared to RMerge, our method offers on average 1.4x (up to 2.5x) speedup and 2.8x (up to 4.9x) speedup for SpSGEMM and on average 1.0x (up to 1.5x) and 2.1x (up to 3.4x) speedup for SpDGEMM on the GTX Titan Black GPU and GTX 980 GPU, respectively.

We can see that the cuSPARSE method outperforms our approach when and only when the input matrices are fairly regular (belong to the first 9 matrices in Table~\ref{spgemm.jpdc.tab.benchmarksuite}). For all irregular matrices and some regular ones, our bhSPARSE is always more efficient. On the other hand, the absolute performance of the CUSP method is very stable since its execution time almost only depends on the number of necessary arithmetic operations. Therefore this approach is insensitive to sparsity structures. Actually this insensitivity may bring better performance on matrices with some specific sparsity structures. However in most cases, the CUSP method suffers with higher global memory pressure. The RMerge method offers significant speedups over the other methods on three matrices (i.e., \textit{Epidemiology}, \textit{M133-b3} and \textit{Mario002}), which are characterized by short rows. However, for the other matrices, RMerge supplies relatively lower performance due to imbalanced workload and high-overhead global memory operations between iterative steps. Further, we can see that since RMerge mainly relies on computational power of the SIMD units, its performance decreases from GTX Titan Black (2880 CUDA cores running at 889 MHz) to GTX 980 (2048 CUDA cores running at 1126 MHz). In contrast, our method also depends on capacity of scratchpad memory. Thus we can see that bhSPARSE obtains better performance while using GTX 980 (1536 kB scratchpad) over GTX Titan Black (720 kB scratchpad).

Compared to Intel MKL on the Intel CPU, our CUDA-based implementation on the nVidia GPUs obtains better SpSGEMM and SpDGEMM performance on all 23 matrices, and delivers on average 2.5x (up to 5.2x) and 2.2x (up to 4.9x) SpSGEMM and SpDGEMM speedup, respectively. Our OpenCL-based implementation on the AMD GPU in the machine 2 obtains better SpSGEMM and SpDGEMM performance on 23 and 18 matrices, and delivers on average 2.3x (up to 4.2x) and 1.9x (up to 3.8x) SpSGEMM and SpDGEMM speedup, respectively. 

\begin{figure*}[h!t]
\centering
\subfloat[Single precision SpGEMM]{\includegraphics[width=2.6in]{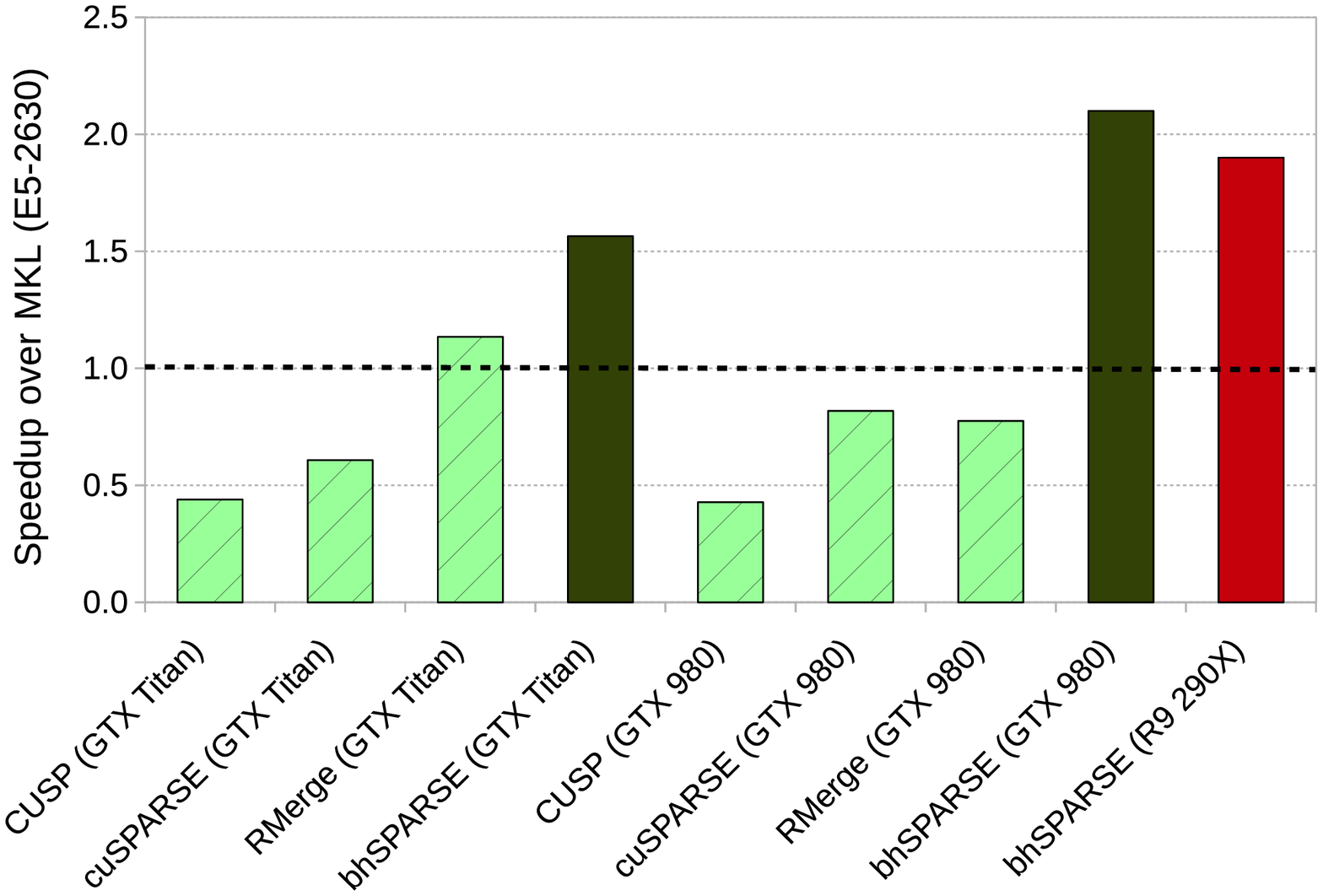}}
\subfloat[Double precision SpGEMM]{\includegraphics[width=2.6in]{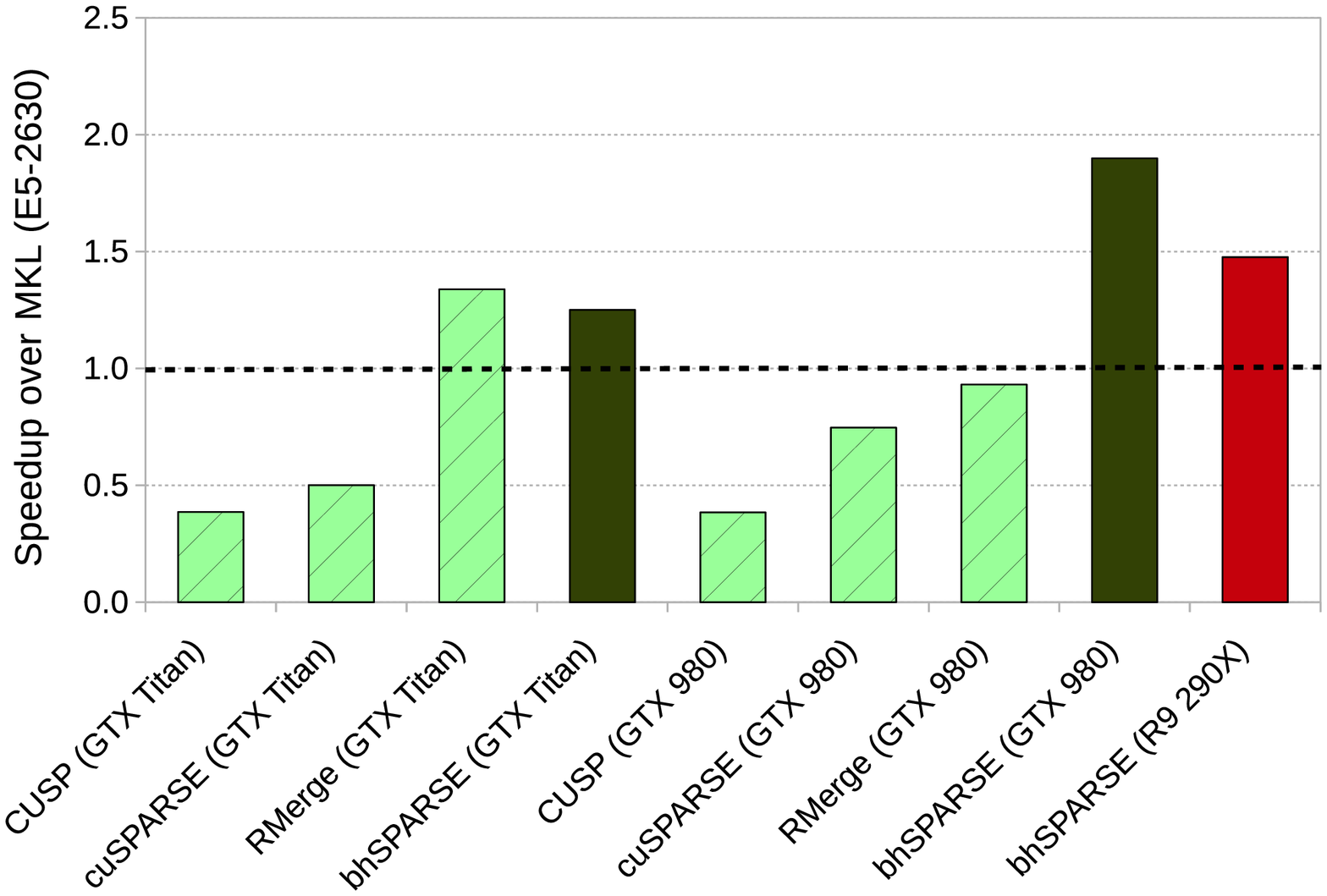}}
\caption{Average (harmonic mean) relative performance comparison of the 23 matrices, using SpGEMM method in MKL on Intel Xeon E5-2630 as a baseline.}
\label{spgemm.jpdc.fig.relativeperf}
\end{figure*}

The relative performance (harmonic mean) of the SpGEMM algorithms that compute $C = A^2$ is shown in Figure~\ref{spgemm.jpdc.fig.relativeperf}. We can see that our method in general delivers the best performance on the used testbeds while running the 23 matrices as a benchmark suite. If we set the Intel MKL SpGEMM performance in this scenario as a baseline, our approach is the only GPU SpGEMM that constantly outperforms well optimized CPU method.

\subsection{Memory Pre-allocation Comparison}
\begin{figure*}[h!t]
\centering
\subfloat[Absolute memory requirement]{\includegraphics[width=5.3in]{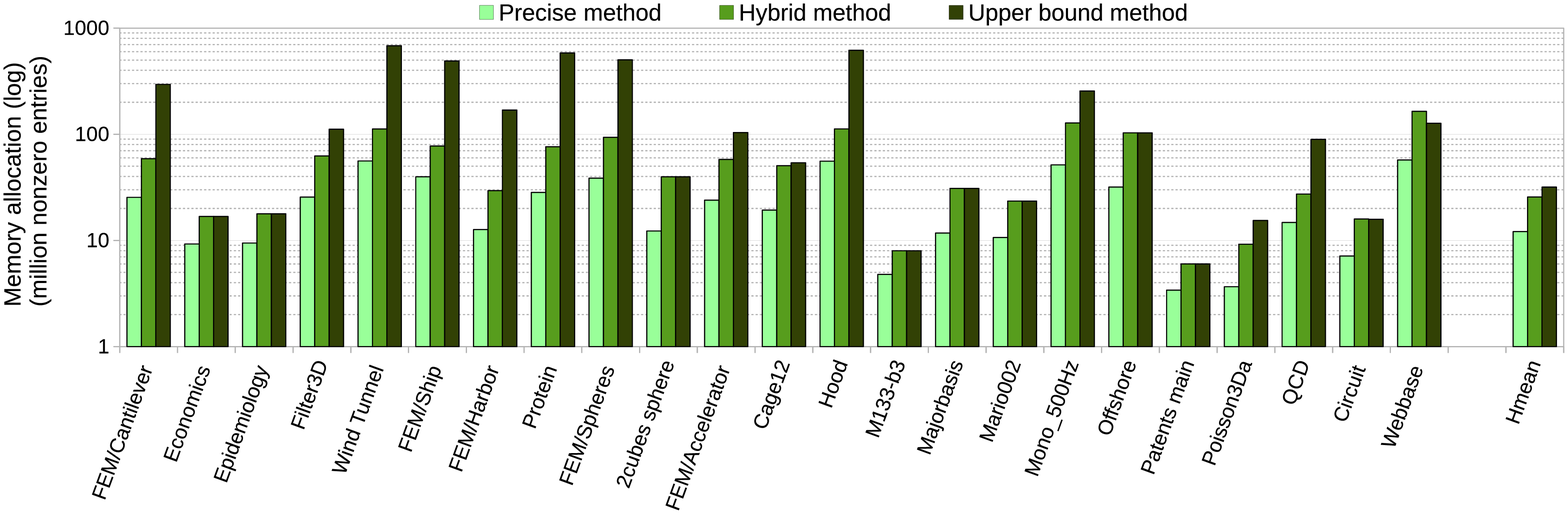}}\quad
\subfloat[Relative memory requirement]{\includegraphics[width=5.3in]{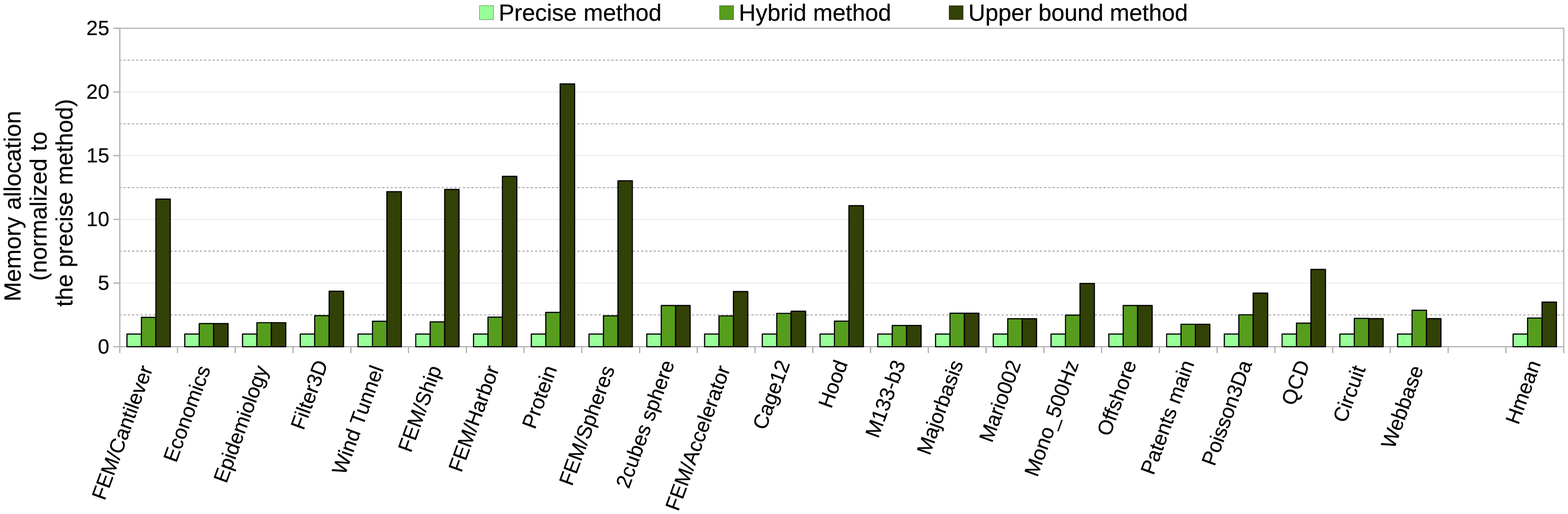}}
\caption{Global memory requirement comparison of the precise method, our hybrid method and the upper bound method, when benchmarking $C=A^2$ on the 23 matrices. The memory requirement of the precise method includes the two input matrices and the resulting matrix. The memory requirements of the other two methods also contain additional intermediate matrices. ``Hmean'' refers to harmonic mean.}
\label{spgemm.jpdc.fig.space}
\end{figure*}

Figure~\ref{spgemm.jpdc.fig.space} shows the comparison of the three memory pre-allocation methods, while benchmarking $C=A^2$. We can see that, for small matrices (e.g., \textit{2cubes\_sphere}), our hybrid method shows exactly the same space requirements as the upper bound method does. However, for large matrices, allocated memory sizes through our hybrid method are much closer to the memory sizes allocated by the precise method. Taking the matrix \textit{Protein} as an example, our hybrid method requires 2.7x memory space over the precise method, while the upper bound method needs 20.6x space requirement. One exception is the matrix \textit{Webbase}, our hybrid method actually allocates more memory space than the upper bound method. The reasons are that the reduced rate of the intermediate matrix $\widehat{C}$ to the resulting matrix $C$ is very low (see Table~\ref{spgemm.jpdc.tab.benchmarksuite}) and our 2x progression mechanism just allocates memory across the upper bound size. But overall, our hybrid method saves space allocation of the upper bound method and execution time of the precise method without introducing any significant extra space requirements.

\subsection{Using Re-allocatable Memory}

For some matrices with relatively long rows in the bin group 5, our method dumps scratchpad data to global memory, allocates a larger memory block, copies the old data to the newly allocated portion, reloads values and continues processing. We have to do the allocation/copy operation pair and pay the overhead since current GPUs are not able to re-allocate memory (i.e., change the size of the memory block pointed to a certain pointer). However, the emerging heterogeneous processors with shared virtual memory (or unified memory) address space deliver a possibility that lets integrated GPUs use system memory, which is re-allocatable from the CPU side. 

\begin{figure*}[h!t]
\centering
\subfloat[Single precision SpGEMM]{\includegraphics[width=2.6in]{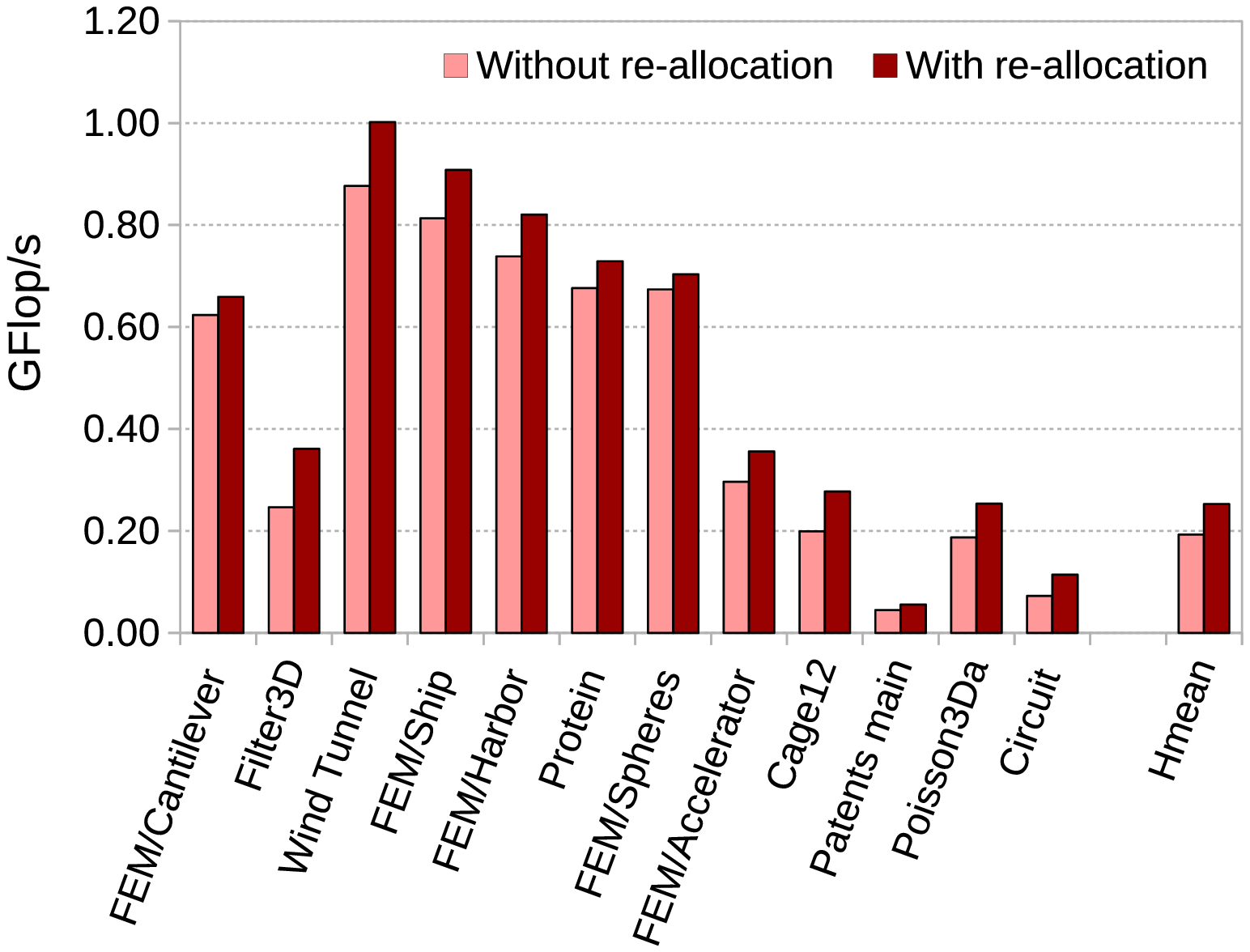}}
\subfloat[Double precision SpGEMM]{\includegraphics[width=2.6in]{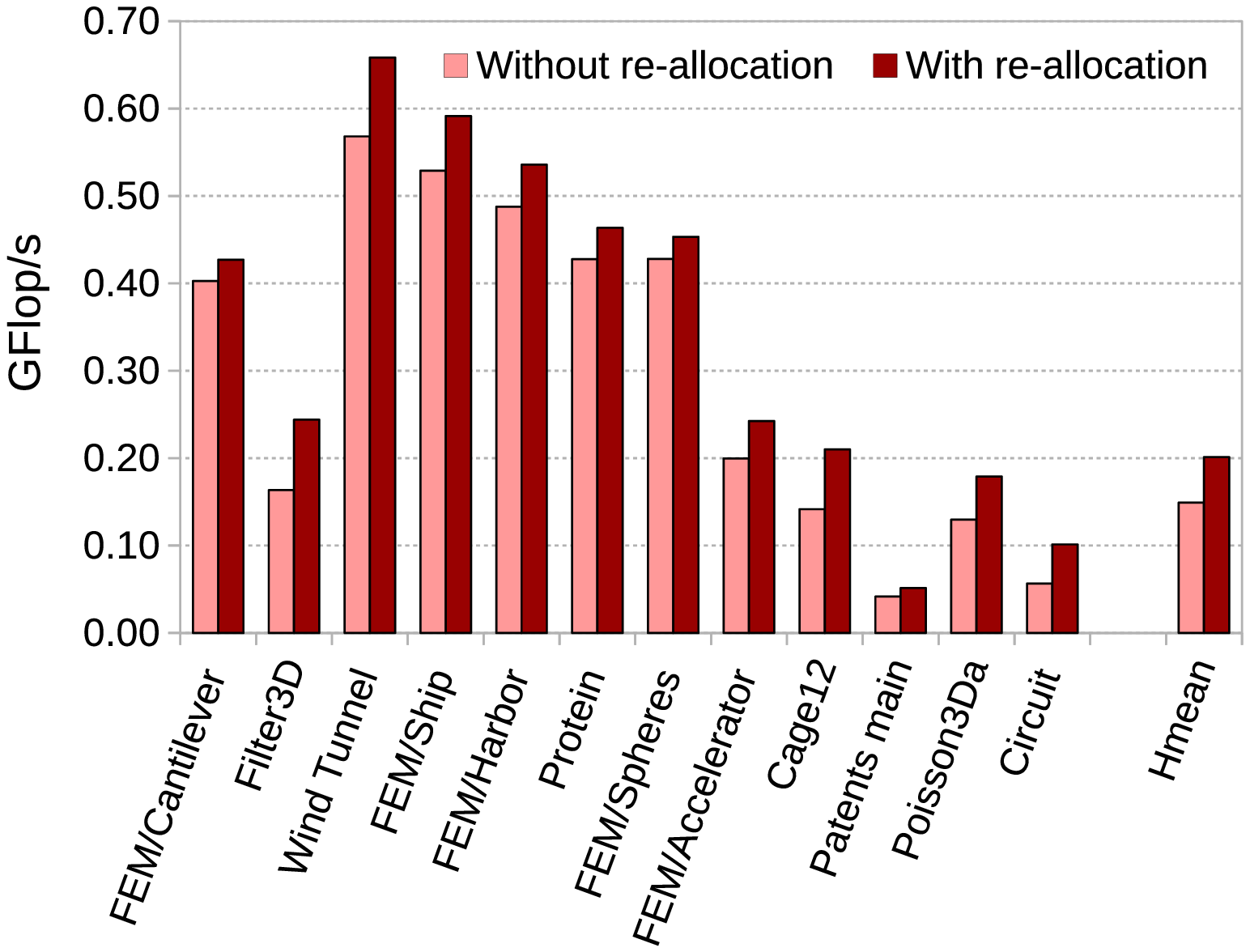}}
\caption{$C=A^2$ performance of bhSPARSE running with and without re-allocatable memory on an AMD A10-7850K APU. Note that only executable matrices that require memory re-allocation are included here. ``Hmean'' refers to harmonic mean.}
\label{spgemm.jpdc.fig.apuperf}
\end{figure*}

We evaluated two memory allocation strategies (i.e., a typical allocation/copy approach and an improved re-allocation approach) of our OpenCL-based SpGEMM algorithm on the GPU part (8 GCN core, 512 Radeon cores running at 1028 MHz, 1052.7 GFlop/s SP peak, 65.8 GFlop/s DP peak) in the AMD A10-7850K APU. Figure~\ref{spgemm.jpdc.fig.apuperf} shows the results. We can see that re-allocatable memory brings on average 1.2x (up to 1.6x) speedup and on average 1.2x (up to 1.8x) speedup for SpSGEMM and SpDGEMM, respectively. Therefore, our GPU SpGEMM method may deliver further performance improvement on future GPUs with re-allocatable memory, or on emerging heterogeneous processors composed of CPU cores and GPU cores. Moreover, both CPU cores and GPU cores can be utilized for Stage 3 in our framework. We leave this heterogenous workload partitioning (similar to the methods described in~\cite{Shen:An, Shen:Improving}) to future work.

\section{Conclusion}
In this paper we demonstrated an efficient SpGEMM framework and corresponding algorithms on GPUs and emerging CPU-GPU heterogeneous processors for solving the three challenging problems in the SpGEMM. In the two experimental scenarios using matrices with diverse sparsity structures as input, our  SpGEMM algorithm delivered excellent absolute and relative performance as well as space efficiency over the previous GPU SpGEMM methods. Moreover, on average, our approach obtained around twice the performance of the start-of-the-art CPU SpGEMM method. Further, we showed that our method obtained higher performance on emerging heterogeneous processors with re-allocatable memory.





\section*{Acknowledgments}
The authors would like to thank Jianbin Fang at the Delft University of Technology for supplying access to the machine with the Intel Xeon CPU. The authors further thank Felix Gremse at the RWTH Aachen University for sharing source code of the RMerge algorithm and a preprint copy of the corresponding paper~\cite{Gremse:GPU}. The authors also thank the anonymous reviewers of JPDC and IPDPS '14 for their insightful feedback on this version and a shorter version~\cite{Liu:An} of this paper.


\begin{thebibliography}{10}
\expandafter\ifx\csname url\endcsname\relax
  \def\url#1{\texttt{#1}}\fi
\expandafter\ifx\csname urlprefix\endcsname\relax\def\urlprefix{URL }\fi
\expandafter\ifx\csname href\endcsname\relax
  \def\href#1#2{#2} \def\path#1{#1}\fi

\bibitem{Liu:An}
W.~Liu, B.~Vinter, {An Efficient GPU General Sparse Matrix-Matrix
  Multiplication for Irregular Data}, in: Proceedings of the 2014 IEEE 28th
  International Parallel and Distributed Processing Symposium, IPDPS '14, 2014,
  pp. 370--381.

\bibitem{Bell:Exposing}
N.~Bell, S.~Dalton, L.~Olson, {Exposing Fine-Grained Parallelism in Algebraic
  Multigrid Methods}, SIAM Journal on Scientific Computing 34~(4) (2012)
  C123--C152.

\bibitem{Gilbert:High}
J.~Gilbert, S.~Reinhardt, V.~Shah, {High-Performance Graph Algorithms from
  Parallel Sparse Matrices}, in: B.~K{\aa}gstr{\"o}m, E.~Elmroth, J.~Dongarra,
  J.~Wa{\'s}niewski (Eds.), Applied Parallel Computing. State of the Art in
  Scientific Computing, Vol. 4699 of Lecture Notes in Computer Science,
  Springer Berlin Heidelberg, 2007, pp. 260--269.

\bibitem{Chan:More}
T.~M. Chan, {More Algorithms for All-pairs Shortest Paths in Weighted Graphs},
  in: Proceedings of the Thirty-ninth Annual ACM Symposium on Theory of
  Computing, STOC '07, 2007, pp. 590--598.

\bibitem{Kaplan:Colored}
H.~Kaplan, M.~Sharir, E.~Verbin, {Colored Intersection Searching via Sparse
  Rectangular Matrix Multiplication}, in: Proceedings of the Twenty-second
  Annual Symposium on Computational Geometry, SCG '06, 2006, pp. 52--60.

\bibitem{Vassilevska:Finding}
V.~Vassilevska, R.~Williams, R.~Yuster, {Finding Heaviest H-subgraphs in Real
  Weighted Graphs, with Applications}, ACM Trans. Algorithms 6~(3) (2010)
  44:1--44:23.

\bibitem{Bell:Implementing}
N.~Bell, M.~Garland, {Implementing Sparse Matrix-Vector Multiplication on
  Throughput-Oriented Processors}, in: Proceedings of the Conference on High
  Performance Computing Networking, Storage and Analysis, SC '09, 2009, pp.
  18:1--18:11.

\bibitem{Liu:CSR5}
W.~Liu, B.~Vinter, {CSR5: An Efficient Storage Format for Cross-Platform Sparse
  Matrix-Vector Multiplication}, in: Proceedings of the 29th ACM on
  International Conference on Supercomputing, ICS '15, 2015, pp. 339--350.

\bibitem{Liu:Speculative}
W.~Liu, B.~Vinter, {Speculative Segmented Sum for Sparse Matrix-Vector
  Multiplication on Heterogeneous Processors}, Parallel Computing (2015) --.

\bibitem{Ortega:FastSpMM}
G.~Ortega, F.~V{\'a}zquez, I.~Garc{\'\i}a, E.~M. Garz{\'o}n, {FastSpMM: An
  Efficient Library for Sparse Matrix Matrix Product on GPUs}, The Computer
  Journal 57~(7) (2014) 968--979.

\bibitem{Vazquez:Fast}
F.~Vazquez, G.~Ortega, J.~Fernandez, I.~Garcia, E.~Garzon, {Fast Sparse Matrix
  Matrix Product Based on ELLR-T and GPU Computing}, in: Parallel and
  Distributed Processing with Applications (ISPA), 2012 IEEE 10th International
  Symposium on, 2012, pp. 669--674.

\bibitem{Demouth:Sparse}
J.~Demouth, {Sparse Matrix-Matrix Multiplication on the GPU}, Tech. rep.,
  NVIDIA (2012).

\bibitem{NVIDIA:CUSPARSE}
NVIDIA, \href{https://developer.nvidia.com/cuSPARSE}{{NVIDIA cuSPARSE
  library}}.
\newline\urlprefix\url{https://developer.nvidia.com/cuSPARSE}

\bibitem{Dalton:CUSP}
S.~Dalton, N.~Bell, \href{http://cusplibrary.github.com}{{CUSP : A C++
  Templated Sparse Matrix Library}}.
\newline\urlprefix\url{http://cusplibrary.github.com}

\bibitem{Dalton:Optimizing}
S.~Dalton, L.~Olsen, N.~Bell, {Optimizing Sparse Matrix-Matrix Multiplication
  for the GPU}, {ACM} Transactions on Mathematical Software 41~(4).

\bibitem{Gremse:GPU}
F.~Gremse, A.~H{\"o}fter, L.~O. Schwen, F.~Kiessling, U.~Naumann,
  {GPU-Accelerated Sparse Matrix-Matrix Multiplication by Iterative Row
  Merging}, SIAM Journal on Scientific Computing 37~(1) (2015) C54--C71.

\bibitem{Intel:MKL}
Intel, \href{http://software.intel.com/en-us/intel-mkl}{{Intel Math Kernel
  Library}}.
\newline\urlprefix\url{http://software.intel.com/en-us/intel-mkl}

\bibitem{Gilbert:Sparse}
J.~Gilbert, C.~Moler, R.~Schreiber, {Sparse Matrices in MATLAB: Design and
  Implementation}, SIAM Journal on Matrix Analysis and Applications 13~(1)
  (1992) 333--356.

\bibitem{Gustavson:Two}
F.~G. Gustavson, {Two Fast Algorithms for Sparse Matrices: Multiplication and
  Permuted Transposition}, ACM Trans. Math. Softw. 4~(3) (1978) 250--269.

\bibitem{Matam:Sparse}
K.~Matam, S.~Indarapu, K.~Kothapalli, {Sparse Matrix-Matrix Multiplication on
  Modern Architectures}, in: High Performance Computing (HiPC), 2012 19th
  International Conference on, 2012, pp. 1--10.

\bibitem{Sulatycke:Caching}
P.~Sulatycke, K.~Ghose, {Caching-Efficient Multithreaded Fast Multiplication of
  Sparse Matrices}, in: Parallel Processing Symposium, 1998. IPPS/SPDP 1998.
  Proceedings of the First Merged International Parallel Processing Symposium
  and Symposium on Parallel and Distributed Processing 1998, 1998, pp.
  117--123.

\bibitem{Yuster:Fast}
R.~Yuster, U.~Zwick, {Fast Sparse Matrix Multiplication}, ACM Trans. Algorithms
  1~(1) (2005) 2--13.

\bibitem{Buluc:On}
A.~Bulu{\c c}, J.~Gilbert, {On the Representation and Multiplication of
  Hypersparse Matrices}, in: Parallel and Distributed Processing, 2008. IPDPS
  2008. IEEE International Symposium on, 2008, pp. 1--11.

\bibitem{Fang:A}
J.~Fang, A.~Varbanescu, H.~Sips, {A Comprehensive Performance Comparison of
  CUDA and OpenCL}, in: Parallel Processing (ICPP), 2011 International
  Conference on, 2011, pp. 216--225.

\bibitem{Saule:Performance}
E.~Saule, K.~Kaya, {\"{U}}.~V. {\c{C}}ataly{\"{u}}rek, {Performance Evaluation
  of Sparse Matrix Multiplication Kernels on Intel Xeon Phi}, in: Proc of the
  10th Int'l Conf. on Parallel Processing and Applied Mathematics (PPAM), 2013.

\bibitem{Williams:Optimization}
S.~Williams, L.~Oliker, R.~Vuduc, J.~Shalf, K.~Yelick, J.~Demmel, {Optimization
  of Sparse Matrix-Vector Multiplication on Emerging Multicore Platforms}, in:
  Supercomputing, 2007. SC '07. Proceedings of the 2007 ACM/IEEE Conference on,
  2007, pp. 1--12.

\bibitem{Buluc:Reduced}
A.~Bulu{\c c}, S.~Williams, L.~Oliker, J.~Demmel, {Reduced-Bandwidth
  Multithreaded Algorithms for Sparse Matrix-Vector Multiplication}, in:
  Parallel Distributed Processing Symposium (IPDPS), 2011 IEEE International,
  2011, pp. 721--733.

\bibitem{Amossen:Better}
R.~R. Amossen, A.~Campagna, R.~Pagh, {Better Size Estimation for Sparse Matrix
  Products}, Algorithmica (2014) 741--757.

\bibitem{Cohen:On}
E.~Cohen, {On Optimizing Multiplications of Sparse Matrices}, in:
  W.~Cunningham, S.~McCormick, M.~Queyranne (Eds.), Integer Programming and
  Combinatorial Optimization, Vol. 1084 of Lecture Notes in Computer Science,
  Springer Berlin Heidelberg, 1996, pp. 219--233.

\bibitem{Pagh:The}
R.~Pagh, M.~St{\"o}ckel, {The Input/Output Complexity of Sparse Matrix
  Multiplication}, in: A.~Schulz, D.~Wagner (Eds.), Algorithms - ESA 2014,
  Lecture Notes in Computer Science, Springer Berlin Heidelberg, 2014, pp.
  750--761.

\bibitem{Branover:Llano}
A.~Branover, D.~Foley, M.~Steinman, {AMD Fusion APU: Llano}, IEEE Micro 32~(2)
  (2012) 28--37.

\bibitem{AMD:Compute}
AMD, {White Paper: Compute Cores} (jan 2014).

\bibitem{Damaraju:22nm}
S.~Damaraju, V.~George, S.~Jahagirdar, T.~Khondker, R.~Milstrey, S.~Sarkar,
  S.~Siers, I.~Stolero, A.~Subbiah, {A 22nm IA multi-CPU and GPU
  System-on-Chip}, in: Solid-State Circuits Conference Digest of Technical
  Papers (ISSCC), 2012 IEEE International, 2012, pp. 56--57.

\bibitem{Keckler:GPUs}
S.~Keckler, W.~Dally, B.~Khailany, M.~Garland, D.~Glasco, {GPUs and the Future
  of Parallel Computing}, Micro, IEEE 31~(5) (2011) 7--17.

\bibitem{nVidia:Tegra}
nVidia, {NVIDIA Tegra K1 A New Era in Mobile Computing}, 1st Edition (jan
  2014).

\bibitem{Qualcomm:Snapdragon}
Qualcomm, {Qualcomm Snapdragon 800 Product Brief} (aug 2013).

\bibitem{HSA:Manual}
HSA Foundation, {HSA Programmer's Reference Manual: HSAIL Virtual ISA and
  Programming Model, Compiler Writer's Guide, and Object Format (BRIG)}, 0th
  Edition (may 2013).

\bibitem{Munshi:The}
A.~Munshi, The OpenCL Specification, Khronos OpenCL Working Group, 2nd Edition
  (mar 2014).

\bibitem{Gregg:Where}
C.~Gregg, K.~Hazelwood, {Where is the Data? Why You Cannot Debate CPU vs. GPU
  Performance Without the Answer}, in: Performance Analysis of Systems and
  Software (ISPASS), 2011 IEEE International Symposium on, 2011, pp. 134--144.

\bibitem{Kim:Sort}
C.~Kim, T.~Kaldewey, V.~W. Lee, E.~Sedlar, A.~D. Nguyen, N.~Satish,
  J.~Chhugani, A.~Di~Blas, P.~Dubey, {Sort vs. Hash Revisited: Fast Join
  Implementation on Modern Multi-core CPUs}, Proc. VLDB Endow. 2~(2) (2009)
  1378--1389.

\bibitem{Ballard:Communication}
G.~Ballard, A.~Bulu{\c c}, J.~Demmel, L.~Grigori, B.~Lipshitz, O.~Schwartz,
  S.~Toledo, {Communication Optimal Parallel Multiplication of Sparse Random
  Matrices}, in: Proceedings of the 25th ACM Symposium on Parallelism in
  Algorithms and Architectures, SPAA '13, 2013, pp. 222--231.

\bibitem{Buluc:Parallel}
A.~Bulu{\c c}, J.~Gilbert, {Parallel Sparse Matrix-Matrix Multiplication and
  Indexing: Implementation and Experiments}, SIAM Journal on Scientific
  Computing 34~(4) (2012) C170--C191.

\bibitem{Gilbert:Ordered}
J.~R. Gilbert, W.~W. Pugh, T.~Shpeisman, {Ordered Sparse Accumulator and its
  Use in Efficient Sparse Matrix Computation}, United States Patent US 5983230
  A (nov 1999).

\bibitem{Satish:Designing}
N.~Satish, M.~Harris, M.~Garland, {Designing Efficient Sorting Algorithms for
  Manycore GPUs}, in: Parallel Distributed Processing, 2009. IPDPS 2009. IEEE
  International Symposium on, 2009, pp. 1--10.

\bibitem{Green:GPU}
O.~Green, R.~McColl, D.~A. Bader, {GPU Merge Path: A GPU Merging Algorithm},
  in: Proceedings of the 26th ACM International Conference on Supercomputing,
  ICS '12, 2012, pp. 331--340.

\bibitem{Kipfer:Improved}
P.~Kipfer, R.~Westermann, {Improved GPU Sorting}, in: M.~Pharr (Ed.), GPU Gems
  2: Programming Techniques for High-Performance Graphics and General-Purpose
  Computation, Addison-Wesley, 2005, Ch.~46, pp. 733--746.

\bibitem{Peters:Comparison}
H.~Peters, O.~Schulz-Hildebrandt, {Comparison-Based In-Place Sorting with
  CUDA}, in: W.-M. Hwu (Ed.), GPU Computing Gems Jade Edition, Morgan Kaufmann,
  2011, Ch.~8, pp. 89--96.

\bibitem{Peters:A}
H.~Peters, O.~Schulz-Hildebrandt, N.~Luttenberger, {A Novel Sorting Algorithm
  for Many-Core Architectures Based on Adaptive Bitonic Sort}, in: Parallel
  Distributed Processing Symposium (IPDPS), 2012 IEEE 26th International, 2012,
  pp. 227--237.

\bibitem{Inoue:AA}
H.~Inoue, T.~Moriyama, H.~Komatsu, T.~Nakatani, {AA-Sort: A New Parallel
  Sorting Algorithm for Multi-Core SIMD Processors}, in: Parallel Architecture
  and Compilation Techniques, 2007. PACT 2007. 16th International Conference
  on, 2007, pp. 189--198.

\bibitem{Davidson:Efficient}
A.~Davidson, D.~Tarjan, M.~Garland, J.~Owens, {Efficient Parallel Merge Sort
  for Fixed and Variable Length Keys}, in: Innovative Parallel Computing
  (InPar), 2012, 2012, pp. 1--9.

\bibitem{Baxter:Modern}
S.~Baxter, \href{http://www.moderngpu.com/}{{Modern GPU Library}}.
\newline\urlprefix\url{http://www.moderngpu.com/}

\bibitem{Kristensen:Bohrium}
M.~R.~B. Kristensen, S.~A.~F. Lund, T.~Blum, K.~Skovhede, B.~Vinter, {Bohrium:
  A Virtual Machine Approach to Portable Parallelism}, in: Proceedings of the
  2014 IEEE International Parallel \& Distributed Processing Symposium
  Workshops, 2014, pp. 312--321.

\bibitem{Buluc:Challenges}
A.~Bulu{\c c}, J.~R. Gilbert, {Challenges and Advances in Parallel Sparse
  Matrix-Matrix Multiplication}, in: Proceedings of the 2008 37th International
  Conference on Parallel Processing, ICPP '08, 2008, pp. 503--510.

\bibitem{Davis:The}
T.~A. Davis, Y.~Hu, {The University of Florida Sparse Matrix Collection}, ACM
  Trans. Math. Softw. 38~(1) (2011) 1:1--1:25.

\bibitem{Shen:An}
J.~Shen, J.~Fang, A.~L. Varbanescu, H.~Sips, {An Application-Centric Evaluation
  of OpenCL on Multi-Core CPUs}, Parallel Computing 39~(12) (2013) 834--850.

\bibitem{Shen:Improving}
J.~Shen, A.~L. Varbanescu, P.~Zou, Y.~Lu, H.~Sips, {Improving Performance by
  Matching Imbalanced Workloads with Heterogeneous Platforms}, in: Proceedings
  of the 28th ACM International Conference on Supercomputing, ICS '14, 2014,
  pp. 241--250.

\end{thebibliography}






\end{document}